\documentclass[manuscript,screen,nonacm]{acmart}

\renewcommand\footnotetextcopyrightpermission[1]{} 
\AtBeginDocument{%
  \providecommand\BibTeX{{%
    \normalfont B\kern-0.5em{\scshape i\kern-0.25em b}\kern-0.8em\TeX}}}

\thanks{This article is accepted in ACM Transactions on Software Engineering and Methodology (TOSEM)}

\PassOptionsToPackage{dvipsnames,table, prologue, svgnames, x11names, usenames}{xcolor}
\usepackage{xcolor}
\definecolor{Gr}{rgb}{0.0, 0.5, 0.0}
\usepackage{amsmath, balance}
\usepackage{scalerel}
\usepackage{tikz}
\usetikzlibrary{svg.path}
\usepackage{tabularx} 

\usepackage{arydshln}
\usepackage{caption}
\usepackage{academicons}
\usepackage{tabularray}  
\usepackage{lipsum}  
\usepackage{adjustbox}
\usepackage{mdframed}

\usepackage{longtable}
\usepackage{subcaption}
\usepackage{multirow}
\usepackage{multicol}
\usepackage{ctable}
\usepackage{threeparttable} 

\usepackage{booktabs}
\usepackage{array}
\usepackage{ragged2e}
\usepackage{soul}

\usepackage[normalem]{ulem}

\usepackage{tikz}
\usetikzlibrary{shapes.geometric, arrows, positioning} 

\tikzstyle{process} = [rectangle, minimum width=3cm, minimum height=1cm, text centered, draw=black, fill=orange!30]
\tikzstyle{decision} = [diamond, minimum width=3cm, minimum height=1cm, text centered, draw=black, fill=green!30]
\tikzstyle{arrow} = [thick,->,>=stealth]

\usepackage[breakable]{tcolorbox} 
\usepackage{enumitem}

\usepackage[normalem]{ulem} 
\definecolor{light-gray}{gray}{0.95}

\definecolor{purple}{HTML}{AF72B0}

\begin{document}

\title{A Survey on LLM-based Code Generation for Low-Resource and Domain-Specific Programming Languages}

\author{Sathvik~Joel}
\authornote{Both authors contributed equally to this research.}
\email{ksjoe30@gmail.com}
\orcid{0009-0006-8716-1430}
\affiliation{%
  \institution{Indian Institute of Technology Madras, Chennai}
  \country{India}
}

\author{Jie JW Wu}
\authornotemark[1]
\email{jie.jw.wu@acm.org}
\orcid{0000-0002-7895-2023}
\affiliation{%
  \institution{University of British Columbia, Kelowna}
  \state{British Columbia}
  \country{Canada}}

\author{Fatemeh~Fard}
\email{fatemeh.fard@ubc.ca}
\orcid{0000-0002-4505-6257}
\affiliation{%
  \institution{University of British Columbia, Kelowna}
  \state{British Columbia}
  \country{Canada}}

\renewcommand{\shortauthors}{Joel, Wu, and Fard}

\begin{abstract}

Large Language Models (LLMs) have shown remarkable capabilities in code generation for popular programming languages. However, their performance in Low-Resource Programming Languages (LRPLs) and Domain-Specific Languages (DSLs) remains a critical challenge. This gap affects millions of developers - with Rust alone having 3.5 million users - who are currently unable to fully leverage LLM capabilities. LRPLs and DSLs face unique challenges, including severe data scarcity and, for DSLs, highly specialized syntax and semantics that are poorly represented in general-purpose datasets. 
Addressing these challenges is crucial as LRPLs and DSLs significantly enhance development efficiency in specialized domains and applications, including financial and scientific works.
While several surveys on LLMs for software engineering and code exist, none comprehensively address the challenges and opportunities specific to LRPLs and DSLs. 
Our survey fills this gap by providing a systematic review of the current state, methodologies, and challenges in leveraging LLMs for code generation in LRPL and DSL. 
We filtered $111$ papers from over $27,000$ published studies from $2020 - 2024$ to understand the capabilities and limitations of LLMs in these specialized domains. We also expanded our literature search to include 5 recent papers from $2024 - 2025$. We report LLMs used, benchmarks, and metrics to evaluate code generation in LRPLs and DSLs, as well as strategies used to enhance LLM performance, and the collected datasets and curation methods in this context.

We identified four main evaluation techniques used in the literature, along with several metrics to assess code generation in LRPL and DSL. 
We categorized the methods used for LLM improvement into six main groups and summarized the novel methods and architectures proposed by the researchers. 
We also classified different approaches used for data collection and preparation.
While different techniques, metrics, and datasets are used, there is a lack of a standard approach and a benchmark dataset to evaluate code generation in several LRPLs and DSLs.
We discuss several distinctions of the studied approaches with the ones used in high-resource programming languages (HRPLs), as well as several challenges unique to these languages, especially DSLs. The challenges stem from the scarcity of data, the unique requirements, and specialized domains, which often need expertise guidelines or domain-specific tools. Accordingly, we provide insights into different research opportunities for the studied aspects. 
This survey serves as a comprehensive resource for researchers and practitioners working at the intersection of LLMs, software engineering, and specialized programming languages, providing a foundation for future advancements in LRPL and DSL code generation. A GitHub repository was created to organize the papers of this survey at \href{https://github.com/jie-jw-wu/Survey-CodeLLM4LowResource-DSL}{https://github.com/jie-jw-wu/Survey-CodeLLM4LowResource-DSL}.

\end{abstract}



\maketitle


\section{Introduction}

Large Language Models (LLMs), models trained on large amounts of data, have introduced a new paradigm in the software development life cycle \cite{hou2023large}. 
Among all software engineering tasks, LLMs or LLMs trained on code datasets are widely used for code generation, given a natural language summary of the desired code functionality~\citep{tian2023chatgpt, githubsurvey}.
Developers can leverage LLMs to generate code in many programming languages to boost their efficiency~\citep{peng2023impact} and even solve complex coding challenges~\citep{alphacode}.
However, the data used to train these models are often absent for low-resource languages~\citep{ID84paper}.  
Accordingly, data for Low-Resource Programming Languages (LRPLs) like Rust and R or Domain-Specific Programming Languages (DSLs) like Ansible and Verilog remains scarce. 
LRPL refers to those programming languages that are characterized by low availability online, and the same is reflected in the training datasets used for LLMs~\citep{lrpl}.
DSLs are languages that are tailored for specific application domains, offering advantages in expressiveness and ease of use compared to general-purpose languages within their domain~\citep{dspl}.
The performance disparity is starkly illustrated by the results of the MultiPL-E benchmark \cite{ID7paper}, with significantly higher scores for Python compared to LRPLs for most LLMs (see Figure~\ref{fig:heatmap}).
This disparity can be attributed to the limited training data available for LRPLs.

LRPLs and DSLs have a large population of developers and are widely used for various applications. 
SlashData's State of the Developer Nation report (25th Edition) highlights the extensive use of LRPLs~\footnote{\url{https://www.developernation.net/resources/reports/state-of-the-developer-nation-25th-edition-q3-20231/}}. 
For example, the developer population for Rust stands at $3.5$ million, Swift at $4.5$ million, Dart at $2.9$ million, Ruby at $2.3$ million, and Lua at $1.6$ million. 
Additionally, languages like COBOL \cite{sammet1978early}, Fortran~\citep{ID62paper}, R~\citep{staples2022expansion}, Rust~\citep{bugden2022rust}, Ansible, and Verilog~\citep{ID5paper, thakur2023verigenlargelanguagemodel} are widely used for applications from IoT to hardware systems.
Despite this significant user base and wide application areas, existing literature mainly focuses on the applications of LLMs for High-Resource Programming Languages (HRPLs)~\citep{hossain2024deepdivelargelanguage, zhang2023surveys1}.
The exclusion of LRPLs and DSLs is also seen in several related literature reviews that study LLMs for Software Engineering (SE) from different perspectives~\citep{zhang2023surveys1,unifyingSURVEY, unifyingSURVEY, OPENSURVEY, FUTURESURVEY, nl2code, hou2024large, watson2021systematicliteraturereviewuse,yang2020surveydeeplearningsoftware}. This is despite the fact that recent studies have highlighted the growing interest in using LLMs for code generation in LRPLs \citep{ID116paper}.

To address this knowledge gap, in this study, we conducted a Systematic Literature Review (SLR) exploring the landscape of code generation with LLMs for LRPLs and DSLs, studying information from $111$ papers from a pool of over $27,000$ papers. 
We investigate strategies, metrics, and benchmarks for enhancing and testing the LLM performance in these specialized contexts, including dataset collection and processing methodologies for less common languages.
Our study contributes valuable insights that complement and extend the knowledge base established by previous works.
Our findings reveal that there is a need for standard benchmark datasets and evaluation metrics, curating LRPL and DSL datasets, and developing new techniques and models to improve code generation for LRPL and DSLs. 
We provide categorization of different metrics and techniques, as well as discussing challenges and future directions of research, providing a roadmap for advancing LLM's code generation for LRPL/DSL. We have also created a GitHub repository where we organize the papers of this survey at \href{https://github.com/jie-jw-wu/Survey-CodeLLM4LowResource-DSL}{https://github.com/jie-jw-wu/Survey-CodeLLM4LowResource-DSL}.


\section{Background and Related Work}\label{sec:background}

\subsection{LRPLs and DSLs}
\label{lrpl-dsl-def}

\begin{figure}[htbp]
    \centering
    \setlength{\fboxrule}{3pt}
    \includegraphics[scale=0.4]{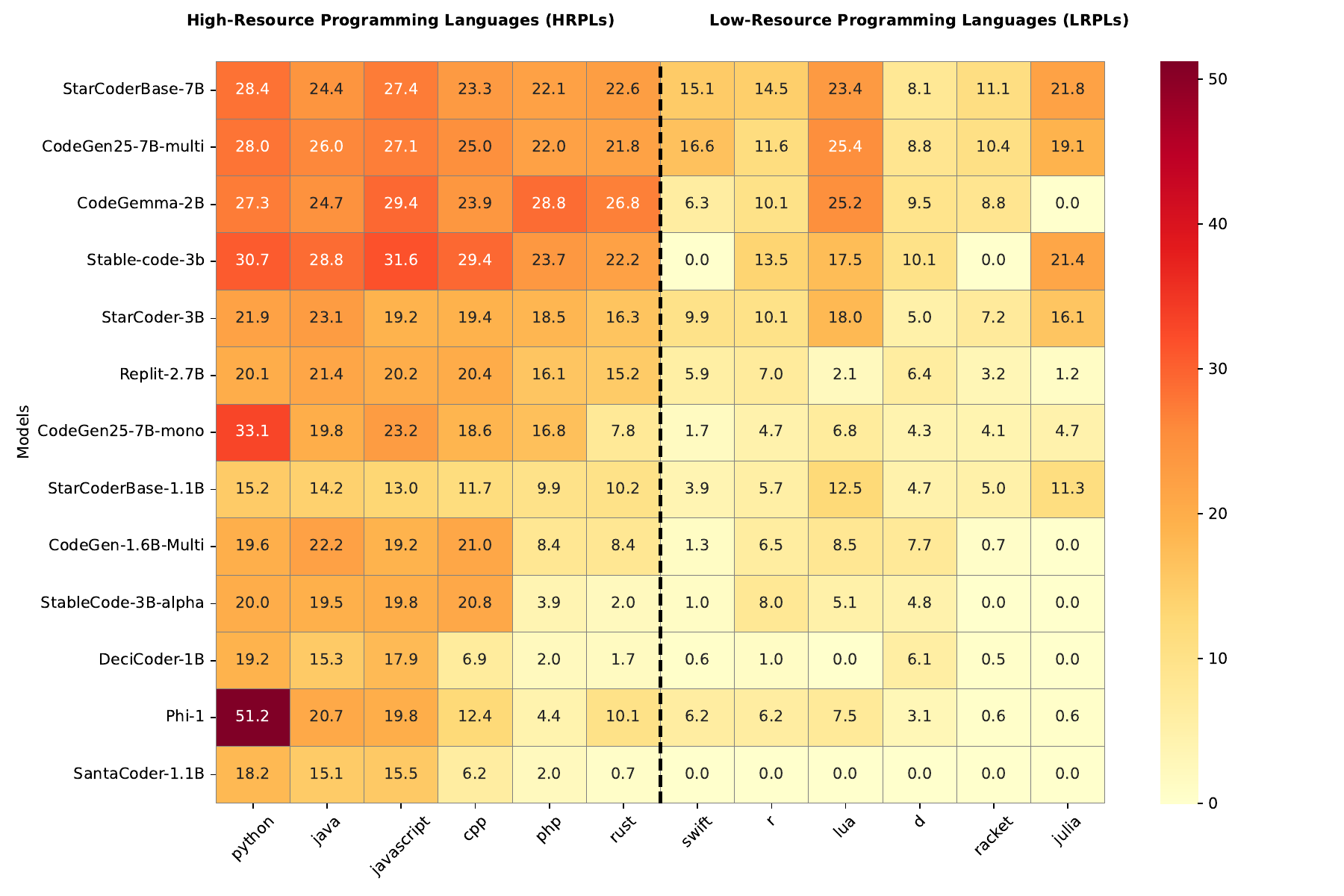}
    \caption{Heatmap of model performance on MultiPL-E benchmark across HRPLs and LRPLs. The vertical dashed line separates HRPLs (left) from LRPLs (right). Darker colors indicate higher performance scores.}
    \label{fig:heatmap}
\end{figure}

Low-Resource Programming Languages (LRPLs) are those programming languages that have limited data available~\citep{ID58paper, ID69paper,ID61paper, ID67paper, ID58paper, ID44paper}.
Their under-representation in LLM training data leads to lower performance compared to high-resource languages~\citep{ID58paper}. 
Figure~\ref{fig:heatmap} shows the performance of different models on MultiPL-E benchmark accessed from the BigCode Models leaderboard\footnote{\url{https://huggingface.co/spaces/bigcode/bigcode-models-leaderboard}} as of September 4th, 2024, across high and low-resource programming languages. This performance disparity is evident across various code generation models, with consistently higher scores for HRPLs like Python and Java compared to LRPLs such as R and Julia. 


Domain-Specific Programming Languages (DSLs) are programming languages that are optimized for specific problem domains, and offer higher abstraction levels and improved efficiency in targeted contexts \citep{JETBRAINS}. 
DSLs are also characterized by data scarcity and have received less attention in LLM research, leading to LLM's performance drop~\citep{ID6paper,ID58paper}.  
DSLs present additional challenges due to their unique syntax, semantics, and use cases, which are often not well-represented in general-purpose code datasets \citep{ID6paper}. 
The lack of specialized datasets and benchmarks \citep{ID67paper}, as well as limited research focus on LRPLs and DSLs, has resulted in fewer advancements in handling these languages effectively \citep{ID6paper, ID75paper}, which hinders the ability of LLMs for LRPLs and DSLs~\citep{IDsnow34paper, ID1paper}.
However, the implications of these challenges are significant, specifically, given the increasing complexity of the application domains in which LRPL/DSL are used, e.g., IoT, Quantum, hardware \citep{ID75paper}. LRPLs/DSLs developers may find minimal utility from LLMs \citep{ID67paper, ID62paper}, and training developers leads to software maintenance costs~\citep{ID62paper}. 
This is despite the fact that several studies have shown code generation models can enhance the developers' productivity~\citep{10.1145/3661145}.

There is a growing need for AI tools for LRPL/DSLs \citep{ID62paper}. 
Such advancements not only can enhance developer productivity but also facilitate the migration or modernization of projects from one programming language to another through improved code translation capabilities \citep{ID62paper, ID51paper}.
In this survey, we examine LRPLs and DSLs together due to their shared challenges and potential synergies in LLM-based code generation~\citep{ID58paper, ID6paper}. 
They often serve specialized use cases or niche developer communities, making them equally crucial for comprehensive AI-assisted software development~\citep{ID62paper, JETBRAINS}. 
Techniques developed for one category may benefit the other. 
By examining LRPLs and DSLs together in a systematic literature review, we aim to provide a comprehensive understanding of current limitations and future directions in expanding LLM capabilities beyond mainstream programming languages.

\subsection{Related Surveys}
\label{section:related-work}


There are several surveys that focus on LLM applications in Software Engineering~\citep{OPENSURVEY,hou2024large,zhang2023surveys1,unifyingSURVEY,watson2021systematicliteraturereviewuse,yang2020surveydeeplearningsoftware}. 
The studies examine different types of LLMs, their architectures, pre-training objectives, and downstream tasks ~\citep{zhang2023surveys1,unifyingSURVEY},
explore the transition from statistical models to pre-trained Transformers and LLMs in code processing ~\citep{unifyingSURVEY},
discuss the role of hybrid techniques combining traditional SE methods with LLMs~\citep{OPENSURVEY}, and the importance of well-curated datasets~\citep{hou2024large}. 
The surveys also highlight the need for reliable evaluation methods, benchmarking, and addressing security and reliability concerns when integrating LLMs into SE workflows~\citep{zhang2023surveys1}. 
Others examine the broader context of deep learning in SE, including automating various SE tasks~\citep{watson2021systematicliteraturereviewuse,yang2020surveydeeplearningsoftware}.

Another category of survey studies focuses on code generation. 
The performance of low-cost language models in Python code generation is evaluated in \citep{espejel2024lowcostlanguagemodelssurvey}.
A survey of 27 LLMs for NL2Code and reviewing benchmarks is provided in \citep{nl2code}.
Methods and metrics for evaluating LLMs in code generation~\citep{chen2024surveyevaluatinglargelanguage}, analyzing performance differences of LLMs across various software engineering tasks \citep{FUTURESURVEY},
solutions for LLMs and LLM-based agents in software engineering \citep{jin2024llmsllmbasedagentssoftware},
and survey on LLM-based agents for software engineering~\citep{liu2024largelanguagemodelbasedagents} are among other studies.
Finally, the benefits of integrating code into LLMs' training data
is discussed in~\citep{codeiswand}. 

\textbf{Differences of the current literature and our work.}
While the current studies have advanced the understanding of LLMs and their applications in software engineering, our work extends this body of knowledge in a critical yet under-explored direction of the application of LLMs for code generation in LRPLs and DSLs. Our approach differs by examining not only the models and evaluation methods, but also the strategies and methodologies proposed to enhance LLMs' performance in these specialized contexts. This includes an investigation of how datasets are collected, processed, and utilized to support code generation tasks in LRPL/DSL. By concentrating on LRPL and DSL domains, our work addresses a significant gap in the current literature, offering insights into the unique challenges and opportunities presented when applying LLMs to more specialized programming environments. This perspective is crucial for expanding the applicability of LLM-based code generation beyond well-resourced languages.

\section{Method} \label{sec:methof}

We conducted a systematic literature review by adopting the filtering approach outlined in \citep{percieved_diversity}.

\subsection{Study Goals}\label{subsec:aim of the study}

We have followed the Goal Question Metric (GQM) approach~\citep{GMQ} to identify the aim of the study. 
We use GQM as it has been employed in several software engineering research studies to ensure a systematic and measurable approach to defining and achieving research objectives. 
Here are the three coordinates of our goal, along with the purpose:


\textbf{Issue:} The underperformance of LLMs in code generation for LRPLs and DSLs due to challenges like data scarcity and specialized syntax and semantics. 
\textbf{Object:} The methodologies, evaluation metrics, datasets, and strategies used in leveraging LLMs for code generation in LRPLs and DSLs.   
\textbf{Viewpoint:} From the perspective of researchers and practitioners in software engineering, machine learning, and specialized programming domains. 
\textbf{Purpose:} To provide insights into the methodologies, performance, and challenges of LLM code generation and evaluation for LRPLs and DSLs, informing future research directions and advancements in this area. 
Summarizing the above, the \textbf{goal} of our study is to systematically investigate and analyze the current state of LLMs in code generation for LRPL and DSL, with the purpose of identifying challenges, evaluating methodologies, and providing insights to guide future research and development from the perspective of researchers and practitioners in the field.

\subsection{Research Questions}\label{subsec:Research Questions}

In this study, we aim to explore how LLMs are used and what techniques are developed to address code generation for LRPLs and DSLs. 
We will answer the following research questions.

--- \textbf{RQ1: Which LLMs, Metrics, and Benchmarks are used to evaluate code generation in LRPL and DSL Domains?}

LLMs have been widely used for code generation, and there are many LLMs mentioned on the LLM leaderboards for code generation. However, most of the research focuses on popular languages, which are dominated by Python for code generation. 
It is not clear which LLMs have been used for LRPLs and DSLs, and what evaluation metrics and benchmark datasets are used to evaluate code generation in these languages. Knowing this information helps understand the capabilities of the LLMs and whether new metrics are needed and developed for LRPLs and DSLs.

--- \textbf{RQ2: What strategies and methodologies have been proposed in the literature to enhance the performance of LLMs for code generation in LRPLs and DSLs?}

Enhancing the performance of LLMs in LRPL and DSL settings is essential for bridging the existing capability gaps that prevent a significant developer population from leveraging AI-assisted coding tools. Specialized domains and resource-constrained languages often present unique challenges, such as limited training data and highly specialized syntax, which general-purpose LLMs may not effectively address. Understanding the strategies and methodologies that have been proposed to overcome these challenges is crucial for identifying effective approaches and guiding future research aimed at improving LLM performance in these specific contexts.

--- \textbf{RQ3: How are datasets for LRPLs and DSLs collected, processed, and utilized to support code generation tasks using LLMs?}

High-quality datasets are fundamental to training LLMs, especially for code generation tasks. However, LRPLs and DSLs often suffer from data scarcity and imbalance, which can significantly hinder the performance of LLMs in these languages. Understanding the methodologies employed to collect, process, and utilize datasets for these specialized languages is critical for addressing data-related challenges and ensuring that LLMs can generate accurate and reliable code. This knowledge is essential for developing robust datasets that adequately represent the unique characteristics of LRPLs and DSLs, thereby enhancing the overall effectiveness of LLMs in these domains.

\subsection{Search Strategy}\label{subsec:Search Stratagy}

To identify relevant papers for our study, following previous work \citep{zhang2023surveys1}, we employed a systematic search strategy. Initially, we conducted a manual search on arXiv using keywords derived from our formulated research questions, such as \texttt{low resource}, \texttt{domain specific} and \texttt{code generation}, leading to an initial list of papers extracted from arxiv~\citep{ID31paper, ID5paper, ID68paper, ID4paper, ID80paper, ID6paper, ID92paper, ID106paper, ID62paper, ID58paper, ID69paper, ID57paper, ID34paper, ID8paper, ID113paper, ID11paper, ID70paper}. 
Given that code generation for LRPLs and DSLs has received comparatively less attention than for HRPLs, it is essential to ensure comprehensive coverage of emerging work. In recent years, arXiv has become a widely used platform for disseminating research in both AI and Software Engineering, particularly due to the rapid pace of developments in these fields. Researchers often publish preprints on arXiv, whether eventually accepted to peer-reviewed venues or not, as a means to quickly share novel findings with the community. Therefore, arXiv was selected as an initial source to identify relevant papers, ensuring inclusivity of the most recent and potentially impactful work. 
 Notably, recent systematic literature reviews also treat arXiv as a legitimate data source~\citep{hou2024large, zhang2023surveys1, 10.1145/3487043}, and a substantial portion of peer-reviewed publications begin as arXiv preprints. 
This further supports the importance of incorporating arXiv to achieve a more complete and timely literature review. Building upon this collection of papers, we noticed the abstracts of relevant papers often contain names of LRPLs and DSLs \citep{ID5paper, ID58paper, ID31paper}, and include \texttt{low-resource} and \texttt{domain-specific} terms \citep{ID80paper, ID58paper}. To ensure comprehensive coverage, we included multiple variations of conceptually similar terms (e.g., \texttt{domain-specific}, \texttt{domain specific}, \texttt{resource poor}, \texttt{resource-poor}) to account for different writing styles.

We then constructed a comprehensive search string incorporating a list of LRPLs and DSLs extracted from the papers identified in our manual search. 
We defined the list of LRPLs from \citep{ID70paper, ID69paper, ID67paper}. 
We also included \textsc{Ruby} in this category, despite its `medium' frequency classification in \citep{ID70paper}, as it is considered low-resource in some studies \citep{ID69paper}. These languages will be collectively referred to as LRPLs in the rest of this survey.
For DSLs, we used the terms and programming languages from our initial papers~\citep{ID31paper, ID5paper, ID68paper, ID4paper, ID80paper, ID6paper, ID92paper, ID106paper}. 
Additionally, we listed general descriptive terms such as \texttt{programming language*}, 
\texttt{multilingual}, 
\texttt{domain adaptive}
and their variations, such as:
\texttt{resource-scare}, 
\texttt{multi-lingual} and
\texttt{domain-adaptive}. 
For \textsc{R} and \textsc{D} languages, to reduce a high number of false positive results, we used \texttt{R software}, \texttt{R programming}, \texttt{R code}, \texttt{D software}, \texttt{D programming}, and \texttt{D code}.
For the code generation search string, we consulted the related literature reviews~\citep{OPENSURVEY, FUTURESURVEY, hou2024large} to gather relevant keywords such as \texttt{code synthesis} and \texttt{large language models} and used wildcards (denoted by *) to gather all variants of a term. 

Following the methodology in \citep{hou2024large}, we divided our keywords into two groups to enhance search precision. This approach was designed to identify papers containing keywords from both groups.
The first keyword group encompassed terms related to LRPLs and DSLs, including both specific language names and general descriptive terms. The second group focused on large language models and associated concepts, including code generation-specific terms and their wildcard variations. The search terms within each group are joined by ``OR'' operators, while the resulting strings from the two groups are connected by an ``AND'' operator.   
Table \ref{tab:keyword_groups} presents the complete list of keywords in each group. 
%

\begin{table}[htbp]
\caption{Keyword groups for literature search}
\label{tab:keyword_groups}
\begin{tabularx}{\textwidth}{|l|X|}
\hline
\textbf{Group} & \textbf{Keywords} \\
\hline
Group 1: & programming language*, domain-specific, domain specific, HPC, humaneval-x, mbxp, domain adaptive, domain-adaptive, multi-lingual, multilingual, resource poor, resource-poor, resource scare, resource-scare, low resource, low-resource, Ansible, mpi, Verilog, Rust, Lua, perl, tex, latex, julia, COBOL, fortran, assembly, coq, hpc, yaml, R software, R programming, R code, D software, D programming, D code, Ruby, Scala, Haskell, fortran, shell, bash, hansl, kotlin, matlab, ocaml, racket, smalltalk, swift, cuda, pascal \\
\hline 
Group 2: & large language model*, llm*, language model*, natural language processing, chatgpt, gpt*, nlp, artificial intelligence, neural network*, transformer, code llm*, AI, code generation, code completion, program synthesis, code synthesis \\
\hline 
\end{tabularx}
\end{table}

Our literature search spanned four carefully selected databases: arXiv \footnote{\url{https://arxiv.org/}}, IEEE Xplore\footnote{\url{https://ieeexplore.ieee.org/Xplore/home.jsp}}, Web of Science\footnote{\url{https://www.webofscience.com/wos/author/search}}, and ACM Digital Library\footnote{\url{https://dl.acm.org/}}, covering papers published from January 1, 2020, to May 15, 2024.
As noted in previous works \citep{nl2code, FUTURESURVEY}, significant works applying language models to code generation began to emerge in 2020, indicating a pivotal shift in the field. Therefore, we used 2020 as the start date.
Our database selection strategy aimed to provide comprehensive coverage of our research domain, balancing cutting-edge preprints with peer-reviewed publications. We chose IEEE Xplore for its strong coverage of software engineering and AI applications, Web of Science for its multidisciplinary scope and citation analysis capabilities, and the ACM Digital Library for its focus on computing and information technology. 
These databases are used in previous survey studies as well~\citep{hou2024large, zhang2023surveys1}.
Additionally, we included arXiv for several reasons. First, the rapid growth of the Artificial Intelligence field means many high-quality research papers are published on arXiv before formal peer reviews. Second, research on LLMs for low-resource scenarios is relatively recent, limiting the number of published papers in traditional venues. Finally, many relevant works might be in the process of being submitted to/published by journals or conferences, making arXiv a valuable source for the most current research. 
It is worth noting that arXiv is used as a database in recent systematic literature reviews as well, and a considerable number of publications were published as arXiv reports~\citep{hou2024large, zhang2023surveys1, 10.1145/3487043}.


\paragraph{Scope of our Survey}
We consider LLMs as language models with parameter counts of one billion or greater. It is important to note that there is no formal consensus on the definition of Large Language Models in the existing literature~\citep{zhao2023surveylargelanguagemodels}. 
However, the use of LLMs began in 2020 according to previous studies~\citep{nl2code}. 
Moreover, the landscape of language models has rapidly evolved, and recent research has introduced a class of smaller yet highly capable models, often referred to as ``small language models''~\citep{phi, gemmamodel}, like Phi ($1.3B$)\citep{phi}, SantaCoder ($1.1B$)~\citep{santacoder}, and Gemma ($2B$)\citep{gemmamodel}, which have demonstrated great performance comparable to some larger models, particularly in code generation tasks. 
To avoid the threat of missing papers, in the final set of papers after the last filtering iteration, we included all papers, though five of them had a smaller number of parameters.


\subsection{Eligibility}\label{subsec:Eligibility}

To ensure the quality of the papers and their relevance, we use the following factors to include or exclude a paper, beyond our filtering process. 
 Our criteria for including a work as relevant in the survey were the following:
 
\begin{itemize}

    \item The paper must investigate or utilize large language models above 1 billion parameters for code generation.
    
    \item The work must focus on applying LLMs in the context of LRPLs and DSLs.
    
    \item The study must present empirical evidence or a formal methodology for the application of LLMs in code generation tasks, as opposed to code analysis or other tangential applications, such as comment generation or code translation.
    
\end{itemize}

To ensure the quality and relevance of our survey, we established a set of exclusion criteria, which are adopted from previous research~\citep{hou2024large, FUTURESURVEY} as follows:

\begin{itemize}
    \item The paper is a grey publication, e.g., a technical report or thesis.
    \item Short papers whose number of pages is less than or equal to 4.
    \item Non-English written literature.
    \item Tool demos and editorial.
    \item Duplicate papers or similar studies with different versions from the same authors.
\end{itemize}

For arxiv papers, however, we included six technical papers due to their high quality and relevance, where the models included languages such as R, Rust, Ruby, and Kotlin, or they were evaluated on multi-lingual benchmarks that included LRPLs. These arxiv papers mention the word `technical report' in their title or abstract \cite{ID82paper, ID100paper, IDsnow34paper, ID11paper, ID18paper, ID76paper}. 

In this survey, we made a deliberate decision to exclude Structured Query Language (SQL) from our analysis. This choice was primarily driven by two factors: the abundance of existing research in the Natural Language to SQL (NL2SQL) domain and our focus on underrepresented languages. The field of NL2SQL has been extensively studied, with several comprehensive surveys~\citep{deng-etal-2022-recent, 10.1007/s00778-022-00776-8, hong2024nextgenerationdatabaseinterfacessurvey} already providing in-depth analyses of methodologies, challenges, and advancements specific to SQL generation using (L)LMs. 
By excluding SQL, we aimed to allocate more attention to languages and domains that have received comparatively less focus in the context of LLM-based code generation, addressing gaps in the literature and contributing novel insights to the field.

\subsection{Screening Process}\label{subsec:Screenign Process}

Our screening process, adapted from \citep{percieved_diversity}, involved a four-iteration approach to ensure a comprehensive and unbiased selection of relevant papers. The initial search across ArXiv, IEEE Xplore, Web of Science, and ACM Digital Library yielded a total of $27,333$ papers. 
We then conducted a systematic screening process as follows:

\begin{enumerate}[label=\roman*., leftmargin=*]
    \item \textbf{Title Screening:} The first author reviewed all $27,333$ paper titles, categorizing them as  \texttt{Include} or \texttt{Exclude} based on our predefined criteria mentioned above and the scope of the study.
    
    \item \textbf{Abstract Screening:} For papers that passed the title screening, the primary labeler examined the abstracts, again applying the  \texttt{Include}, \texttt{Exclude} categorization.
    Papers labeled as \texttt{Uncertain} were investigated more carefully in the next step.
    
    \item \textbf{Preliminary Content Review:} The first author conducted a cursory examination of the full text of papers that were marked as \texttt{include} in the previous iteration, to further assess their relevance, considering the \texttt{Include}, \texttt{Exclude} categorization.
    
    \item \textbf{Final Full-Text Review:} We consolidated all papers that passed the previous stages and reviewed the titles and abstracts to remove duplicates. This duplication removal was applied to the papers from all databases, including arXiv.
    The primary labeler then conducted a thorough full-text review of these remaining papers to make final inclusion decisions. This step involved a comprehensive assessment of each paper's relevance to our research questions and a quality check to ensure the studies met our methodological standards, selecting the ones for final review. 
    
\end{enumerate}

To ensure reliability and minimize bias, the second and third authors independently reviewed the categorizations at each iteration. Any papers deemed uncertain or where disagreements arose were discussed among all three authors.
Furthermore, we employed both forward and backward \textit{snowballing} approaches to expand our literature base~\citep{hou2024large}. 
For forward snowballing, we meticulously examined the reference lists of the initially identified papers. Backward snowballing was conducted using Google Scholar to identify papers that had cited our final set of publications. 
This process yielded an additional 36 papers that met our selection criteria, significantly enriching our corpus.
It is worth noting that papers with over 300 citations were excluded from the backward snowballing process. Due to the high citation counts of several seminal papers in our field, we were unable to conduct comprehensive backward snowballing for some works: \citep{codellama, ID83paper, ID84paper, ID117paper, ID76paper, ID18paper, ID82paper, nijkamp2023codegenmodel}. 

In our screening process, we made some decisions about three DSLs. First, First Order Logic is not considered a programming language in all papers. However, we included related papers that generated First Order Logic in our study, because it is considered a formal language with syntax and semantics, and it is commonly used in software verification and theorem proving~\cite{ID43paper}. Additionally, in the work of~\citet{ID43paper}, First Order Logic is experimented as natural language to formal expressions using LLMs along with Regex and Linear-time Temporal Logic (LTL)--LTL is the foundation for hardware specification languages. 
Similarly, in~\cite{ID81paper}, the authors focused on creating CAD sketches, using a designed sequential language; therefore, we considered this work in our study. 
Additionally, automated theorem proving is considered in our analysis~\cite{ID39paper, IDsnow22paper}, as they use a formal language, Lean. 
Finally, we included the work of \citet{ID112paper}, where a new DSL is designed to answer user questions about chemical processes.

\begin{table}
    \centering
    \begin{tabular}{@{} l c c c c c c @{}}
    \toprule
    Database Name & Iter 1 & Iter 2 & Iter 3 & Iter 4 & Iter 5 & Final Count \\
    \midrule
    ArXiv & $8,345$ & $202$ & $126$ & $116$ & $113$ & $63$ \\
    Web of Science & $11,460$ & $140$ & $35$ & $35$ & $35$ & $10$ \\
    IEEE & $6,468$ & $55$ & $21$ & $20$ & $20$ & $9$ \\
    ACM & $1,057$ & $109$ & $22$ & $21$ & $21$ & $4$ \\
    \midrule
    Combine and remove duplicates & -- & -- & -- & -- & -- & $75$ \\
    Snowballing & -- & -- & -- & -- & -- & 36 \\
    \midrule
    \textbf{Total} & $27,330$ & $506$ & $204$ & $192$ & $189$ & $111$ \\
    \bottomrule
    \end{tabular}
    \caption{Number of papers filtered in each iteration}
    \label{tab:iter_count}
\end{table}

\subsection{Extracted Papers} \label{sec:extracted_papers}

\begin{figure}[htbp]
    \centering
    \includegraphics[width=\linewidth]{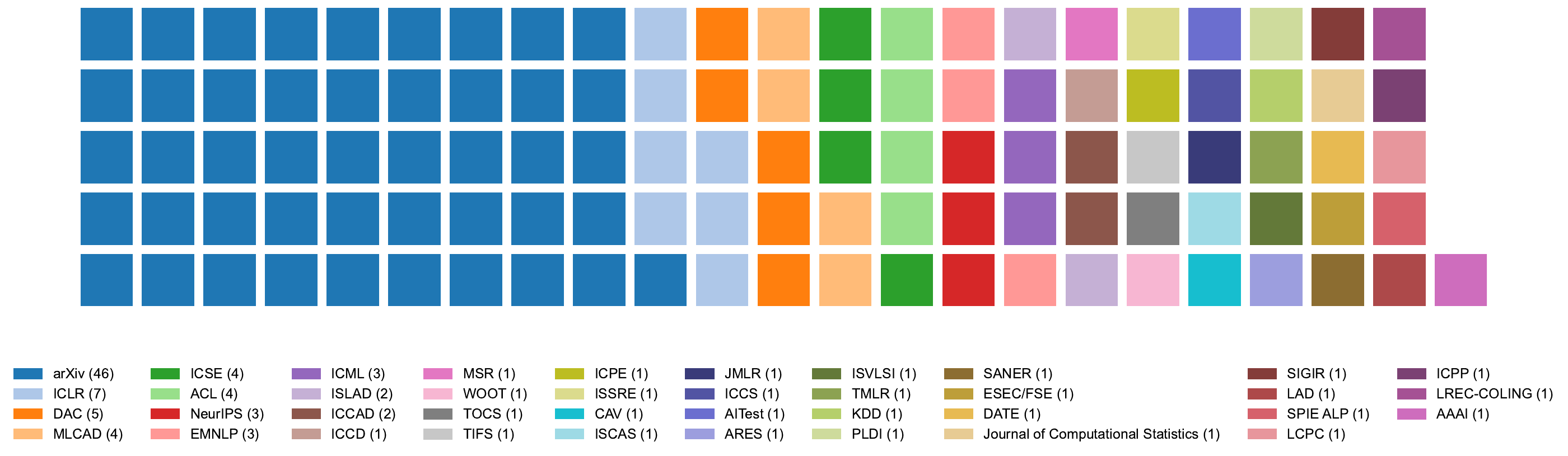} 
    \caption{Venue distribution}
    \label{fig:waffel_chart}
\end{figure}

Table~\ref{tab:iter_count} details the number of papers we initially found and after each of the filtering steps, separated by the database.
In total, out of $27,330$ papers, we filtered $111$ papers to read in the `final list'. 
Due to the rapid pace of research dissemination in this domain, many impactful contributions initially appear as preprints (e.g., on arXiv) and are later published in peer-reviewed venues. To ensure a high-quality and stable foundation for analysis, we revisited and updated all citations in Summer 2025 to reflect the peer-reviewed versions.
We read all the papers in this list to extract the information and answer each of the RQs.
There has been an upward trajectory in research output pertaining to LLMs for LRPL and DSL code generation over the past four years. The data reveal a modest initial output of one and four papers in 2020 and 2021, respectively, followed by a notable increase to 13 publications in 2022, 44 papers in 2023, and reaching a zenith of 49 publications just in the first half of 2024.
There are 51 papers addressing LRPLs, 59 papers focusing on DSLs, and one paper~\citep{IDsnow42paper} covering both.
The waffle chart in Figure~\ref{fig:waffel_chart} illustrates the distribution of all papers analyzed in our survey across 39 different conferences and journals. Each colored square represents one paper, with colors corresponding to specific venues. 

\textbf{Expanded Literature Search for Recent Publications.} Our initial systematic search covered publications from January 1, 2020, to May 15, 2024. However, given the rapid pace of research in this domain and the time elapsed during the review and publication process, we conducted an expanded search to capture more recent contributions. Specifically, we extended our search to include full papers published in major AI, NLP, and SE venues (NeurIPS, ICML, ICLR, ACL, EMNLP, NAACL, ICSE, FSE, and ASE), from May 2024 to July 2025. This targeted expansion yielded an additional 5 papers~\cite{extend1_codem,extend2_mora2024synthetic,extend3_kon2024iac,extend4_shah2024stackeval,extend5_bhatia2024verified} that met our inclusion criteria about code generation in LRPLs and DSLs. These newly identified papers are integrated throughout our analysis and findings; however, they are not reflected in the quantitative statistics presented in Table 2 and Figure 2, which represent the results of our primary systematic search methodology to maintain methodological consistency and transparency in our reporting.


\section{LLMs, Metrics and Benchmarks for LRPL and DSL}
\label{sec:metrics_benchmarks}

In this section, we provide a comprehensive overview of the current landscape in evaluating code generation capabilities of LLMs for LRPLs and DSLs. 

\subsection{LLMs Used}
\label{subsec:llms_used}

\begin{figure}[htbp]
    \centering
    \begin{subfigure}{\textwidth}
        \centering
        \setlength{\fboxrule}{3pt}
        \includegraphics[width=\textwidth]{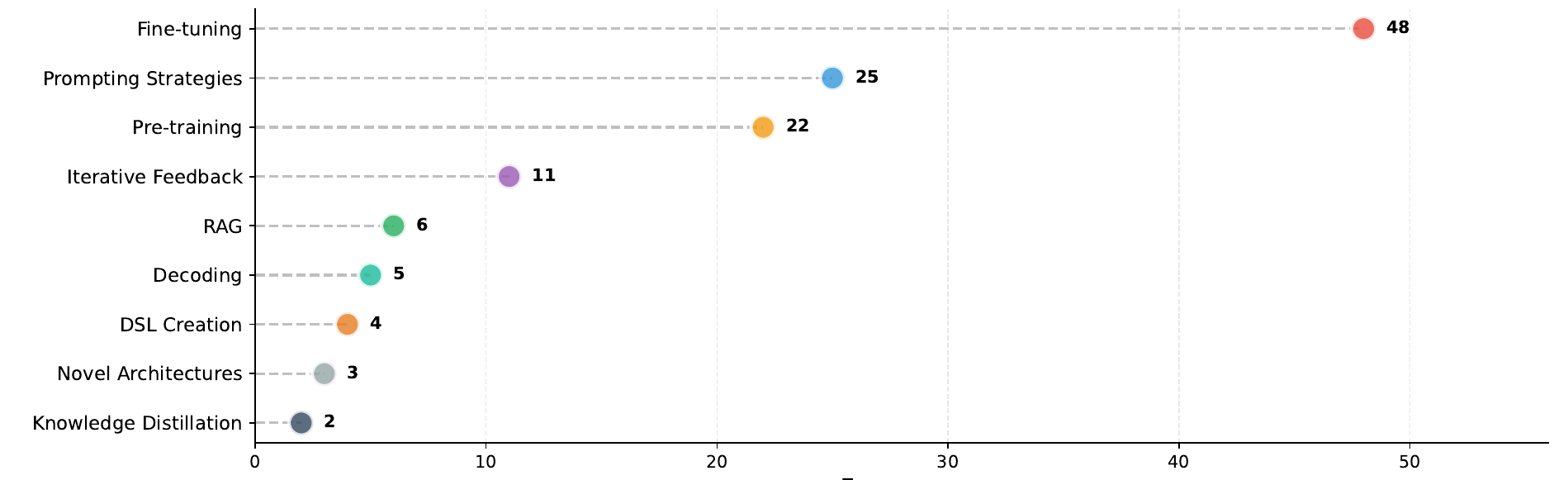}
        \caption{Techniques used for enhancing the performance of LRPL/DSL code generation}
    \end{subfigure}%

    \vspace{0.7em} 


    \begin{subfigure}{0.6\textwidth}
        \centering
        \includegraphics[width=\textwidth]{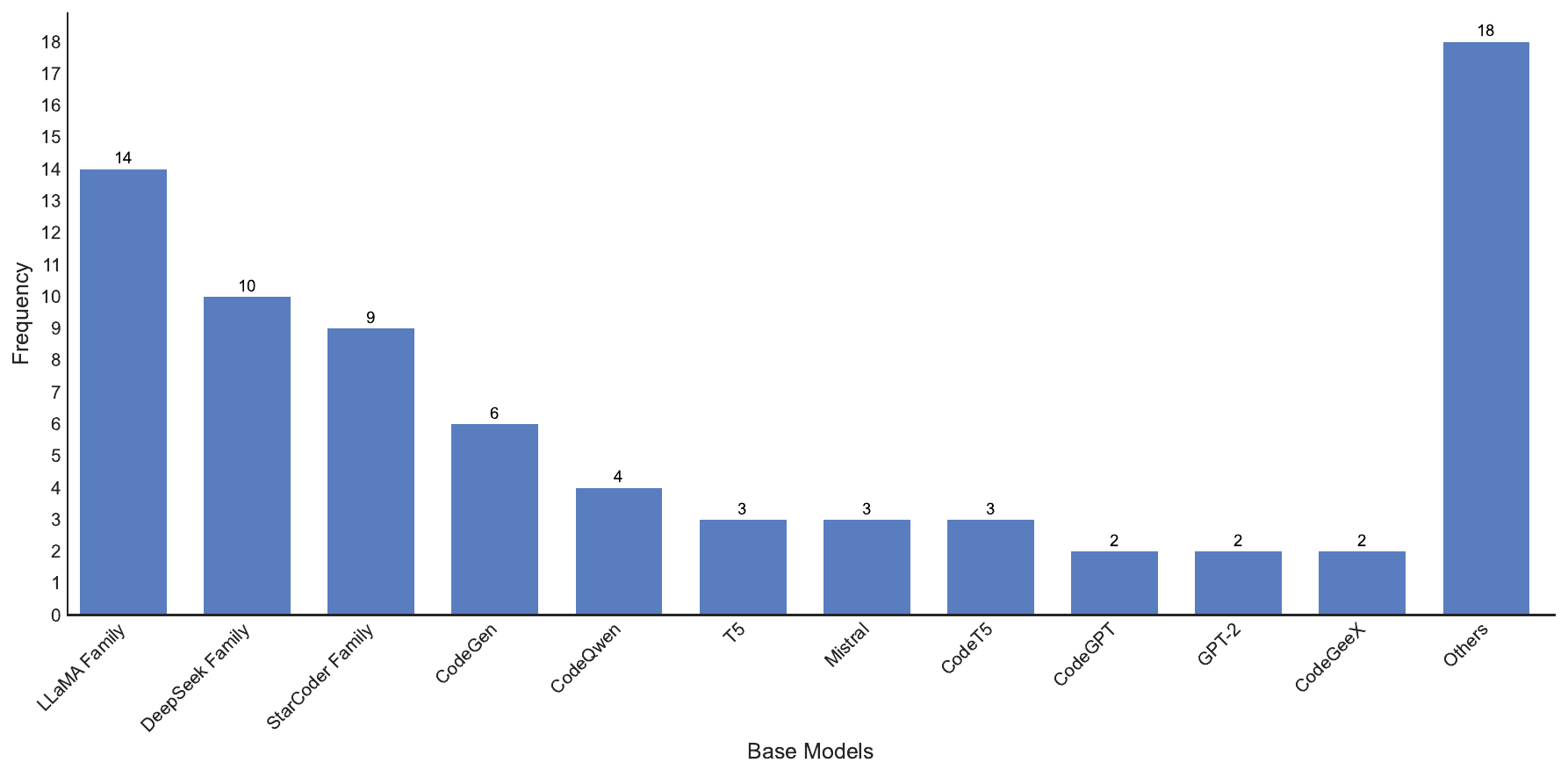}
        \caption{Frequency distribution of base models used for fine-tuning}
    \end{subfigure}
    \hfill
    \begin{subfigure}{0.35\textwidth}
        \centering
        \setlength{\fboxrule}{3pt}
        \includegraphics[width=\textwidth]{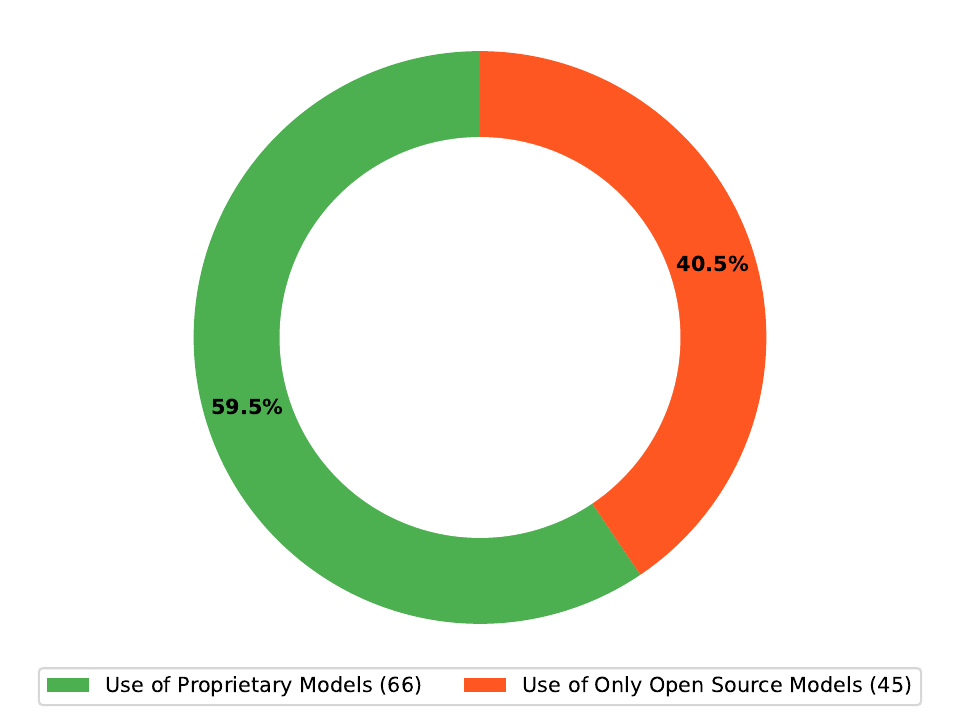}
        \caption{Proprietary vs. open-source models}
    \end{subfigure}%

    \caption{(a) Techniques used for enhancing the performance of LRPL/DSL code generation. Note that papers may use multiple techniques. (b) Frequency distribution of base models used for fine-tuning. The x-axis shows individual base models, while the y-axis represents the frequency of their use. Models with a frequency of 1 are aggregated into the `Others' category to improve readability. This visualization highlights the most commonly used base models and provides insight into the diversity of model choices. (c) Proportion of proprietary models versus open-source models used. }
    \label{fig:finetuned_models_frequency}
\end{figure}

\textbf{LLM Usage For Fine-tuning and Other Approaches.} Figure~\ref{fig:finetuned_models_frequency}(a) shows a breakdown of the number of papers in techniques for enhancing the performance of LRPL/DSL code generation. We can see that fine-tuning has the most frequency, so we chose to calculate the frequency of LLM base models used in fine-tuning. Other LLMs used beyond fine-tuning include GPT-3.5, GPT-4, ChatGPT variants, Claude, Gemini, Phi-1, SantaCoder, CodeGeeX, Replit, Stable Code, CodeGen, Mistral, and T5. These models are primarily employed for prompting (See Section \ref{subsec:prompting} for details), RAG (Section \ref{subsec:retrieval}), evaluation (Section \ref{subsec:benchmarks}), synthetic data generation (Section \ref{subsec:dataset-curation}), and iterative feedback approaches (Section \ref{subsec:prompting}). Unlike fine-tuning, which predominantly uses LLaMA, DeepSeek, and StarCoder families, other techniques rely heavily on proprietary models like GPT variants for their superior reasoning capabilities in complex LRPL and DSL tasks. We refer readers to the corresponding sections for further details. 

Figure~\ref{fig:finetuned_models_frequency}(b) illustrates the frequency of various base models used for fine-tuning to generate code in LRPLs or DSLs. The LLaMA Family emerges as the most popular choice~\citep{ID65paper, ID66paper, ID105paper, IDsnow11paper, IDsnow31paper}, used in 14 instances, closely followed by the DeepSeek Family with 10 occurrences~\citep{IDsnow29paper, IDsnow37paper, IDsnow15paper, ID105paper} and then the StarCoder family~\citep{ID58paper, ID7paper, ID84paper}. 
The LLaMA models, developed by Meta AI, include models called Code LLaMA~\citep{codellama} specialized for code generation, based on the LLaMA 2 architecture. It offers three main variants: Code LLaMA (general-purpose), Code LLaMA - Python (Python-specific), and Code LLaMA - Instruct (instruction-tuned), each available in 7B, 13B, 34B, and 70B parameter sizes. The LLaMA family of LLMs is a common choice because they perform well on most common benchmarks~\citep{ID58paper}. Deepseek-Coder is also a frequent choice for fine-tuning given its strong coding performance on benchmarks like HumanEval, MPPP, and code contests~\citep{deepseek-coder}. StarCoder~\citep{ID84paper} is a publicly accessible model boasting a significant 15-billion parameter count. It has been fine-tuned on a carefully selected subset of the Stack dataset, covering 86 programming languages, which ensures its versatility and proficiency across a wide array of coding tasks. Other common choices include CodeGen~\citep{codegen2} and CodeQwen~\citep{qwen}, followed by T5, CodeT5~\citep{wang-etal-2021-codet5}, Mistral~\citep{jiang2023mistral7b}, GPT-2~\citep{gpt2model}, and CodeGeeX~\citep{codegeex}. 

\textbf{Analyzing the dominance of the LLaMA family.} The popularity of the LLaMA family began with its February 2023 release, arriving 2.5 months before StarCoder \cite{ID84paper} and 8 months before DeepSeek-Coder \cite{deepseek-coder}, allowing researchers to access these models while alternatives remained unavailable.  Other contributing factors could include its open-source availability and the wide range of model sizes (7B-70B parameters), which make it easier for researchers to use. Code LLaMA's specialized variants \cite{codellama} offered solutions for different coding tasks, and the architecture's compatibility with parameter-efficient fine-tuning (LoRA/QLoRA) enabled domain specialization with minimal resources for LRPLs/DSLs. These practical advantages, combined with a robust ecosystem including community fine-tuning, created network effects sustaining LLaMA's dominance despite newer models showing superior benchmark performance.


\textbf{Open Source and Closed Source Models.}  
In contrast to closed-source models, open-source models are publicly available, allowing full access to their code, architecture, and pre-trained weights~\citep{wu2024geb13bopenlightweightlarge}. 
Figure~\ref{fig:finetuned_models_frequency}(c) illustrates the distribution of research papers using proprietary versus open-source models. 
According to the data, $60.4\%$ of the surveyed papers report using proprietary models in their research, while $39.6\%$ exclusively use open-source models. Many works~\citep{ID52paper, ID92paper, ID93paper, ID33paper} utilize proprietary models such as GPT-3.5/4 for evaluation purposes and to compare their approach with the performance of these proprietary models. 
In contrast, other studies~\citep{ID65paper, IDsnow15paper, IDsnow31paper} employ open-source models like LLaMA and DeepSeekCoder, fine-tuning them on LRPL/DSL data.

\subsection{Metrics}
\label{subsec:metrics}

This section explores various metrics employed to assess the quality of the generated code by LLMs. 
We divided the evaluation metrics into three groups of \textit{Automatic Evaluation} metrics, including Pass@k and BLEU, \textit{User Centric Evaluation} that focus on metrics that prioritize the end-user experience, \textit{Domain Specific Evaluation} tailored to assess performance in particular programming contexts or applications, and finally \textit{Manual Evaluation} techniques implemented by researchers to provide human-driven assessments of generated code quality. 



\textbf{Automatic Evaluation.}
Pass@K, BLEU, ROUGE, Edit Similarity, and METEOR are the common automatic evaluation metrics used in several studies. 
Table \ref{tab:metrics-languages-refs} provides an overview of these metrics and the languages of the studies. The last column cites the papers in which these metrics are used. Among these metrics, Pass@K calculates the percentage of problems for which a model produces at least one correct solution in its top k predictions.
The other metrics, however, consider the similarity of the generated code with the ground truth. Besides the above metrics, in~\cite{extend4_shah2024stackeval}, LLM Judge is used in evaluating coding assistant tasks, though such a method tends to have reliability concerns in automated judgment.







\begin{table}[htbp]
\centering
\scriptsize
\caption{Commonly used automatic evaluation metrics}
\label{tab:metrics-languages-refs}
\begin{tabular}{|p{0.11\textwidth}|p{0.6\textwidth}|p{0.2\textwidth}|}
\hline
\textbf{Metrics} & \textbf{Languages} & \textbf{Ref} \\
\hline
pass@k & Awk, Bash, Codon, CoffeeScript, Crystal, D, Dart, Elixir, Erlang, Fortran, Go, Groovy, Haskell, Julia, Kotlin, Lean, Lua, Nim, OCaml, Pascal, Perl, PHP, PowerShell, R, Racket, Ruby, Rust, Scala, Scheme, Swift, Tcl, Verilog, VHDL, Vim script, F\#, Terraform & \citep{ID70paper,ID52paper,ID105paper,ID73paper,ID8paper,ID92paper,ID66paper,ID113paper,ID77paper,ID67paper,ID93paper,ID58paper,ID2paper,ID157paper,ID84paper,ID27paper,ID39paper,IDsnow4paper,IDsnow7paper,IDsnow8paper,IDsnow15paper,IDsnow27paper,IDsnow29paper,IDsnow31paper,IDsnow43paper,IDsnow33paper,IDsnow34paper,IDsnow37paper,IDsnow38paper,IDsnow39paper,IDsnow40paper,IDsnow41paper} \citep{extend1_codem,extend3_kon2024iac}\\
\hline
BLEU & Ansible, Assembly, Bash, Codon, Crystal, CQL, D, Fortran, GitHub Actions YAML, Haskell, Kotlin, LLVM IR, Nim, PowerShell, Ruby, Rust, Scala, Swift, Verilog & \citep{ID59paper,ID105paper,ID6paper,ID133paper,ID60paper,ID75paper,ID27paper,IDsnow11paper,IDsnow23paper,IDsnow25paper, ID130paper, ID141paper} \\
\hline
ROGUE & Assembly, Codon, Crystal, D, Fortran, Haskell, Julia, Kotlin, Lua, Nim, PowerShell, R, Ruby, Rust, Scala, Swift & \citep{ID59paper,ID105paper,ID58paper,ID133paper,IDsnow23paper} \\
\hline
Edit Similarity & Kotlin, Rust, Scala, Ruby, Haskell, Bash & \citep{ID59paper,ID57paper,IDsnow23paper, ID130paper} \\
\hline
Exact Match & Kotlin, Rust, Scala, Ruby, Regex, FOL, LTL, Ansible, Assembly, LLVM IR, CAD, Haskell, R, OCL, CQL, Bash, YAML & \citep{ID59paper,ID43paper,ID6paper,ID147paper,ID133paper,ID60paper,ID81paper,ID27paper,ID57paper,ID34paper,IDsnow20paper,IDsnow25paper,IDsnow42paper} \\
\hline
METEOR & Kotlin, Rust, Scala, Ruby, Assembly, PowerShell & \citep{ID59paper,ID133paper,IDsnow23paper} \\
\hline
Accuracy by LLM Judge & 25 programming languages & \citep{extend4_shah2024stackeval} \\
\hline
\end{tabular}
\end{table}

\textbf{User Centric Evaluation.}
A few works~\citep{ID5paper, ID59paper, ID34paper} have evaluated the generation capabilities from the user's perspective. 
\textit{Acceptance Rate} measures the proportion of model-generated completions that users actually incorporate into their code, providing a direct measure of the model's practical utility \citep{ID59paper, ID5paper}.
\textit{N-day User Retention} is used to evaluate the ongoing engagement and value of a service or application over time. It measures the percentage of users who return to use the product N days after their initial interaction or installation \citep{ID5paper}.
Finally, \textit{\#attempt\_k} is another metric that represents the average number of attempts required for a user to generate a satisfactory code solution using an LLM, with a maximum of k attempts allowed per task \citep{ID34paper}.

\textbf{Domain Specific Evaluation.}
Common metrics like Pass@k might not be suitable for evaluating all the aspects of specialized languages due to their unique structures and purposes. These languages often require more nuanced evaluation methods that can capture their domain-specific semantics and functionality. Table \ref{tab:code-gen-metrics} presents a comprehensive overview of these metrics.

\begin{table}[h]
\caption{Domain-specific evaluation metrics for code generation}
\label{tab:code-gen-metrics}
\small
\begin{tabular}{p{0.3\linewidth}|p{0.3\linewidth}|p{0.3\linewidth}}
\hline
Metric & Language(s) & Ref \\
\hline
Ansible Aware Metric & Ansible & \citep{ID6paper,IDsnow42paper} \\
Schema Correct Metric & Ansible & \citep{ID6paper,IDsnow42paper} \\
Command Accuracy (CMD Acc) & Bash & \citep{ID27paper,IDsnow42paper} \\
Accuracy per Sequence & Multiple & \citep{ID43paper} \\
Semantic Accuracy & Regex & \citep{ID43paper,semantic-accuracy} \\
Pass@(scenario) & Verilog & \citep{ID8paper} \\
Syn-VCS, Syn-DC & Verilog & \citep{ID77paper} \\
Accuracy & DSL for Simulation Process, Verilog, R, LTL, FOL, CQL, MapReduce, Network packet processing, TACO, Tensor processing & \citep{ID112paper,ID99paper,ID71paper,IDsnow10paper,IDsnow12paper,IDsnow25paper} \citep{extend5_bhatia2024verified} \\
Execution Accuracy & Bash, CQL & \citep{ID103paper,IDsnow25paper} \\
Execution @k & M & \citep{IDadd1paper} \\
CQL BLEU & CQL & \citep{IDsnow25paper} \\
FOL BLEU & FOL & \citep{IDsnow11paper} \\
Logical Equivalence (LE) & FOL & \citep{IDsnow11paper} \\
Syntax Score, Semantic Score & DSL for Qore-Base & \citep{ID147paper} \\
Power-Performance-Area (PPA) & Verilog & \citep{ID4paper,IDsnow2paper,IDsnow5paper,ID160paper} \\
Area-Delay Product (ADP) & Verilog & \citep{IDsnow3paper} \\
Pass Rate & ODSL & \citep{ID72paper} \\
Qualitative Assessment & ST & \citep{ID65paper} \\
Number Generated & XDL, SVA & \citep{ID32paper,ID12paper} \\
VCS Simulation Time, SVA Generation Time & SVA & \citep{ID12paper} \\
Perplexity, Negative Log Likelihood & System Verilog, Verilog, VHDL & \citep{ID49paper} \\
E\_syntax, E\_stat & CAD Sketches & \citep{ID81paper} \\
MRR, Recall@5 & R & \citep{ID129paper} \\
Cosine Similarity, Syntactic Validity & OCL & \citep{ID74paper} \\
Program Accuracy, Execution Accuracy, Validity, Diversity, Retrosynthesis Score, Membership & SMILES, PDDL & \citep{ID46paper} \\
Accuracy@k & YAML & \citep{ID75paper} \\
Compilation Rate, Simulation Rate, Correctness Rate & SVA & \citep{IDsnow18paper} \\
Average Correctness (AC) & Ansible & \citep{IDsnow16paper} \\
Verify@k & F$^\ast$ & \citep{IDsnow19paper} \\
Validity@k & OCL & \citep{IDsnow20paper} \\
SketchMatch & Excel Formulas & \citep{ID42paper} \\
Hit@5 & Bash & \citep{ID130paper} \\
Semantic Score & UCLID5 & \citep{extend2_mora2024synthetic}\\

\hline
\end{tabular}
\end{table}

\textbf{Manual Evaluation.}
Manual evaluation plays a crucial role in assessing the quality and effectiveness of AI-generated code and proofs, especially in complex domains where automated metrics may fall short. 
A manual evaluation metric was introduced in \cite{ID35paper, ID20paper} for assessing the quality and correctness of GitHub Copilot's code suggestions for high-performance computing (HPC) numerical kernels. The metric rates code suggestions on a scale from 0 to 1, with five distinct levels: non-knowledge (0), novice (0.25), learner (0.5), proficient (0.75), and expert (1). These levels are determined by the code's adherence to the requested programming model and consistency across multiple suggestions. This metric was applied to evaluate Copilot's performance across various programming languages (Fortran, and Julia), different HPC kernels (such as AXPY, GEMV, GEMM), and multiple parallel programming models (like OpenMP, CUDA, and OpenACC).
In \cite{ID31paper}, manual inspections were conducted to compare outputs from different models, focusing on key aspects such as correctness, diversity of proof strategies, and adherence to Coq syntax. The evaluation involved comparing outputs from a fine-tuned model against those from prominent LLMs like ChatGPT 4, Google Gemini, and Starcoder2-15b. Researchers assessed the models' ability to generate valid proofs, recognize erroneous lemmas, and produce Coq-specific responses.

In the field of chemistry, authors \cite{ID32paper} employed manual inspection by expert chemists to evaluate the quality of generated plans. This approach was chosen because automated metrics alone cannot fully capture the nuances and correctness of chemistry procedures, and expert judgment was necessary to assess whether the generated plans were scientifically sound and practically executable.
Manual evaluation was conducted to assess the efficacy of their GPT-4-powered tool in automating proof generation for program verification \cite{ID63paper}. Similarly, the execution accuracy of the generated OCL constraints was manually evaluated, where experts were asked to verify 
if constraints were correctly implemented given specifications, and adhered to model semantics \cite{ID74paper}. Another similar example is the empirical evaluation conducted by authors to get Semantic Scores for UCLID5 in~\cite{extend2_mora2024synthetic}.

In \cite{ID75paper}, automated metrics were complemented with a manual validation process to assess the semantic correctness of generated workflows. Researchers randomly sampled 30 valid workflows for each model and mode's best-performing BLEU configurations, totaling 270 workflows, and computed the percentage of semantically correct steps in each workflow.
Lastly, an evaluation of subjective quality attributes to assess ChatGPT-generated R code was conducted in \cite{ID34paper}. A trained evaluator manually assessed each solution across eight dimensions: accuracy, completeness, structuredness, conciseness, logic of code, parameter coverage, readability, and depth of explanation, using a 5-point Likert scale. The evaluation process involved examining both code and explanations, comparing them against reference solutions and predefined criteria.

\subsection{Benchmarks Used to Evaluate LRPL and DSL Generated Code}
\label{subsec:benchmarks}

Table \ref{tab:pl-benchmarks} catalogues the benchmarks that are used in the literature to evaluate the performance of LRPL and DSL Code generation from LLMs.      
The Research Utilization column provides guidance on how each benchmark should be used in research contexts. It categorizes benchmarks by their primary research applications, such as Cross-language and baseline studies (e.g., MultiPL-E), Domain-specific evaluation (e.g., RTLLM and VerilogEval), Specialized application testing (e.g., TLDR and InterCode), Formal reasoning assessment (e.g., FIMO and miniF2F). The column serves as a guide to help researchers and practitioners select appropriate benchmarks based on their specific study objectives and the type of programming language capabilities they want to evaluate.
MultiPL-E~\citep{ID70paper} is extensively used in the literature~\citep{ID67paper, ID58paper, ID66paper, ID7paper, ID105paper, ID18paper, ID82paper} to evaluate the code generation abilities of models in LRPLs such as Bash, Lua, Perl, R, Ruby, Racket, D, Go, Julia, Rust, Scala, Swift. 
MultiPL-E is one of the first massive multi-lingual benchmarks that includes low-resource languages. It contains compilers to translate function signatures and unit tests, translates types and type inferences, preserves comments and translates doctests, and translates the Python-specific terminology to equivalent terminology in the target language. A similar approach is used in \cite{ID113paper}, where the test cases and prompts are translated to the target language. 
CodeScope~\cite{IDsnow43paper} develops a code execution engine to support execution-based evaluations of the generated codes in multiple languages. 
xCodeEval is an executable multilingual multitask benchmark with 2.5K unique problems in 11 programming languages, including LRPLs such as Kotlin, Ruby, and Rust \citep{ID93paper}.
It is a predecessor of CodeScope that developed a multilingual code execution engine for unit test execution and evaluation of generated code in 11 different languages.

Another study modified the HumanEval benchmark by translating these problems to 18 other programming languages~\citep{ID70paper}, adapting function signatures, unit tests, and comments while preserving the core problem structure and evaluation criteria.
HumanEval is also manually translated to other languages like Kotlin \cite{IDsnow34paper} and Haskell, a functional programming language \cite{ID57paper}. Another benchmark called StackEval was constructed in 25 programming languages for coding assistant tasks. 

\citet{ID56paper} proposed Intercode framework to address the limitations of static code generation tasks, which often fail to capture the iterative nature of real-world programming. InterCode standardizes interactive coding as a reinforcement learning environment, where code serves as actions and execution feedback as observations. The benchmark encompasses three interactive coding environments for Bash, SQL, and Python, leveraging data from existing static datasets like NL2Bash, Spider, and MBPP. 
They also constructed TLDR, which is a dataset driven from the TLDR community-driven project for over 2.5k Bash commands. The dataset contains 1,879 Bash commands and 9,187 NL2Bash pairs.
\citet{ID103paper} designed and evaluated an execution based environment for Bash commands, where they set up an environment to run test cases, with containers that should be examined (as they are test-specific with a variety of commands) and evaluated.

\begin{table}[htbp]
\centering
\caption{Programming language benchmarks and their respective languages with research utilization guidelines}
\label{tab:pl-benchmarks}
\footnotesize
\renewcommand{\arraystretch}{1.1}
\begin{tabular}{p{0.18\textwidth}p{0.28\textwidth}p{0.15\textwidth}p{0.35\textwidth}}
\hline
\textbf{Benchmark Name} & \textbf{Languages} & \textbf{Reference} & \textbf{Research Utilization} \\
\hline
MultiPL-E & Bash, Lua, Perl, R, Ruby, Racket, D, Go, Julia, Rust, Scala, Swift & \citep{ID70paper, ID67paper, ID58paper, ID66paper, ID7paper, ID105paper, ID18paper, ID82paper} \citep{extend1_codem} & Cross-language comparison; knowledge transfer studies; baseline metrics \\

MuliPL-MBPP & Bash, Lua, Perl, R, Ruby, Racket, D, Go, Julia, Rust, Scala, Swift & \citep{ID58paper} & Alternative to MultiPL-E; additional test coverage \\

MBXP, MathQA-X & Ruby, Kotlin, Scala, Swift, Perl & \citep{ID113paper} & Math problem solving; algorithm implementation \\

multilingual human eval & Ruby, Kotlin, Scala, Swift, Perl & \citep{ID73paper, ID113paper} & Standard LRPL baseline; cross-language benchmarking \\

BabelCode & Dart, Lua, Rust, C\#, R, Julia, and Haskell & \citep{ID67paper} & Language diversity studies; syntax transfer analysis \\

CodeScope & Ruby, Kotlin, D, Perl, Rust, Delphi & \citep{IDsnow43paper} & Code execution testing; functional correctness \\

XCodeEval & Kotlin, Ruby, Rust & \citep{ID93paper} & LRPL-focused evaluation; executable code testing \\

HumanEval-Kotlin & Kotlin & \citep{IDsnow34paper} & Kotlin-specific evaluation; Android development \\

HumanEval-Haskell & Haskell & \citep{ID57paper} & Functional programming; type safety evaluation \\

Humaneval-X & Rust & \citep{IDsnow7paper} & Rust-specific evaluation; systems programming \\

TLDR & Bash & \citep{ID27paper, IDsnow42paper} & Command generation; system administration \\

InterCode & Bash & \citep{ID56paper} & Interactive environments; iterative feedback \\

Exec NL2Bash & Bash & \citep{ID103paper} & Executable Bash commands; real system testing \\

RTLLM & Verilog & \citep{ID160paper, ID4paper, ID77paper, IDsnow4paper, IDsnow31paper, IDsnow15paper, IDsnow29paper} & Hardware design; RTL generation \\

Thakur-et-al & Verilog & \citep{IDsnow4paper} & Hardware benchmarking; complements other Verilog datasets \\

VerilogEval & Verilog & \citep{ID92paper, ID4paper, ID77paper, IDsnow37paper, IDsnow27paper, ID2paper, IDsnow38paper, IDsnow39paper, IDsnow41paper, IDsnow31paper, IDsnow15paper, IDsnow29paper} & Standard Verilog benchmark; hardware LLM evaluation \\

VHDL-Eval & VHDL & \citep{IDsnow38paper} & VHDL code generation; FPGA development \\

FOL-mnli, FOL-codesc & FOL & \citep{ID43paper} & Logic specification; formal reasoning \\

FOLIO & FOL & \citep{IDsnow11paper, IDsnow12paper} & Natural language to logic; reasoning tasks \\

ProofWriter & FOL & \citep{IDsnow12paper} & Proof generation; mathematical reasoning \\

LogicNLI & FOL & \citep{IDsnow11paper} & Logic inference; reasoning evaluation \\

LTL-pattern, LTL-synthesis & LTL & \citep{ID43paper} & Temporal logic; system verification \\

JigsawM & M & \citep{IDadd1paper} & M language evaluation; database programming \\

SketchGraphs dataset & CAD Sketches & \citep{ID81paper} & CAD design; geometric modeling \\

Regex-synthetic, Regex-turk & Regex & \citep{ID43paper} & Pattern matching; text processing \\

FIMO, miniF2F & Lean & \citep{ID39paper, IDsnow22paper} & Theorem proving; formal math \\

nvBench & Vega-lite & \citep{IDsnow32paper} & Data visualization; chart generation \\

TCQL & CQL & \citep{IDsnow25paper} & Query generation; corpus search \\

MCEval & 40 PLs & \citep{IDsnow8paper} & Large-scale multilingual testing; broad coverage \\

NL-to-ansible & Ansible & \citep{IDsnow42paper} & Infrastructure automation; DevOps tasks \\

UCLID5 benchmarks & UCLID5 & \citep{extend2_mora2024synthetic} &  Formal verification languages generation\\
IaC-Eval & Terraform HCL & \citep{extend3_kon2024iac} & Cloud Infrastructure-as-Code (IaC) generation\\
StackEval & 25 PLs & \citep{extend4_shah2024stackeval} & Coding assistance tasks\\
4 distinct DSL benchmarks& MapReduce, Network packet processing, TACO, tensor processing & \citep{extend5_bhatia2024verified}& Verified code transpilation\\
\hline
\end{tabular}
\end{table}

For Verilog, standard benchmarks like VerilogEval~\citep{ID92paper} and RTLLM~\citep{ID160paper} are proposed. 
Other studies \cite{ID92paper} prepared VerilogEval in two formats: VerilogEval-machine (generated by LLMs) and VerilogEval-human (manually converted). 
The VerilogEval-machine format was created using GPT-3.5 to generate problem descriptions from existing Verilog code, while VerilogEval-human involved the manual conversion of problem descriptions from HDLBits into a text-only structure, addressing ambiguities and translating visual elements like circuit diagrams and state transition graphs into textual formats. \citet{ID160paper} created 30 diverse digital designs and provides automated assessment of generated RTL across syntax correctness, functional accuracy, and design quality metrics.
\citet{ID8paper} collected Verilog datasets from GitHub and Verilog textbooks, along with an evaluation test bench that evaluates the functional correctness and syntax of Verilog code. 
Authors of \cite{IDsnow15paper} use a new dataset augmentation methodology that pairs Verilog with natural language comments so LLMs learn the mapping of RTL with high-level semantics. This data augmentation is used for generating large-scale RTL code. 
Other studies translate VerilogEval benchmark to VHDL code snippets using ICARUS Verilog tool. The problem descriptions and self-verifying tests are manually crafted, so the dataset has test cases for each problem. They also augment the data with VHDL tutorials with varying complexities~\cite{IDsnow38paper}. 

Other than these standard benchmarks, many works~\citep{ID133paper, ID112paper, IDsnow36paper, ID8paper, ID74paper, ID6paper} have developed their own custom evaluation datasets, such as Verilog problem sets \citep{ID8paper}, 15 UML models with 168 OCL specifications \citep{ID74paper},
a set of 40 coding tasks across four categories (Data Science, Games, Security, and Simple Algorithms) \citep{ID1paper}, 
a dataset of 351 test cases extracted from accredited R programming textbooks~\citep{ID34paper},
15 UML class models with a total of 168 English specifications for OCL \citep{IDsnow20paper},
and a collection of 100 prompts across 10 categories for control logic code  \citep{ID142paper}.
\citet{ID35paper} evaluated AI-assisted code generation for six fundamental HPC kernels (AXPY, GEMV, GEMM, SpMV, Jacobi Stencil, and CG) across multiple programming languages and parallel programming models.
Ansible-YAML dataset is collected by extracting data from Google BigQuery and Ansible Galaxy \cite{IDsnow42paper}. IaC-Eval is a benchmark for Cloud Infrastructure-as-Code (IaC) generation proposed by~\citet{extend3_kon2024iac}. In~\cite{extend5_bhatia2024verified}, four distinct DSL benchmarks are experimented for code transpilation tasks.
 

\subsection{Distinctive Features for LRPLs and DSLs} \label{sec:rq1-distinction}

\textbf{Domain-specific Evaluation Needs. } Most HRPLs' code generations are open-ended problems, while the DSLs are focusing on tasks such as command synthesis, configuration snippets, formula completion, and chemistry plan synthesis. These tasks, though narrow, are specialized and require expert-aware metrics to assess LLMs’ generated code.
While there are some similarities between the evaluation metrics in LRPL and DSLs and the HRPLs, such as automatic evaluations of Table 3 and accuracy, user-centric, domain-specific, and manual evaluations differ significantly. Even for similar metrics, such as the pass rate that is also used for HRPLs, the metric should be adapted to the studied language. 
For example, the pass rate metric~\cite{ID72paper} used to evaluate code generation in ODSL is defined as the percentage of test cases where the system generates a program that correctly fulfills the user’s request. 
Another example is qualitative assessment used in \cite{ID65paper}, where the generated PLC programming code outputs are evaluated by domain experts based on three key criteria: correctness, maintainability, and adherence to industry coding standards. 
CQLBLEU, is a combination of BLEU and semantic similarity metrics to evaluate the similarity between the CQL generated by the model and the reference CQL \cite{IDsnow25paper}. FOL BLEU \cite{IDsnow11paper} is also a special version of BLEU score adapted for First-Order Logic statements, which uses a custom tokenizer that splits quantifiers, operators, and terms into tokens. Even though there are similarities between these metrics and BLEU score, adaptation is required based on the domain language to use it in practice.

\textbf{Functional Correctness Complexity Beyond Run-and-Check Evaluation.} Another difference in the evaluation metrics is the \textit{functional correctness} aspect. The functional accuracy in HRPLs is assessed through metrics like pass@k. The HRPLs have abundant test cases in the benchmark datasets. However, DSLs may represent hardware, chemistry, or configurations where test oracles must be built with simulators or domain solvers, applied on a dataset, or require a controlled environment, making “run‑and‑check” costlier. 
For example, in the Verify@k metric \citep{IDsnow19paper}, a `verified definition' means that the generated code successfully passes the F* type-checker and satisfies the given specification (type). 
The Validity@k \citep{IDsnow20paper}, refers to the percentage of valid OCL constraints, where an OCL constraint is considered `valid' if it conforms with the syntactical structure and formatting rules of OCL.
The execution accuracy metric \cite{ID103paper} is defined for an execution based evaluation (EE) system for Natural Language to Bash (NL2Bash) translation. It assesses the correctness of Bash commands generated by large language models in response to natural language prompts by executing them in a controlled container environment and comparing the actual output to the expected result. Another study \cite{IDsnow25paper} also used Execution Accuracy to evaluate how well the generated CQL query executes on the corpus engine, determining whether the generated query executes correctly and returns the desired result. The Execution @k metric \citep{IDadd1paper}, on the other hand, measures the percentage of queries for which at least one of the top K generated code candidates produces the correct output when executed on the input dataset. Similarly, Compilation Rate, Simulation Rate, and Correctness Rate metrics \citep{IDsnow18paper} are used for evaluating SVA generation. The compilation rate is the percentage of generated assertions that successfully compile. The simulation rate represents the proportion of compiled assertions that can be functionally simulated without errors or timeouts. The correctness rate determines the percentage of simulated assertions that match the behavior of golden reference assertions.

\textbf{Performance and Physical Constraint Evaluation Needs for DSLs. }
Some DSL metrics require evaluating other metrics \textit{beyond functional correctness}, such as run-time performance. For example, Syn-DC metric \citep{ID77paper} employs Synopsys Design Compiler to conduct a more rigorous syntactic evaluation, extending beyond basic syntax checking to verify the physical synthesizability of the RTL design. 
Power-Performance-Area (PPA) \citep{ID4paper, IDsnow2paper, IDsnow5paper, ID160paper}, used for evaluating Verilog code generation, combines three key components: power consumption (measured in microwatts), performance (represented by clock speed in picoseconds, ps or nanoseconds, ns), and area (measured in square micrometers). 
Area-Delay Product (ADP) \cite{IDsnow3paper}, also used for evaluating Verilog code generation, is a performance estimation metric derived from synthesis results of the generated Verilog code. It combines the physical area occupied by a circuit synthesized by the Verilog code generated by the model (measured in square micrometers) with its signal propagation delay (measured in picoseconds).  VCS Simulation Time and SVA Generation Time, metrics for SVA generation, consider the time taken by the simulation process to verify the generated assertions, and the initial time taken by the framework to generate the System Verilog Assertions. 
For chemistry and CAD sketches, the metrics developed should capture physical world constraints and be chemically feasible. These are defined in metrics like NumberGenerated \cite{ID32paper} and E\_syntax and E\_stat \cite{ID81paper}. E\_stat evaluates how well the statistical properties of generated sketches match those of ground truth data, using Earth Mover’s Distance between normalized histograms of various sketch properties. For molecule generation, validity quantified the proportion of chemically feasible structures, diversity evaluated structural variability, Retrosynthesis Score estimated synthesizability, and membership gauged adherence to target molecular classes \cite{ID46paper}.

\textbf{Semantic Equivalence Handling and Tooling for DSLs.} 
Considering the \textit{semantic equivalence} of the generated code, HRPLs use metrics such as CodeBLEU, a metric that considers syntactic and semantic features of code through the abstract syntax tree and data flow. However, DSLs frequently admit multiple syntactically different yet semantically equal forms (e.g., Regex, First Order Logic, spreadsheet formulas) or face challenges when assessing semantic and syntax correctness.
For example, Bash commands can be syntactically different but semantically equivalent or they can have correct syntax but incorrect semantic~\cite{ID103paper}.  
Therefore, specific metrics are developed to ignore irrelevant surface differences to reward latent intent alignment. Semantic Accuracy, Sketch Match, and Logical Equivalence are among these metrics. 
Semantic Accuracy \cite{ID43paper, semantic-accuracy} is developed to assess Regex translations, addressing the flexibility in Regex formulation where syntactically distinct patterns can be functionally equivalent. This metric uses an equivalence check, accounting for predictions that diverge syntactically but maintain semantic equivalence. 
Sketch Match \citep{ID42paper} is used for evaluating formula completion tasks. It is similar to exact match but ignores differences in cell references and constants. Consider a formula completion example: \texttt{=B2<=EDATE(TODAY(),-33)}, where the number \texttt{-33} is treated as a token type rather than a specific value, allowing for structural comparison while disregarding context-dependent constants. 
Logical Equivalence \cite{IDsnow11paper} is a sophisticated metric designed to quantify the similarity between two First-Order Logic (FOL) rules by comparing their truth tables. The process involves parsing the FOL rules, matching their literals using a greedy search algorithm based on edit distance, generating truth tables for each rule, and then comparing these tables row by row.
While there are several tools for HRPLs to assess syntax and semantic features, the semantic equivalence of DSLs requires specific metrics and techniques that are specialized to that language and can barely be adopted for other languages.

\textbf{Human Usability and Expert Judgment for DSLs.} In DSL evaluations, human usability is considered with specific metrics defined for it, while in HRPLs, manual evaluations are occasionally conducted. For Ansible and PLC, domain experts care about ordering, module aliases, conventions, and industry standards that tests miss \cite{ID65paper}. Therefore, the custom composite scores mimic expert judgment. The Ansible Aware Metric \citep{ID6paper, IDsnow42paper} is a custom evaluation metric designed to assess the semantic correctness and practical usability of generated Ansible code. It analyzes the structure of Ansible tasks and playbooks, considering the order of key-value pairs, optional elements, module names, keywords, and parameters. The metric accounts for Ansible-specific features such as equivalent modules and flexible ordering, aligning closely with how an Ansible user would evaluate code quality. This score reflects how well the generated code matches the expected content and structure from a practical, user-oriented perspective. 
The Schema Correct Metric is a binary evaluation metric that determines whether generated Ansible code adheres to the formal Ansible schema. Using the Ansible linter’s schema as a reference, it validates the structural correctness of the generated code without comparing it to a target. This metric ensures that the output will be accepted by Ansible tools and linters, identifying potential structural issues that might not be apparent from other evaluation methods. Similarly, in~\cite{extend2_mora2024synthetic}, empirical evaluation was conducted manually for UCLID5 formal modeling tasks.

To summarize, for exact match and similarity, DSLs often need custom variants of similar metrics used in HRPLs. DSLs rely on custom validators or simulators and need domain-specific executors or symbolic checks for syntax validity and semantic correctness. Even for functional correctness, DSLs may lack a runtime engine or require hardware simulations (e.g., Verilog), which complicates the automated evaluation. In metrics like CMD Acc, the evaluation metrics emphasize domain-relevant primitives (e.g., commands, assertions) rather than general syntax. The human evaluations often require expert domains, and resource constraints and logical equivalence are also used for some DSLs. 

\textbf{Evaluation Paradigm Differences and LLM-as-Judge Limitations.}
In HRPLs, most evaluation metrics are based on similarity scores or functional correctness, such as exact match, BLEU, ROUGE, METEOR, CodeBLEU, and Pass@k~\citep{jiang2024survey}. The human evaluations often focus on certain aspects, such as code styles and code quality. Some studies focus on alignment of the generated code with human values through reinforcement learning to overcome the human evaluation challenges, and recent studies use LLMs as a judge for the generated code. 
While these techniques are valuable, the scarcity of data and the specialized domain of the LRPLs and DSLs make it difficult to use LLMs-as-judge or provide enough data to train models that align with coding standards and domain expert values.

\textbf{Domain Expert Requirements for Manual Assessment. } Manual evaluations in LRPLs and DSLs also differ from the evaluations conducted in HRPLs. In contrast to manual evaluation in HRPLs that can be done by general software engineers or being crowd-sourced, manual evaluations in LRPLs and DSLs require domain experts (e.g., chemists, HPC engineers, formal methods researchers) due to domain-specific syntax/semantics. They focus on domain semantics, adherence to specialized models (e.g., OpenMP), and logical soundness, in addition to functional correctness and usage patterns. Additionally, the rubrics often involve task-specific or domain-adapted rubrics, such as proof validity check and adherence to model semantics, while in HRPL, the metrics are adopted from software engineering best practices, such as readability or maintainability of the generated code. Furthermore, while in HRPLs the whole function is investigated, in LRPL and DSL, sub-components and different aspects of the generated code can be assessed, such as proof steps, constraint formulations, or workflow stages. Finally, the tool supports in LRPLs and DSLs is not as extensive as in HRPLs, thus, reducing manual workload is not necessarily available.

\textbf{Limited Benchmark Diversity and Manual Translation Efforts.} 
The construction of benchmark datasets mostly covers LRPLs and some DSLs like Verilog. However, the diversity of benchmarks in other languages or even their coverage is low.
Techniques like collecting data from textbooks and tutorials, automated data augmentation, domain-specific data augmentation, and code translation are used~\citep{IDsnow15paper, IDsnow38paper, ID8paper}.
For many LRPLs, approaches like MulriPL-E and xCodeEval develop frameworks with compilers, DSL representations, and execution engines, to translate benchmarks to other languages ~\cite{ID70paper, pldist, ID93paper}. 
Many of the existing variations of HumanEval benchmark are also manual translations. Similar to the above-mentioned points, developing benchmarks, specifically for DSLs requires special techniques and tools and domain expertise, which is different than how benchmarks are developed for HRPLs.

\textbf{Specialized Validation Infrastructure Requirements.} The validation and quality assurance processes for LRPL/DSL datasets require domain-specific tools that don't exist for general-purpose languages. Unlike HRPLs where syntax checking provides basic validation, these domains integrate specialized validators directly into dataset pipelines. Verilog datasets use synthesis tools to verify design correctness, Bash command datasets execute in sandboxed environments \cite{ID103paper}, and formal methods datasets incorporate proof checkers to ensure mathematical validity. This executable validation provides quality signals unavailable for mainstream languages but requires significant infrastructure investment and domain expertise. To summarize, what makes the creation of benchmarks for LRPLs and DSLs is the scarcity of data and the necessity of special techniques tailored for each language for many of these languages. This is specifically more pronounced for DSLs. The efforts to create these benchmarks are beyond mere data collection or manual translation efforts. For example, execution environment is set up for Bash~\cite{ID103paper}, compilers or execution engines are developed~\cite{ID70paper, pldist, ID93paper}, and toolchains are used for code translation~\cite{IDsnow38paper}.


\begin{tcolorbox}[
  breakable,
  colback=gray!10,
  colframe=gray!40,
  boxrule=0.5pt,
  title=Summary RQ1,
  coltitle=black
]
\begin{enumerate}
\item The LLaMA family emerged as the most popular choice for LRPL and DSL code generation, followed by DeepSeek and StarCoder families. We also examined the distribution of papers using proprietary models in the surveyed papers.
\item We categorized the \textit{evaluation} done in the literature into four categories: Automatic Evaluation, User Centric Evaluation, Domain Specific Evaluation, and Manual Evaluation. Additionally, we summarized all the \textit{metrics} used in the literature. While Automatic Evaluation Metrics such as Pass@k are prevalent, many researchers have used Domain Specific Evaluation Metrics to measure the code generation ability of LLMs.
\item We cataloged the common \textit{benchmarks} used in the literature to evaluate the code generation abilities of Language Models in LRPL and DSLs. Due to the unavailability of standardized benchmarks in many DSLs, researchers have created their own evaluation datasets.

\item While the evaluation metrics and some benchmarks are similar to HRPLs, specifically for LRPLs, there are several differences and challenges that needs to be addressed for DSLs. Even for LRPLs, different techniques such as code translation and developing compilers are adopted to make the benchmark datasets.  
\end{enumerate}
\end{tcolorbox}

\section{Strategies and Methodologies for Enhancing LLM Performance }
\label{sec:rq2}

Despite the remarkable advancements in LLMs, significant challenges persist in their ability to generate high-quality parallel code and handle specialized programming tasks~\cite{ID88paper} or formally prove any statements from the International Mathematical Olympiad~\citep{ID39paper}. 
Struggling with LRPLs compared to HRPLs in xCodeEval \citep{ID93paper}, missing core problems or specific hardware design concerns \citep{ID120paper}, weak visualization code for R \citep{ID34paper}, and struggling with COQ syntax and semantics \citep{ID31paper} are among some issues. 
Studies report multiple iterations (117 rounds) required to produce correct and implementable code \citep{ID99paper}, low-quality outputs \citep{ID35paper}, security concerns \citep{ID75paper}, and requiring human intervention to fix errors for generating comprehensive testbench for hardware design and verification~\citep{ID33paper}. 
To address the challenges faced by LLMs in LRPL/DSL, researchers have developed various strategies and techniques. 
Table~\ref{tab:techniques} presents the main techniques we found in the literature.

\begin{table}[ht]
\centering
\small
\begin{tabular}{p{0.25\linewidth}p{0.35\linewidth}p{0.3\linewidth}}
\toprule
\textbf{Main Category} & \textbf{Sub-method} & \textbf{References} \\
\midrule
Model adaptation techniques & Pre-training & \cite{ID60paper, ID157paper, ID49paper, ID6paper, ID67paper, ID113paper, ID19paper, ID84paper, codellama, ID18paper, ID82paper, ID117paper, ID76paper, IDsnow23paper, IDsnow30paper} \\
\cline{2-3}
& Cross lingual transfer & \cite{ID105paper} \\
\cline{2-3}
& Fine-tuning (including domain-specific fine-tuning) & \cite{ID73paper, ID43paper, ID31paper, ID92paper, ID107paper, ID66paper, ID6paper, ID93paper, ID57paper, IDsnow4paper, IDsnow8paper, IDsnow19paper, IDsnow22paper, IDsnow23paper, IDsnow25paper, IDsnow31paper, IDsnow34paper, IDsnow37paper, IDsnow36paper, IDsnow38paper} \cite{extend1_codem}\\
\cline{2-3}
& Parameter-Efficient Fine-Tuning (PEFT) \newline methods & \cite{ID65paper, ID7paper, ID75paper, IDsnow11paper, ID15paper, IDsnow27paper, IDsnow29paper, IDsnow39paper} \\
\midrule
Prompting and iterative \newline techniques & Prompting strategies & \cite{ID56paper, ID8paper, ID113paper, ID147paper, ID71paper, ID74paper, ID46paper, ID75paper, ID39paper, ID78paper, IDsnow38paper, IDsnow2paper, IDsnow5paper, IDsnow7paper, IDsnow10paper, IDsnow11paper, IDsnow14paper, IDsnow18paper, IDsnow20paper, IDsnow25paper, IDsnow32paper, IDsnow40paper} \\
\cline{2-3}
& Iterative feedback & \cite{ID33paper, ID65paper, ID99paper, ID4paper, ID32paper, ID12paper, ID63paper, IDsnow5paper, IDsnow26paper, IDsnow33paper, IDsnow40paper} \cite{extend2_mora2024synthetic,extend5_bhatia2024verified}\\
\midrule
Novel architectural and objective approaches & Novel objectives for training & \cite{ID77paper, ID133paper} \\
\cline{2-3}
& Novel architectures & \cite{ID2paper, ID133paper} \\
\midrule
Input, output and token processing & Prefix generation and variable replacing & \cite{ID129paper} \\
\cline{2-3}
& Tokenization & \citep{ID80paper} \\
\cline{2-3}
& Decoding  & \cite{ID48paper, ID46paper, IDsnow3paper, IDsnow24paper, IDsnow42paper} \\
\midrule
Retrieval-augmented generation & & \cite{ID72paper, ID27paper, ID78paper, IDsnow19paper, IDsnow29paper, IDsnow42paper} \\
\midrule
Others & Own-DSL creation  & \cite{ID112paper, ID147paper, ID72paper, ID81paper} \\
& Temperature mixing  & \cite{IDadd1paper} \\
\cline{2-3}
& Early stopping &  \cite{ID129paper} \\
\cline{2-3}
& Knowledge distillation &  \cite{IDsnow15paper, IDsnow33paper} \\
\bottomrule
\end{tabular}
\caption{Techniques for code generation in LRPLs and DSLs}
\label{tab:techniques}
\end{table}



\subsection{Model Adaptation Techniques}
\label{subsec:model_adaptation_tech}

\textbf{Pre-Training.}
Several studies demonstrate the effectiveness of adapting pre-training approaches to specialized contexts, leading to performance improvement. 
These studies have made modifications to account for the unique characteristics and constraints of LRPLs and DSLs.
A 7B parameter model from scratch was pre-trained to optimize code, introducing two auxiliary tasks alongside producing the list of optimizing passes: generating instruction counts of the code before and after optimization, and generating the output Intermediate Representation (IR) after the optimization is applied \citep{ID60paper}. 
To bridge the gap between open-source and closed-source models in code generation performance, a 2.7B parameter multilingual model, PolyCoder,
was pre-trained on data in 12 programming languages~\citep{ID157paper}.

Similarly, other studies pre-trained different models for hardware design called Hardware-Phi-1.5B, specifically trained on datasets relevant to hardware design \citep{ID49paper}, sketch language to convert CAD sketches into token sequences \citep{ID81paper}, to generate code for Ansible-YAML \citep{ID6paper}, and PowerShell code \citep{IDsnow23paper}.
SketchGen, a novel approach using a specialized sketch language to convert CAD sketches into token sequences, was developed in~\citep{ID81paper}. SketchGen generates both primitives and constraints, preserving structural relationships critical for CAD functionality. 
For PowerShell code generation, the pre-training addressed the models' struggle with zero-shot learning, where the model often reverted to generating Python code instead~\citep{IDsnow23paper}. The researchers introduced a pre-training phase using a large corpus of general PowerShell code. This domain adaptation step was designed to familiarize the models with PowerShell's unique syntax and conventions, providing a stronger foundation for the subsequent fine-tuning on offensive security tasks.

FLAME is a fairly small model (60M) trained for Excel formulas, based on T5 architecture that is pre-trained exclusively on Excel formulas~\cite{ID42paper}. 
Despite its relatively small size, FLAME competed effectively with much larger models like Codex (175B parameters) on domain-specific tasks such as last-mile repair and formula completion.
ShellGPT and Equivalent Command Learning (ECL), a pre-training technique addressing limited high-quality data challenges for Shell scripting, was also introduced in~\cite{ID130paper}. 
ECL teaches models to recognize functionally equivalent commands, such as \texttt{cat file.txt | grep 'pattern'} and \texttt{grep 'pattern' file.txt}. By learning these equivalences, ShellGPT enhances its understanding of command intent, leading to more flexible and robust Shell script generation. This approach effectively expands the training dataset and improves the model's ability to generate appropriate commands across different environments.

To provide a balanced distribution of languages in the training corpus, Unimax algorithm was used to limit the maximum number of times a language's data can be duplicated, leading to $66.48\%$ improvement in pass@k metric for low-resource languages, while seeing only a $12.94\%$ decrease on high-resource languages~\citep{ID67paper}.
The authors found that a more balanced distribution of languages in the training corpus significantly improved performance on LRPLs, with only a minimal impact on high-resource programming languages. 
On the other hand, `knowledge spillover,' where language models demonstrated capabilities in programming languages they were not explicitly trained on, was investigated in \citep{ID113paper}.
This effect is attributed to the presence of cross-language data in the training corpus, such as code snippets embedded in comments or multi-language projects. The spillover effect explains how models can acquire incidental knowledge of various programming languages during pre-training, contributing to their ability to generalize across languages and perform tasks in low-resource or unseen programming languages.

\textbf{Cross-Lingual Transfer.}
\citet{ID105paper} utilized the LLVM compiler's Intermediate Representation (IR) as a shared interlingual to ground heterogeneous source code languages, aligning languages to an IR. By creating a parallel dataset of nearly $4$ million pairs of source code and corresponding LLVM IR across $12$ programming languages and continuation of pre-training a code language model, they enhanced cross-lingual transfer for code generation, showing consistent performance gains both for high-resource and low-resource languages, notably for languages like D, Ruby, and Swift.
A key insight was that the IR provided a semantically rich representation that could capture higher-level programming concepts like control and data flow, which the authors hypothesized were not fully grasped by existing code LLMs despite their large-scale pre-training. Importantly, the authors observed that this IR-based transfer learning did not exhibit the negative interference between languages often seen in multilingual models, suggesting that the IR effectively served as a universal anchor for code understanding across diverse programming languages.

\textbf{Fine-Tuning on Specialized Datasets.}
Many Works have gathered high-quality data in specific LRPLs like Rust~\citep{ID73paper}, Kotlin~\citep{IDsnow34paper}, PowerShell~\citep{IDsnow22paper} and DSLs like Verilog~\citep{IDsnow37paper, IDsnow31paper}, Regex~\citep{ID43paper}, FOL~\citep{ID43paper}, LTL~\citep{ID43paper}, COQ~\citep{ID31paper} and fine-tuned models on these datasets. This process has led to better performance compared to the base model, i.e., non-fine-tuned models on these languages, sometimes performing better or on par with much larger models like GPT-4/GPT-3.5 \citep{IDsnow22paper, ID92paper, IDsnow23paper, IDsnow25paper, IDsnow37paper}. 
Table~\ref{tab:model_improvement} showcases examples of the improvement in performance metrics from base models to their fine-tuned counterparts across various LRPLs and DSLs. 
Other studies also reported that fine-tuning improved the performance of the models, such as reducing syntax errors and handling new variable names and operation descriptions. 
These studies include instruction tuning using git commits for Rust \citep{ID73paper}, or fine-tuning on Kotlin exercises \citep{ID34paper}, DSLs like Regex, FOL, and LTL for Regex \citep{ID43paper}, COQ dataset \citep{ID31paper}, and Verilog \citep{ID92paper, IDsnow31paper, IDsnow37paper, ID141paper}.
Fine-tuning has improved automated theorem proving \citep{IDsnow22paper} and the generation of more accurate and contextually appropriate PowerShell code for security applications \citep{IDsnow23paper}.

\begin{table}[htbp]
\centering
\small
\begin{tabular}{@{}
    p{0.04\linewidth}
    p{0.24\linewidth}
    p{0.10\linewidth}
    p{0.15\linewidth}
    p{0.10\linewidth}
    p{0.07\linewidth}
    p{0.07\linewidth}
    @{}}
\toprule
Paper & Benchmark & Lang. & Model & Metric & Base & Fine-tuned \\
\midrule
\citep{ID73paper} & HumanEval & Rust & Starcoder & pass@1 & 21.8 & 23.4 \\
\citep{ID43paper} & Regex-turk & Regex & T5 & acc. & 58.0 & 64.2 \\
\citep{ID43paper} & FOL-mnli & FOL & T5 & acc. & 46.9 & 53.9 \\
\citep{ID43paper} & FOL-codesc & FOL & T5 & acc. & 58.6 & 59.0 \\
\citep{ID43paper} & LTL-syhnthesis & LTL & T5 & acc. & 87.5 & 87.9 \\
\citep{ID66paper} & MultiPL-E & Rust & CodeLLaMA-PY & pass@1 & 27.0 & 40.3 \\
\citep{ID57paper} & HumanEval-Haskell & Haskell & CodeGPT & ExMatch & 23.2 & 40.0 \\
\citep{IDsnow4paper} & Thankur-et-al. & Verilog & LLaMA2 & pass@5 & 41.2 & 70.6 \\
\citep{IDsnow8paper} & MCEval & Rust & CodeQwen-1.5 & pass@1 & 47.2 & 67.9 \\
\citep{IDsnow34paper} & HumanEval-Kotlin & Kotlin & CodeLLAMA & pass@1 & 26.1 & 42.2 \\
\citep{IDsnow34paper} & HumanEval-KOTLIN & Kotlin & DeepSeek & pass@1 & 41.0 & 55.3 \\
\bottomrule
\end{tabular}
\caption{Performance improvement from base models (i.e., not fine-tuned) to fine-tuned models across various benchmarks and languages}
\label{tab:model_improvement}
\end{table}


Fine-tuning on carefully filtered datasets like KStack-clean resulted in reduced syntax error rates and enhanced code completion capabilities, highlighting the importance of data quality over quantity in the fine-tuning process for Kotlin code generation tasks~\citep{ID34paper}. For DSLs, \citet{ID43paper} fine-tuned T5 model on Regex, FOL, and LTL could handle new variable names, nouns, and operator descriptions not present in the training data, implying that LLMs maintain their generalization capabilities to DSLs. 
For fine-tuning on Verilog, a synthetical problem-code pair was generated by leveraging LLMs to produce descriptions for existing Verilog modules from GitHub data~\citep{ID92paper}. Similarly, a large synthetic dataset of 8 million formal statements with proofs was generated and used for fine-tuning for theorem proving~\citep{IDsnow22paper}.

For Verilog generation, \citet{IDsnow31paper} proposed CodeV employing multi-level summarization to generate high-quality instruction-tuning data from real-world Verilog code. Similarly, the dataset used by \citet{IDsnow37paper} in BetterV was instrumental in fine-tuning their model and improving its performance for Verilog generation. Their approach involved collecting and rigorously filtering open-source Verilog repositories, extracting individual modules, and ensuring syntactic correctness. The inclusion of C program translations of Verilog code facilitated knowledge transfer from general programming to hardware description languages. This carefully curated dataset, combined with their data augmentation techniques, enabled effective domain-specific instruction tuning and contributed to BetterV achieving state-of-the-art performance on the VerilogEval benchmark~\citep{ID92paper}. \citet{ID141paper} employed a two-step fine-tuning process to improve the model's performance. First, they conducted large-scale fine-tuning containing 90 repositories, which helps the model learn general patterns and structures in Verilog. This is followed by small-scale fine-tuning derived from 10 repositories, allowing the model to capture more localized and project-specific code patterns. This approach combines the benefits of learning from a broad dataset with the ability to adapt to specific coding styles and conventions.

In summary, several studies have used fine-tuning on LRPL or DSL and demonstrated its impact on improving code generation for LRPLs and DSLs. These efforts have not only improved model performance in their respective domains but have also shown that carefully curated, domain-specific datasets can enable smaller, specialized models to compete with or even outperform much larger, general-purpose language models. Table~\ref{tab:gpt_comparison} focuses on example works where fine-tuning has led to results that are competitive with or superior to those achieved by GPT-3.5 or GPT-4. This comparison is particularly noteworthy, as it demonstrates that smaller, specialized models can often match or outperform much larger, general-purpose models across various benchmarks and evaluation metrics, when trained on high-quality, domain-specific data.

\begin{table}[htbp]
\centering
\footnotesize
\begin{tabular}{@{}
    p{0.04\linewidth}
    p{0.14\linewidth}
    p{0.08\linewidth}
    p{0.12\linewidth}
    p{0.10\linewidth}
    p{0.10\linewidth}
    p{0.12\linewidth}
    p{0.10\linewidth}
    @{}}
\toprule
Paper & Benchmark & Lang. & Metric & GPT Model & GPT Perf. & Fine-tuned Model & Model Perf. \\
\midrule
\citep{IDsnow22paper} & FIMO & Lean & \#Solved & GPT-4 & 0/148 & DeepSeek-FT & 5/148 \\
\citep{IDsnow22paper} & miniF@F-test & Lean & pass@k & GPT-4 & 27.5 & DeepSeek-FT & 52.0 \\
\citep{ID92paper} & VerilogEval & Verilog & pass@1 & GPT-3.5 & 46.7 & CodeGEn & 46.2 \\
\citep{IDsnow23paper} & curated-test & PowerShell & BLEU-4 & GPT-3.5 & 0.07 & CodeGPT-Ft & 0.21 \\
\citep{IDsnow23paper} & curated-test & PowerShell & ROUGE-L & GPT-3.5 & 0.2 & Code-T5+ & 0.38 \\
\citep{IDsnow23paper} & curated-test & PowerShell & ED & GPT-3.5 & 0.33 & CodeGPT & 0.5 \\
\citep{IDsnow23paper} & curated-test & PowerShell & METEOR & GPT-3.5 & 0.22 & CodeT5+ & 0.47 \\
\citep{IDsnow25paper} & EnWiki & CQL & CQLBLEU & GPT-4 & 89.9 & BART-English & 82.1 \\
\citep{IDsnow31paper} & Verilog-Machine & Verilog & pass@1 & GPT-4 & 60.0 & Codellama & 78.1 \\
\citep{IDsnow31paper} & Verilog-Human & Verilog & pass@1 & GPT-4 & 43.5 & CodeQwen & 53.2 \\
\citep{IDsnow37paper} & Verilog-Machine & Verilog & pass@1 & GPT-4 & 60.0 & CodeQwen & 68.1 \\
\citep{IDsnow37paper} & Verilog-Human & Verilog & pass@1 & GPT-4 & 43.5 & CodeQwen & 46.1 \\
\bottomrule
\end{tabular}
\caption{Models performing better than or on par with GPT-3.5/GPT-4 across various benchmarks and languages. Performance is abbreviated as Perf. in the table.}
\label{tab:gpt_comparison}
\end{table}

\textbf{Parameter Efficient Fine-Tuning (PEFT).}
PEFT techniques like Low-Rank Adaptation (LoRA)~\citep{lora} allow for effective domain adaptation and task-specific tuning ~\citep{ID7paper} and are used to improve code generation in various LRPL and DSL code generation tasks. 
Many works have used LoRA~\citep{ID7paper, ID65paper,ID75paper, IDsnow11paper, IDsnow15paper} for LRPL and DSL code generation, including YAML-based GitHub workflow generation \citep{ID75paper}, Programmable Logic Controllers (PLCs) and Targeting Structured Text (ST) language \citep{ID65paper}, and First-Order Logic (FOL) \citep{IDsnow11paper}. 
For generating FOL from natural language, the fine-tuned LlaMA model called LOGICLlaMA, encompassed tasks such as direct translation, naive correction, and Chain of Thought correction via both supervised fine-tuning and Reinforcement Learning. LoRA's key benefit was enabling the fine-tuning process on a single GPU, making it more accessible and cost-effective~\citep{IDsnow11paper}.

A few studies have used other approaches like Quanitized LoRA (QLoRA)~\citep{IDsnow39paper}, while others have proposed new PEFT methods like FLoRA (Fast LoRA)~\citep{ID7paper}.
FLoRA allows each input example in a minibatch to be associated with unique low-rank adaptation weights. This innovation enables efficient batching of heterogeneous requests, potentially improving throughput in serving diverse user queries in real-world applications while maintaining the parameter efficiency of LoRA~\citep{ID7paper}.

\subsection{Prompting Strategies and Iterative Techniques}
\label{subsec:prompting}


\textbf{Prompting Strategies.}
Various prompting strategies were employed for LRPL and DSL generation from LLMs.
Table~\ref{tab:prompting_methods} provides a list of different types of prompting methods that were used, including zero-shot, few-shot, and other techniques like Chain of Thoughts.
High-level, pseudo-code-like prompts for Verilog \citep{ID36paper}, 
persona-based system prompts with details for workflow generation \citep{ID75paper}, and highly detailed prompts for hardware security \citep{IDsnow18paper}, are examples of different prompting used to improve the models' performance. 
These studies collectively underscore the importance of comprehensive context in prompts, particularly for specialized tasks; while also suggesting that the optimal level of detail may vary depending on the specific task and desired output~\citep{ID36paper, IDsnow18paper, ID75paper}.

Several works have also reported the superiority of few-shot prompting over zero-shot approaches in LLMs~\citep{ID8paper, ID74paper, IDsnow32paper, ID147paper, IDsnow38paper}.
When using few-shot prompting, the examples should be crafted carefully and can result in better accuracy compared to zero-shot in tasks related to chart generation~\citep{IDsnow32paper}. 
Interestingly, they noted that zero-shot performed better on some specific tasks like scatter plots, suggesting that the effectiveness of few-shot prompting can vary depending on the particular subtask or domain.
Similarly, for VHDL code generation, marginal improvements were observed using few-shot compared to zero-shot prompting~\citep{IDsnow38paper}.
The authors attributed this limited improvement partly to the context length constraints of the models and the inherent verbosity of VHDL code, which restricted the number and complexity of examples that could be included in the prompt.
The results of these works suggest that examples may not always generalize well across all tasks, emphasizing the need for careful consideration when selecting and crafting few-shot examples.

In contrast, \citet{ID74paper} observed significant improvements in generating OCL constraints when using few-shot prompting with Codex. By including examples of natural language specifications along with their corresponding correct OCL constraints, they guided the model to produce more syntactically valid and semantically accurate constraints.
Furthermore, adding comments to few-shot examples improved GPT-4's ability to generate complete invariants, particularly for specifying vector sizes in verification tasks \citep{ID113paper}. This highlights how thoughtful example design can guide the model's attention to crucial aspects of the task. 
Another study dynamically selected examples in their prompts for natural language to first-order logic translation, which helped demonstrate the desired format and complexity while reducing the risk of overfitting~\citep{IDsnow11paper}.

Other works \citep{ID56paper} have explored domain-specific prompting strategies such as Try Again, ReAct~\citep{Yao2022ReActSR}, and Plan \& Solve~\citep{Wang2023PlanandSolvePI}. 
\citet{ID56paper} has shown that more flexible frameworks like ReAct, which do not enforce a particular logical formula or roadmap, are more conducive to eliciting a broader set of reasoning capabilities compared to imperative reasoning paradigms like Plan \& Solve. Their study found that ReAct generally achieves higher success rates with fewer turns and more admissible commands compared to other strategies like Try Again. This suggests that prompting strategies that allow for more dynamic and adaptable reasoning processes can lead to improved performance on interactive coding tasks.

Inspired by the existing Program of Thought (PoT) prompting strategy, Multi-PoT was proposed \citep{ID71paper} where the model is prompted to generate code in multiple programming languages in few-shot setting, including low-resource languages such as R. The generated code in each language is executed to get an answer for the problem at hand, and finally it integrates the results obtained from executing code in each language through a voting mechanism, capitalizing on the strengths of different languages while increasing solution diversity. This technique consistently outperformed single-language prompting approaches across various tasks and models, demonstrating its effectiveness in enhancing the reasoning capabilities of large language models.

Grammer Prompting for DSL code generation \citep{ID46paper}, Intermediate Target Prompting for knowledge transfer to LRPLs \citep{IDsnow7paper}, multi-stage approach using a prompt manager for hardware design tasks \citep{IDsnow2paper} and hardware verification \citep{IDsnow14paper}, `self-planning' for RTL generation tasks \citep{ID160paper},
Hierarchical Prompting for LLM-based Chip Design for generating complex hardware description language (HDL) \citep{IDsnow40paper},
and structured prompting for hardware design~\citep{IDsnow5paper}, are among other proposed techniques.

For DSL generation, \citet{ID46paper} proposed Grammar Prompting, which enhances few-shot domain-specific language generation by augmenting examples with minimally sufficient specialized grammars. During inference, the LLM predicts a grammar for the input, then generates output following those rules, incorporating domain-specific constraints. This technique has proven effective across diverse domains, including semantic parsing, molecule generation, and planning tasks, showcasing its versatility for various DSLs and applications.

To transfer knowledge from HRPLs to LRPLs, \citet{IDsnow7paper} proposed Intermediate Target Prompting. First, the model is prompted to generate a solution in an intermediate language, which could be another programming language, natural language, or pseudo-code. Then, the model is prompted again with both the original task description and the intermediate solution to generate code in the target language. They observed that larger models tend to benefit more from this prompting technique.

Breaking down complex tasks into smaller, more manageable steps has been shown to significantly improve the performance of language models in various domains. This approach, often referred to as multi-stage or hierarchical prompting, allows for a more structured and controlled generation process. 
For instance, in hardware design tasks, a multi-stage approach using a prompt manager is implemented, which generates a sequence of interconnected prompts rather than a single prompt~\cite{IDsnow2paper}. This method proved particularly effective for complex software metadata processing. 
Similarly, \citet{IDsnow14paper} demonstrated the efficacy of this approach in generating assertions for hardware verification by decomposing the task into three main steps: specification analysis, signal mapping, and assertion generation. Each step utilized a specialized language model, allowing for more focused and effective prompts at each stage. By breaking down complex tasks in this manner, researchers have found that language models can better handle intricate problems, leading to improved accuracy and more reliable outputs in specialized domains.

Building on this trend of multi-stage approaches, \citet{ID160paper} proposed a novel prompt engineering technique called `self-planning' for RTL generation tasks. The technique involves a two-step process: first, the LLM generates a plan and reasoning steps for the design, along with syntax error prevention advice. Then, both the original design description and this generated plan are used to produce the final RTL design. This approach allows the model to strategize before implementation, potentially reducing errors and improving design quality. The authors found that applying self-planning to GPT-3.5 significantly boosted its performance, bringing it close to the state-of-the-art results achieved by GPT-4.

Similarly, \citet{IDsnow40paper} proposed hierarchical prompting for LLM-based Chip Design for generating complex Hardware Description Language (HDL) code. This approach breaks down complex hardware designs into simpler submodules, allowing LLMs to generate code for each part sequentially. The authors presented two methods: Human-Driven Hierarchical Prompting (HDHP), where humans provide the structure, and Purely Generative Hierarchical Prompting (PGHP), where the LLM generates both the hierarchy and code. Results showed improvements across all tested LLMs. Hierarchical prompting also reduced generation time by up to $73.73\%$ for complex modules.
In another work, \citet{IDsnow5paper} implemented a structured prompting approach that shares similarities with hierarchical prompting. Their method breaks down complex baseband hardware designs into simpler submodules, using a five-part prompt structure for each component: identity, I/O specifications, function description, examples, and cautions. This structured approach guides the LLM to generate accurate Verilog code for individual modules. The framework then uses higher-level prompts to connect these submodules, mirroring the bottom-up design process common in hardware engineering.

\begin{table}[htbp]
\centering
\small
\begin{tabular}{@{}p{0.4\linewidth}p{0.35\linewidth}p{0.2\linewidth}@{}}
\hline
Prompting Method & Languages & Reference \\
\hline
\multirow{3}{*}{Zero-shot} & Verilog & \citep{ID8paper} \\
 & GitHub-YAML & \citep{ID75paper} \\
 & OCL & \citep{IDsnow20paper} \\
\hline
\multirow{8}{*}{Few-shot} & Verus & \citep{ID63paper} \\
 & GitHub-YAML & \citep{IDsnow10paper} \\
 & SVA & \citep{IDsnow18paper} \\
 & CQL & \citep{IDsnow25paper} \\
 & Lean  & \citep{IDsnow36paper} \\
 & Lean & \citep{ID39paper} \\
 & Ruby, Kotlin, Scala, Swift, Perl & \citep{ID113paper} \\
\hline
\multirow{4}{*}{Zero-shot and Few-shot} & OCL & \citep{ID74paper} \\
 & Fortran & \citep{ID78paper} \\
 & Vega-lite & \citep{IDsnow32paper} \\
 & VHDL & \citep{IDsnow38paper} \\
\hline
\multicolumn{3}{@{}l}{\textit{Other Prompting Methods}} \\
ReAct, Plan and Solve, Try Again & Bash & \citep{ID56paper} \\
Program of Thought (PoT) & R & \citep{ID71paper} \\
Grammar Prompting & SMILES, PDDL, GeoQuery, Overnight-Blocks, SMCalFlow & \citep{ID46paper} \\
Prompt Manager &  Verilog & \citep{IDsnow2paper} \\
Intermediate Target Prompting & Rust & \citep{IDsnow7paper} \\
Chain of Thought (CoT) & FOL & \citep{IDsnow11paper} \\
Multi Step Prompting & SVA & \citep{IDsnow14paper} \\
Self Planning & Verilog & \citep{ID160paper}\\
\multirow{2}{*}{Hierarchical Prompting} &  Verilog & \citep{IDsnow40paper} \\
 &  Verilog & \citep{IDsnow5paper} \\
\hline
\end{tabular}
\caption{Prompting methods, languages, and their corresponding references}
\label{tab:prompting_methods}
\end{table}

\textbf{Iterative Feedback.}
Large language models, while powerful, can sometimes produce code with syntax errors or functional issues, especially in the case of DSLs~\citep{IDsnow5paper}. To overcome this issue, various works have utilized iterative feedback that improves the quality and correctness of the generated code~\citep{ID33paper, ID65paper, ID99paper, ID4paper, ID32paper}. 
In this approach, the LLM is repeatedly prompted with error messages asking it to rectify the error. 
This iterative process mirrors how human experts might debug and refine code~\citep{IDsnow5paper}. 
Many works have shown that this feedback loop helps in generating not only syntactically correct code, but also leads to functionally executable code~\citep{ID32paper, ID63paper}. 
To provide a comprehensive overview of these iterative feedback approaches, we present a summary in Table \ref{tab:feedback-summary}.
Iterative feedback is used in various domains, including hardware designs and testbenches in Verilog~\citep{ID33paper}, 
Verilog Code generation \citep{ID4paper}, 
verifiable code for Programmable Logic Controllers (PLCs) in Industrial Control Systems \citep{ID65paper}, automated hardware description code generation \citep{ID99paper}, 
generating proofs for Rust programs verified by Verus \citep{ID63paper},
and Planning Domain Definition Language (PDDL) domains \citep{IDsnow26paper}.

For hardware designs and testbenches in Verilog, \citet{ID33paper} proposed an iterative feedback that involves a cycle of prompting the LLM, followed by compiling the generated code using iVerilog, and providing error feedback to the model for correction. Initially, a simple feedback using the prompt ``please provide fixes'' is provided. If the error still persists, Human Feedback is given. The authors categorize human feedback into three levels of increasing specificity. Simple Human Feedback involves stating the general type of Verilog error causing the issue. Moderate Human Feedback provides more directed information to identify the specific error. Advanced Human Feedback precisely points out the error location and suggests a method for fixing it. The iterative process showed improvements in design quality, particularly for GPT-4, which was able to correct many errors with simple human input. However, the study revealed that even advanced models often required multiple iterations and sometimes substantial human guidance, especially for complex testbench creation.

In the same domain, for Verilog Code generation, \citet{ID4paper} proposed VeriRectify that involves ICARUS Verilog simulator for automatic syntax checking and functionality verification using comprehensive testbenches. In their method, the diagnostic data from these tools is incorporated into an enriched prompt for the language model. The process continues until either no errors are detected or a predefined iteration limit is reached. 
Similarly, \citet{IDsnow5paper} uses an iterative feedback mechanism after each generation step. Both syntax checks and functional checks are performed using the compiler and pre-written test benches on the produced code. When errors are detected, the system feeds back specific error messages and locations to the LLM, prompting it to correct and regenerate the problematic sections. This approach significantly enhances the quality and correctness of the generated hardware descriptions.


To generate verifiable code for Programmable Logic Controllers (PLCs) in Industrial Control Systems, researchers~\cite{ID65paper} proposed an iterative feedback methodology called LLM4PLC. In this approach, LLMs are progressively used in conjunction with external verification tools, including grammar checkers and formal verifiers, to progressively refine the generated code. The pipeline incorporates user feedback and automated error detection to guide the LLM through successive iterations, significantly improving both the success rate of code generation and the quality of the output.

For automated hardware description code generation for neuromorphic computing, \citet{ID99paper} proposed an approach involving multiple conversation rounds with the language model, systematically refining the generated Verilog code based on emerging errors and additional requirements. The authors report that this iterative process led to significant improvements, particularly in complex areas such as overflow/underflow management and adherence to Verilog best practices.

\citet{ID32paper} proposed CLARIFY, an approach that combines an LLM-based generator with a rule-based verifier, which checks the syntactic correctness of the generated plans and provides specific error feedback. The iterative process allows the system to refine its output, significantly improving the quality and correctness of the generated plans. Notably, the study found that including both the erroneous plan from the previous iteration and a detailed list of errors in the feedback loop resulted in the highest success rate for plan generation. This methodology demonstrated particular effectiveness in generating chemical experiment procedures in the Chemical Description Language (XDL), outperforming existing methods in terms of both successful plan generation and expert preference. The iterative feedback approach not only improves output quality but also potentially enables LLMs to adapt to unfamiliar DSLs without extensive fine-tuning or domain-specific training. Similarly, in~\cite{extend5_bhatia2024verified}, a program verifier with Boolean feedback is used for verified code transpilation in four distinct DSL domains.

\citet{ID63paper} used an iterative feedback loop for generating proofs for Rust programs verified by Verus. In this work, GPT-4's initial proof attempts are validated by Verus, and any error messages are fed back to the model for refinement. This process significantly enhanced the system's performance, with the number of fully automated proofs (requiring no human intervention) increasing from $3$ out of $20$ programs to $14$ out of $20$ when using an updated GPT-4 model and allowing for multiple refinement attempts.

\citet{IDsnow26paper} investigated PDDL (Planning Domain Definition Language) domains. Their method employs an iterative feedback loop to improve the quality of LLM-generated planning models. The pipeline includes initial domain generation, consistency checks, and an error correction loop that prompts the LLM to fix identified issues. This iterative process significantly reduces the number of errors in the generated domains, leading to more reliable and executable planning models. The authors demonstrated the effectiveness of their approach on both classical and custom planning domains, showing improved plan generation success rates, especially for simpler scenarios. While human intervention may still be needed for complex domains, the proposed method substantially reduces the effort required compared to manually creating PDDL models from scratch.

In summary, iterative feedback methodologies have shown significant improvements in the quality and correctness of LLM-generated code across various domains. These approaches generally involve a cycle of prompting the LLM, evaluating the generated code, and providing error feedback for correction. Different works have employed various tools for error detection and feedback, as noted above. 

\begin{table}[htbp]
\centering
\scriptsize
\caption{Summary of iterative feedback approaches in LLM-based code generation}
\label{tab:feedback-summary}
\begin{tabular}{>{\raggedright\arraybackslash}p{0.05\linewidth}>{\raggedright\arraybackslash}p{0.35\linewidth}>{\raggedright\arraybackslash}p{0.15\linewidth}>{\raggedright\arraybackslash}p{0.35\linewidth}}
\toprule
\textbf{Ref.} & \textbf{Language} & \textbf{LLM(s) Used} & \textbf{Feedback Tool(s)} \\
\midrule
\citep{ID32paper} & Chemical Description Language (XDL) & GPT-3 & Rule-based verifier (syntax checker and static analyzer) \\
\citep{ID63paper} & Verus & GPT-4 & Verus \\
\addlinespace
\citep{ID12paper} & System Verilog Assertions (SVA) & GPT-4  & VCS simulator \\
\addlinespace
\addlinespace
\citep{ID65paper} & Structured Text (ST) & Code Llama 34B-FT & MATIEC, NuXmv \\
\citep{IDsnow26paper} & PDDL (via JSON-like intermediate) & GPT-4  & Custom consistency checker, FastDownward planner, Custom dependency analysis tool \\
\citep{IDsnow5paper}& Verilog & GPT-4 & ICARUS Verilog (iVerilog), custom testbenches \\
\citep{ID99paper} & Verilog &  GPT-4 & Human Feedback \\
\citep{ID33paper} & Verilog &  GPT-3.5, GPT-4 & ICARUS Verilog (iVerilog) \\
\addlinespace
\citep{ID4paper}& Verilog & GPT-3.5, GPT-4 & ICARUS Verilog simulator, Synopsys Design Compiler \\
\citep{extend2_mora2024synthetic}& UCLID5 & GPT-3.5, GPT-4 &  Compiler feedback \\
\citep{extend5_bhatia2024verified}& MapReduce, Network packet processing, TACO, tensor processing & GPT-4 &  Program verifier (boolean feedback) \\
\bottomrule
\end{tabular}
\end{table}

\subsection{Novel Architectural and Objective Approaches}
\label{subsec:novelarch}

\textbf{Novel Architectures.}
To further improve code generation capabilities, researchers have explored architectural modifications to LLMs. These innovations aim to address specific challenges in DSLs and LRPLs that may not be fully resolved through other approaches, such as pre-training or prompting alone. 
\citet{ID2paper} proposed  Multi-Expert Verilog LLM (MEV-LLM) that incorporates multiple expert LLMs, each fine-tuned on datasets corresponding to distinct design complexity levels (i.e., basic, intermediate, advanced, and expert). The complexity level of the input prompt is assessed through an LLM classifier, which then selects the appropriate expert model. This approach enables more nuanced and targeted code generation, addressing diverse requirements across different complexity levels.

For Shellcode generation, \citet{ID133paper} proposed Adjust QKNorm~\citep{QKnorm} that modifies the self-attention mechanism to consider both numerical differences and directional similarities between vectors. 
The motivation stems from the fact that traditional self-attention computations and methods like QKNorm~\citep{QKnorm} primarily focus on directional similarity, which may not capture all relevant information in the limited data context of Shellcode. Their method modifies the self-attention mechanism in the Transformer architecture. Specifically, they introduce a zero-mean adaptation to the Query (Q) and Key (K) matrices before L2 normalization and dot product calculation. This adaptation subtracts the mean values from the last dimension of Q and K, ensuring that the sum of all elements in that dimension equals zero. By doing so, Adjust QKNorm enables the model to capture more nuanced relationships in the low-resource Shellcode data.

MultiCoder approach is another work that combines multi-programming-lingual (MultiPL) pre-training with a novel Mixture-of-Experts (MoE) architecture~\cite{ID69paper}, with its key feature having PL-level MoE routing strategy (PL-MoE). PL-MoE assigns exclusive expert parameters to each programming language while maintaining shared experts for language-agnostic features. 
This design allows the model to effectively learn both language-specific and language-agnostic knowledge. The architecture consists of dense layers at the bottom for learning transferable surface information, followed by MoE layers at the top for capturing exclusive semantic features of each programming language. This combination aims to mitigate the capacity bottleneck often encountered in traditional multilingual models and enhance performance through cross-lingual knowledge transfer.

\textbf{Novel Objectives for Training.}
Researchers have developed novel training objectives to overcome limitations for code generation in DSL/LRPL. 
DualSC, leveraging dual learning, was introduced to address limited data and domain-specific vocabulary challenges for Shellcode generation \citep{ID133paper}. 
DualSC treats Shellcode generation and summarization as complementary tasks, effectively doubling training data utility. 
By exploiting this bidirectional relationship, DualSC enables a more comprehensive understanding of code-comment interactions, making it particularly effective in low-resource scenarios.

For Register Transfer Level (RTL) code generation, Maximum Likelihood Estimation loss combined with a comparative loss term was proposed in \citep{ID77paper}.
This approach generates multiple code candidates for each instruction during training, scoring them based on quality metrics like syntax correctness. The new loss function encourages higher probabilities for higher-quality code samples, penalizing the model when it favors lower-quality outputs. This method enables the model to learn from both reference code and the relative quality of its own generations, aiming to reduce discrepancies between training and inference behaviors and ultimately produce more robust and accurate RTL code.


\subsection{Input, Output and Token Processing}
\label{subsec:input_output_tokenprocessing}

\textbf{Tokenization Techniques.}
TOKOMPILER, a tokenization method specifically designed for HPC code generation that focuses on code structure and language primitives, ensuring all tokens are meaningful for specific code-language comprehension, was proposed in \citep{ID80paper}. 
The method anonymizes variable names, numbers, and strings, generates an Abstract Syntax Tree (AST), and converts the modified AST back to code while discarding extraneous details. This approach reduces the total number of tokens compared to other tokenizers, allowing for smaller model sizes and improved training times, specifically for performance improvement in Fortran. 

\textbf{Prefix Generation and Variable Replacement Naming.}
\citet{ID129paper} proposed prefix generation and variable replacing techniques for code completion in R. 
Prefix generation tackles the mismatch between training and inference contexts, where users may invoke autocompletion mid-token, by allowing the model to consider partial inputs more effectively. 
Variable replacing addresses the limitations imposed by Transformer models' quadratic complexity and the potential negative impact of rare variable names. 
By substituting infrequent variable names with placeholders, this method not only reduces input sequence length, thereby expanding the effective context window, but also simplifies the language modeling task, allowing the model to focus on predicting more meaningful token sequences. 

\textbf{Decoding.}
Some researchers explored various decoding strategies to enhance code generation for LRPLs and DSLs. These approaches include monitor-guided decoding using a static analysis tool \citep{ID48paper}, constrained decoding algorithm complementing grammar prompting \citep{ID46paper}, Monte Carlo Tree Search (MCTS) decoding algorithm using Markov Decision Process \citep{IDsnow3paper}, and coroutine-based constrained decoding guided by a context-free grammar \citep{IDsnow24paper}. Each method aims to improve the quality and correctness of generated code by leveraging different techniques in decoding.

\citet{ID48paper} introduced Monitor-Guided Decoding (MGD), which uses a monitor that observes the generated code and queries static analysis tools at specific trigger points. Subsequently, it uses the resulting suggestions to reshape token probabilities during decoding. This method allows language models to leverage repository-wide context and type information that may not have been encountered during training. The research demonstrated that MGD improved compilation rates and agreement with ground truth across models of varying sizes, with smaller models like SantaCoder-1.1B outperforming much larger models like text-davinci-003.

\citet{ID46paper} proposed constrained decoding to complement their grammar prompting approach. This method employs specialized Backus-Naur Form (BNF) grammars as an intermediate step in the generation process. The authors developed a tailored constrained decoding algorithm in which multiple tokens are speculatively generated ahead and, if invalid, use an Earley parser to extract the longest valid prefix and possible continuations. The algorithm then leverages LLM probabilities to select the best valid option, ensuring syntactic validity of the generated programs. While grammar prompting showed improvements even without constrained decoding, the addition of constraints further enhanced results in most cases, particularly for semantic parsing tasks ,where both program and execution accuracy improved across multiple datasets. However, this came at the cost of increased LLM API calls, approximately tripling the number compared to unconstrained decoding. 

To address the limitations of traditional decoding methods like greedy search and beam search, which often fail to produce functionally correct and optimized Verilog code, the decoding process is formulated as a Markov Decision Process (MDP)~\cite{IDsnow3paper}. The Monte Carlo Tree Search algorithm used in this approach balances exploration and exploitation in token selection, guided by a policy that combines average reward with an Upper Confidence Tree (UCT) term. The method incorporates domain-specific optimizations, such as pruning unnecessary paths like comment tokens, to reduce the search space. Results demonstrate that MCTS decoding significantly outperforms traditional methods in generating functionally correct Verilog code for various modules (adders, multipliers, MAC units). While MCTS decoding proves highly effective, the authors note it is more time-intensive than traditional methods, suggesting a trade-off between generation quality and computational efficiency.

Similarly, \citet{IDsnow24paper} presents a constrained decoding approach for improving DSL generation. The proposed method, implemented in the YieldLang framework, utilizes a coroutine-based content generation constraint system guided by a context-free grammar. By employing asynchronous mechanisms and generating guidance instructions, the system maintains internal DSL parsing and generation states, enabling more accurate and efficient token production. Experimental results demonstrate significant improvements in generating various DSLs such as JSON, Mermaid flowcharts, and function call expressions,  with JSON generation samples reduced to $16.5\%$ of the benchmark in optimal conditions. The approach achieved $100\%$ parsing success for generated JSON strings across extensive tests. This method not only enhances LLM performance in DSL generation but also conserves computational resources through immediate error detection. 

In summary, while each of these decoding approaches offers unique advantages in improving code quality, syntactic validity, and functional correctness, they also present trade-offs in computational efficiency and resource utilization, highlighting the ongoing challenge of balancing these factors in code generation systems.

\subsection{Retrieval Augmentation Generation (RAG)}
\label{subsec:retrieval}

RAG augments LLMs with external knowledge, addressing limitations such as models' hallucination.
Analysis-Retrieval Method (ARM) addressed token limit constraints in LLM-based program synthesis by utilizing entity classification and context analysis for selecting the most relevant ODSL samples within the available token budget~\citep{ID72paper}. 
DocPrompting mimics how human programmers refer to documentation when writing code~\citep{ID27paper}. 
DocPrompting creates a pool of documentation for each programming language and uses a retriever to select relevant documents based on the given natural language intent. The authors experimented with both sparse (i.e., Elasticsearch with BM25) and dense (i.e., neural encoders trained with contrastive learning) retrieval methods. This approach aims to bridge the gap between user intent and code syntax, particularly for unfamiliar or newly introduced libraries and functions.

RAG was integrated with Feature Query Language (FQL) generation in S3LLM framework by retrieving relevant information from technical documents to enhance scientific software analysis~\citep{ID78paper}. 
The RAG component processes and retrieves relevant information from technical documents, while the FQL generation capability allows for precise querying of source code features. This synergy enables S3LLM to leverage contextual information from documentation to inform and improve FQL query generation, resulting in more accurate and context-aware source code analysis.

RAG is also used for the generation of programs and their proofs in F* \citep{IDsnow19paper}, for RTL by a RAG containing example retriever and knowledge retriever \citep{IDsnow29paper}, 
and for languages like Ansible, YAML, and Bash commands \citep{IDsnow42paper}.
For the generation of programs and their proofs in F*, a proof-oriented programming language, RAG is used~\citep{IDsnow19paper}. The authors calculated similarities between input types and training examples to retrieve related code snippets, and selected premises likely to be used in the definition body. While the file context proved to be the most crucial component, the premise selection model did not significantly impact performance. However, the researchers identified potential for substantial improvements with an ideal selection model. Their analysis also revealed that higher overlap between identifiers in the prompt components and the ground truth solution correlated with better performance, suggesting that refined retrieval techniques could further boost code generation capabilities.

\citet{IDsnow29paper} implemented a domain-specific RAG scheme consisting of two types of retrievers: example retriever and knowledge retriever. Example retriever is used to search for similar problems, while knowledge retriever is tasked to extract design principles and explanations of key terms. This method effectively addresses the hallucination issue in LLMs, particularly when the output is grammatically correct but violates RTL design principles. Authors observed that traditional LLMs tend to overuse software-like constructs, such as excessive for loops, in RTL code. Knowledge retriever helps mitigate this problem by providing RTL-specific guidelines. To build effective retrievers, contrastive learning was used with automatically generated positive and negative sample pairs. Experiments show that combining both retrievers yields the best results in most cases. Example retriever is particularly effective for Verilog coding on VerilogEval dataset. Authors suggest continuous expansion of the example database with realistic designs could further enhance RAG system's effectiveness.

Finally, \citet{IDsnow42paper} proposed RAG approach for structured domain-specific languages like Ansible, YAML, and Bash commands. The DocCGen framework consists of two stages. In the first stage, an information retrieval system identifies the most relevant library documentation for a given natural language query, using either sparse (BM25) or dense (ColBERTv2) retrieval methods. The second stage leverages the retrieved documentation to extract structured schema information and grammar rules. DocCGen performs well in out-of-domain and low-resource scenarios across multiple metrics and model sizes. 

Offering a contrasting approach, others focused on post-processing techniques to improve code generation, rather than extensive retrieval~\citep{IDadd1paper}. While they employed a lightweight retrieval mechanism called KATE (Knn-Augmented in-conText Example selection) for few-shot prompting, their primary contributions lie in post-processing, which includes semantic filtering, semantic interleaving, output-based score tuning, and temperature mixing. Their method demonstrates that significant improvements in code generation can be achieved through sophisticated output refinement strategies without heavy reliance on external knowledge retrieval.

These diverse approaches offer several advantages, including improved generalization capabilities, incorporation of up-to-date information, and enhanced accuracy in generated outputs. However, they also face limitations such as dependence on the quality of the retrieved information, potential for information loss during document segmentation, and additional computational requirements.

\subsection{Other Techniques}
\label{subsec:other}

\textbf{DSL Creation.}
A few works have developed their custom DSLs for enhancing the performance of LLMs in solving some specialized tasks \citep{ID112paper, ID72paper, ID81paper}. By providing a more structured, constrained, and domain-appropriate interface, DSLs bridge the gap between natural language queries and complex system operations.
This approach leads to significant improvements in model efficiency and accuracy across various domains, from database management to CAD sketch generation. 
Office Domain Specific Languages (ODSL) \citep{ID72paper} and DSL for CAD sketches \citep{ID81paper} are two of these studies. 
\citet{ID72paper} argues that existing commanding APIs in general-purpose programming languages lead to suboptimal performance for LLM-based program synthesis. These languages' broad scope causes LLMs to hallucinate or forget critical details, while their multiple approaches to solving the same task hinder consistent code generation. In contrast, their custom domain-specific language, ODSL, is designed to be more constrained, uniform, and tailored to the application domain. This design allows for easier learning by LLMs, more consistent and ultimately improved performance in program synthesis. 

In another work, a DSL for CAD sketches was created, which was an important factor in improving the performance~\citep{ID81paper}. By designing a carefully structured language with simple syntax, the authors were able to effectively represent the heterogeneous elements of CAD sketches, both primitives and constraints, in a uniform manner suitable for Transformer models. This structured representation, along with the explicit use of syntax as additional input to the Transformer, helped the model better interpret and generate complex sketch sequences. Quantitative results demonstrated significant improvements over baseline methods, with ablation studies confirming the positive impact of the syntax-based input sequences derived from the DSL.

Similarly, \citet{ID112paper} developed a custom DSL as part of their approach to simulation question-answering. This DSL was designed to be more concise than full-featured programming languages while retaining the ability to define variables, use loops, and include conditional statements. The authors argue that this tailored DSL helped reduce the search space of possible programs, potentially leading to better convergence in their neural program synthesis approach. While the DSL itself is not directly compared to alternative representations, it plays a crucial role in the overall methodology, enabling the model to generate executable code that simulates the described chemical processes and answers complex questions about them.

In another work, \citet{ID147paper} created a new DSL for data management tasks on the Qore-Base platform to simplify complex operations, making it easier for GPT to generate correct output. This tailored DSL significantly improved token efficiency, reducing usage by $77\%$ compared to the original API payload format. Adapting elements from Prisma Schema Language, the DSL effectively expressed Qore-Base's unique features, while remaining intuitive for the model. This intermediate representation bridged the gap between natural language queries and the complex API. When combined with optimal parameters, the DSL-based approach resulted in the successful implementation of database schemas in $90\%$ of experiments. While not the sole factor in improving performance, the DSL played a pivotal role in enhancing the system's effectiveness for Qore-Base data management tasks.

\textbf{Temperature Mixing.}
Temperature mixing is a key technique introduced in \citep{IDadd1paper} to improve code generation performance. 
The authors observed that different sampling temperatures in language models lead to trade-offs between output diversity and accuracy. 
The authors proposed generating programs at both low and high temperatures to address diversity and accuracy trade-offs, combined by reranking methodologies. 
This approach allows the system to capture both high-quality, deterministic outputs and more diverse, potentially novel solutions. It is particularly effective when the model is uncertain about the output, such as with ambiguous queries or unfamiliar languages. In their experiments, temperature mixing showed improvements in top-1 accuracy for the M language. The technique's effectiveness is most pronounced for languages or tasks where the model has less prior knowledge, making it a valuable tool for addressing limitations in code generation for LRPLs and DSLs.

\textbf{Early Stopping.}
The early stopping mechanism used in \citep{ID129paper} addresses the challenge of maintaining output quality and inference speed in full-line code completion tasks, particularly when using smaller model sizes. Observing that their model tended to produce unreliable predictions after generating only 1-2 program tokens, the authors implemented an early stopping routine in their beam search process that prevents this issue by applying a lexer after each beam search iteration and halting the process if more than one program token is detected. 
They also introduced a hyperparameter $k$ to set an upper limit on the number of complete tokens to generate. 
The implementation of this technique yielded significant improvements in inference speed without compromising the model's quality metrics.

\textbf{Knowledge Distillation.}
Knowledge distillation is a technique where a smaller, more efficient model (the student) is trained to mimic the behavior of a larger, more complex model (the teacher). 
This process involves transferring the knowledge embedded in the teacher model to the student model~\citep{IDsnow33paper}, addressing the challenge of limited high-quality datasets in specialized programming languages~\citep{IDsnow15paper}. 
Claude3-Haiku is leveraged as the teacher model and DeepSeek-Coder-7B-Instruct as the student model, implementing a novel ``code-to-code augmentation'' methodology for generating and correcting Verilog code \citep{IDsnow15paper}. 
This process involved filtering open-source RTL code samples, using Claude3-Haiku to generate detailed descriptions, regenerating RTL code from these descriptions, and verifying the code with the iVerilog compiler. Failed compilations were used to create an error-correction dataset, where erroneous code samples, compiler error messages, and their corrected versions were collected to train the model's self-reflection capabilities. 
The authors found that training on smaller, optimized subsets of the augmented dataset was more effective than using the entire dataset at once. This approach significantly improved DeepSeek-Coder-7B-Instruct's capabilities, achieving $76.1\%$ and $51.4\%$ pass@1 on VerilogEval Human and Machine categories, respectively; and $87.3\%$ syntactic correctness on VerilogFixEval, surpassing GPT-4 Turbo by $18.1\%$ in self-reflection capabilities. Ablation studies confirmed that the code-to-code augmentation was primarily responsible for the model's outstanding performance across multiple benchmarks, demonstrating the effectiveness of this knowledge distillation technique in transferring RTL code generation and error correction capabilities from advanced commercial models to open-source alternatives.

The ReflectionCoder introduced in \citep{IDsnow33paper} employs a knowledge distillation process to enhance one-off code generation performance. The method utilizes a teacher model fine-tuned on reflection sequence data and code instruction tuning data, distilling knowledge to a student model targeted for improvement. 
The distillation process comprises two key components: Reflection Self-Distillation and Dynamically Masked Distillation. In Reflection Self-Distillation, two input samples are constructed for each reflection sequence: a teacher sample containing \texttt{[Reflection Instruction, Reflection Sequence, Instruction, Final Code]}, and a student sample with \texttt{[Instruction, Final Code]}. The Dynamically Masked Distillation technique gradually increases the masking of the reflection sequence during training. The ReflectionCoder method demonstrated significant improvements across multiple benchmarks and also enhanced performance on smaller models, sometimes surpassing larger baseline models. The method showed effectiveness across various programming languages in the MultiPL-E benchmark, highlighting its versatility and potential for improving code generation tasks without requiring additional instructions during inference.

\subsection{Comparison and Analysis of the Techniques}
Model adaptation techniques such as fine-tuning require significant computational resources but provide persistent improvements over prompting techniques. Fine-tuning on 1K+ Verilog examples consistently outperforms few-shot prompting across multiple studies \cite{ID92paper, IDsnow4paper}, with improvements in Pass@1, Pass@10, Pass@25 for Rust generation \cite{ID105paper}.
Regarding the trade-off between RAG and Fine-tuning: RAG excels in documentation-heavy domains like Ansible, where specifications evolve rapidly, providing access to current information that training data cannot capture \cite{ID27paper, IDsnow42paper}. Fine-tuning performs better in pattern-heavy domains like Verilog, where consistent structural patterns are more important than evolving documentation \cite{ID2paper, IDsnow37paper}. The effectiveness depends on information volatility: RAG wins when knowledge changes frequently (APIs, standards), while fine-tuning dominates stable domains with clear patterns. Cost structures differ significantly; RAG incurs ongoing retrieval costs per inference, while fine-tuning has high upfront training costs but lower inference costs. On the other hand, PEFT methods like LoRA provide 80-90\% of full fine-tuning benefits at 10-20\% of the computational cost \cite{ID7paper, IDsnow27paper}, making them ideal for resource-constrained scenarios. However, full fine-tuning consistently achieves superior performance when computational resources allow, particularly for complex DSLs like Verilog, where subtle patterns matter \cite{IDsnow37paper}.

Compared with standard approaches, novel architectural approaches like Multi-Expert Verilog LLM \cite{ID2paper} show promise but require substantial development resources and lack the empirical validation of established techniques. Regarding the agent architecture, to further compare Single-Pass Generation and iterative feedback, iterative feedback approaches significantly improve correctness and compilation rates for DSLs with available validators (Verilog simulators, formal verifiers) \cite{ID4paper, IDsnow5paper}. However, they increase inference costs proportionally to iteration count and require robust feedback systems. Single-pass generation remains preferable for cost-sensitive applications or domains lacking automated validation tools. Finally, custom tokenization for HPC code \cite{ID80paper} reduces token count and improves training efficiency, while constrained decoding ensures syntactic validity for formal languages. So tokenization and decoding techniques provide complementary benefits rather than standalone solutions.

\subsection{Distinctive Features for LRPLs and DSLs}

\textbf{Domain-Aware Synthetic Generation and Transfer Learning Approaches.} 
While similar techniques are used in HRPLs, DSLs, and LRPLs require tailored approaches for data synthesis and transfer learning. 
Emerging solutions tailored to LRPL/DSL domains demonstrate innovative approaches to overcome the challenges in data accessibility and availability for LRPLs and DSLs. Domain-aware synthetic data generation has shown particular promise, where models like GPT-4 generate training data using domain-specific constraints and validation tools. For Verilog, this involves incorporating hardware design principles and simulation-based validation \cite{IDsnow37paper, IDsnow39paper}, while Lean datasets use formal structure to generate mathematically valid examples \cite{IDsnow22paper}. Cross-domain transfer learning through intermediate representations, such as using LLVM IR to bridge multiple programming languages \cite{ID105paper}, effectively multiplies available training data by leveraging similarities between related languages. Multi-modal approaches, exemplified by CAD sketch datasets that combine visual and textual information \cite{ID81paper}, represent novel strategies where non-textual domain knowledge augments traditional code datasets.

\textbf{Community-Driven Strategies.}
Access constraints have driven the development of specialized curation strategies unique to LRPL/DSL domains. Template-based augmentation extracts structural patterns from existing code to generate variations while preserving domain semantics, particularly effective for configuration DSLs like Ansible \cite{IDsnow42paper}. Community-driven approaches mobilize specialized expert networks rather than general crowdsourcing, as seen in COQ datasets leveraging the formal methods community \cite{IDsnow23paper}. 
Privacy-preserving could be a challenge, needing new techniques, as some LRPL/DSL code often cannot be shared due to confidentiality constraints. These domain-specific innovations represent a departure from traditional dataset creation and model tuning paradigms and highlight the need for specialized approaches that account for the unique characteristics and constraints of LRPLs and DSLs.

\textbf{Specialized Adaptation Strategies for DSLs and LRPLs.}
While most of the techniques used to enhance the performance of the models, e.g., fine-tuning, iterative feedback, and prompting, are extensively used for HRPLs, the approaches have several differences when adopted for DSLs and LRPLs. For example, the models require extensive human guidance and/or several rounds of iterations, likely due to a lack of specialized training in DSL domain~\citep{ID120paper, ID34paper, ID33paper}.
Another example is InterCode, which uses Docker with RL for code generation~\cite{ID56paper}.
The approach proposed in \cite{IDsnow42paper} for structured languages, BASH  and ANsible YAML, uses documents to detect correct libraries from user prompts, and then extracts schema rules from the docs and enforces schema adherence based on the extracted rules in a constrained decoding process. 
The abundance and accessibility of public code in HRPLs like C++ and Python, lead to higher quality outputs compared to Julia or Fortran~\citep{ID35paper}.
Formal theorem-proving languages also pose unique challenges for LLMs. Models like Mistral-7b-Instruct, ChatGPT 4, Google Gemini, and Starcoder2-15b struggle with COQ syntax and semantics, often failing to provide syntactically correct responses or generating incorrect proofs for simple lemmas~\citep{ID31paper}.
In the realm of hardware design and verification using Verilog, models show poorer performance, frequently failing to produce compliant designs or testbenches, requiring human intervention to fix errors~\citep{ID33paper}.
Similarly, for prompting techniques, researchers developed specialized domain-specific prompting techniques or incorporated pseudo-code in the prompts~\citep{IDsnow7paper}. 
Others provide additional tools or detailed instructions to guide the process~\citep{ID36paper, IDsnow18paper, ID75paper, ID36paper, ID75paper}.
Other studies highlight that few-shot prompting generally outperforms zero-shot approaches, especially for complex tasks~\citep{ID8paper, ID74paper, IDsnow32paper, ID147paper, IDsnow38paper}. However, depending on the task and domain, sometimes zero-shot prompting was found to achieve higher performance compared to few-shot prompting~\citep{IDsnow38paper}. 
The approaches using iterative feedback often require domain knowledge and tools employed for error detection and feedback, such as iVerilog and ICARUS simulators for hardware design \citep{ID33paper, ID4paper}, formal verifiers for PLC code \citep{ID65paper}, and Verus for Rust program verification \citep{ID63paper}. 
Finally, knowledge distillation studies intend to address the challenge of limited high-quality datasets in specialized programming languages~\citep{IDsnow15paper}.

\begin{tcolorbox}[colback=gray!10,colframe=gray!40,boxrule=0.5pt,title=Summary RQ2,coltitle=black]
\begin{enumerate}

    \item We categorized the methods used to improve the LLMs' performance into 6 Main Categories: Model adaptation techniques, Prompting and iterative techniques, Novel architectural and objective approaches, Input, output, and token processing, Retrieval-augmented generation, and Others. We further subdivided these methods. Subsequently, we summarized how these methods were applied to improve the code generation performance in LRPL and DSLs.
    
    \item Fine-tuning and Prompting are the most prevalent methods for improving performance. We also categorized different prompting strategies and iterative feedback tools used in the literature.

    \item Novel methods are used in the literature to improve the performance of LLMs' code generation, such as the Creation of DSL, Temperature Mixing, Early Stopping, and Knowledge Distillation.

    \item Similar to our previous findings, the methodologies used to improve the code generation performance of models should be adapted to adhere to the requirements of the target language, specifically DSLs. In many cases, human guidance is required or domain-specific strategies and tools are developed and used. 
\end{enumerate}
\end{tcolorbox}

\section{Dataset curation and Preparation }
\label{sec:rq3}

The quality of a dataset affects the performance of the models; for example, a smaller model trained on high-quality data can achieve a performance that is comparable to larger models \citep{phi, IDsnow37paper, ID8paper}. 
The LRPLs and DSLs suffer from having high-quality large datasets: (i) Data for LRPLs is limited due to a lack of public repositories, licensing restrictions, or the languages being relatively new or specialized~\citep{ID62paper}. (ii) DSL data is scarce because they are used in specific domains or problem areas. 
Hence, in the following, we investigate approaches used in the literature for data curation and data processing in DSL/LRPL, and how they differ from techniques used for HRPLs.


\subsection{Dataset Curation}
\label{subsec:dataset-curation}

We have primarily identified three approaches used by the researchers to tackle the challenges of dataset acquisition for LRPL and DSL studies. These approaches are as follows and detailed in the next subsections:

\begin{enumerate}[label=(\roman*)]
    \item \textbf{Curated datasets}: This approach entails curating datasets from different sources, such as GitHub, textbooks, forums, and other relevant sources.
    We further divide curated datasets into three categories.
    \begin{itemize}
        \item Existing Datasets: This approach involves the direct use of pre-existing datasets in their original form. Researchers utilize these datasets as-is for their studies, without modifications \citep{ID57paper}.
        
        \item Modified Existing Datasets: In this category, researchers derive datasets from existing ones, but the datasets are modified toward the goal of the study \citep{IDsnow42paper, ID133paper}. Examples of these modifications include additional pre-processing to ensure the data quality~\citep{ID80paper}, removing non-compilable files~\citep{ID65paper}, and grounding the dataset to make it work in a specified environment~\citep{ID56paper}.

        \item Collected Datasets: In this group, researchers collect or combine data from various sources such as code repositories on GitHub, academic literature, online forums, and other relevant platforms \citep{IDsnow34paper, ID18paper, ID82paper}.
        
    \end{itemize}
    
    \item \textbf{Synthesized datasets}: In this approach, researchers have used powerful conversational large language models like ChatGPT to create and annotate their datasets~\citep{IDsnow11paper, IDsnow16paper, IDsnow22paper, IDsnow39paper, IDsnow27paper}.

    \item \textbf{Manual creation}: In this approach, researchers have manually created their datasets. These datasets are usually high quality and are often used for evaluation purposes, considered as benchmark datasets~\citep{IDsnow34paper, IDsnow38paper, IDsnow8paper}.
    
\end{enumerate}

\begin{table}[htbp]
\begin{threeparttable}
\centering
\scriptsize
\caption{Sources of data for code generation in low-resource programming languages}
\label{tab:lrpl_datasets}
\begin{tabular}{p{0.05\textwidth}p{0.10\textwidth}p{0.25\textwidth}p{0.35\textwidth}p{0.15\textwidth}}
\toprule
\multirow{34}{*}{\rotatebox[origin=c]{90}{\textbf{Curated}}} & \textbf{ID} & \textbf{LRPL Language(s) Improved} & \textbf{Source} & \textbf{Dataset Type}\\
\midrule
& \citep{ID157paper} & PHP, Ruby, Rust, Scala & GitHub & Collected \\
& \citep{ID113paper} & Ruby, Kotlin, Scala, Swift, Perl & GitHub & Collected \\
& \citep{ID69paper} & Ruby & CodeSearchNet & Modified Existing \\
& \citep{IDsnow34paper} & Kotlin & GitHub & Collected \\
& \citep{IDsnow19paper} & F$^{\ast}$ & GitHub & Collected \\

& \citep{ID57paper} & Haskell & Blastwind Dataset & Existing\\
& \citep{ID67paper} & PHP, Dart, Lua, Rust, R, Julia, Haskell & Publicly Available Code Data & Collected \\
& \citep{ID7paper} & D, Perl, Ruby, Rust, Racket, Swift & Stack~\citep{ID90paper} & Existing \\
& \citep{ID93paper} & Kotlin, PHP, Ruby, Rust & Codeforces Website & Collected \\
& \citep{ID105paper} & Codon, Rust, Haskell, Fortran, D, Ruby, Crystal, Nim, Swift & Programming contest problems, GitHub, OpenWebMath Dataset, PeS2o, Stack Dataset & Collected \\

& \citep{ID59paper} & Kotlin, Rust, Scala, Ruby & Auto-completion User Data & Collected \\

& \citep{IDsnow7paper} & Rust & HumanEval-X~\citep{codegeex} & Existing\\
& \citep{IDsnow33paper} & Rust & CodeFeedback-Filtered-Instruction & Modified Existing  \\
& \citep{ID73paper} & Rust & GitHub Commits, OpenAssistant Dataset, xP3x Dataset & Collected \\

& \citep{ID80paper} & Fortran & HPCorpus~\citep{HPCorpus} & Modified Existing \\

& \citep{ID27paper} & Bash & TLDR Project, Stack Overflow & Collected \\
& \citep{ID130paper} & Bash & Ubuntu Man Pages, Proprietary Bash histories, explainshell project\tnote{z}  & Collected \\
& \citep{IDsnow42paper} & Bash & TLDR~\citep{ID27paper}, NL2Bash~\citep{lin-etal-2018-nl2bash}, Linux Man pages\tnote{t} & Modified Existing \\
& \citep{ID56paper} & Bash & NL2Bash Dataset\citep{lin-etal-2018-nl2bash} & Modified Existing \\
& \citep{ID70paper} & Bash, D, Julia, Lua, Perl, R, Racket, Rust, Scala, Swift, Ruby & HumanEval~\citep{codex} & Modified Existing \\

& \citep{ID133paper} & Assembly & Shell Storm\tnote{c} and Exploit Database~\citep{liguori-etal-2021-shellcode} & Modified Existing\\
& \citep{ID107paper} & SKILL & Proprietary SKILL Repositories, GitHub & Collected \\
& \citep{IDsnow23paper} & PowerShell & GitHub, Atomic Red Team, Stockpile, Empire, Hacktricks, Red team recipe, infosec matter & Collected \\

& \citep{ID18paper} & 49 PLs & GitHub, Stack, StarCoder Data & Collected \\
& \citep{IDsnow43paper} & 14 PLs & CodeForces Website & Collected \\
& \citep{ID82paper} & 17 PLs & StarCoderData, GitHub, Stackoverflow & Collected\\
& \citep{codegeex} & 23 PLs & Pile, CodeParrot & Collected \\
& \citep{ID117paper} & 24 PLs & GitHub & Collected \\
& \citep{ID76paper} & 24 PLs & GitHub & Collected \\
& \citep{ID84paper} & 32 PLs & GitHub & Collected \\

& \citep{ID1paper} & Julia, R, Perl, Ruby, Smalltalk & University Websites, platforms that provide coding challenges for technical interviews & Collected \\

& \citep{ID35paper} & Julia, Fortran & Well-known HPC Kernels & Collected \\
& \citep{ID20paper} & Julia, Fortran & Well-known HPC Kernels & Collected \\

& \citep{ID34paper} & R & R Programming Textbooks & Collected \\
& \citep{ID44paper} & R & Open Science Framework & Collected \\
& \citep{ID129paper} & R & GitHub & Collected \\

& \citep{extend4_shah2024stackeval} & 25 PLs & Stack Overflow & Collected \\
\bottomrule
\end{tabular}
\end{threeparttable}
\end{table}

\begin{table}[htbp]
\begin{threeparttable}
\centering
\scriptsize
\caption{Sources of data for code generation in domain-specific languages}
\label{tab:dsl_datasets}
\begin{tabular}{p{0.05\textwidth}p{0.10\textwidth}p{0.20\textwidth}p{0.40\textwidth}p{0.15\textwidth}}
\toprule
\multirow{41}{*}{\rotatebox[origin=c]{90}{\textbf{Curated}}} & \textbf{ID} & \textbf{Languages} & \textbf{Source} & \textbf{Dataset type} \\
\midrule
& \citep{ID2paper} & Verilog &  GitHub & Collected \\
& \citep{ID92paper} & Verilog & GitHub & Collected \\
& \citep{IDsnow29paper} & Verilog & GitHub & Collected \\
& \citep{IDsnow31paper} & Verilog & GitHub & Collected \\
& \citep{IDsnow37paper} & Verilog & GitHub & Collected \\
& \citep{IDsnow39paper} & Verilog & GitHub & Collected\\
& \citep{ID141paper} & Verilog & GitHub & Collected \\
& \citep{ID92paper} & Verilog & HDLBits website &  Collected \\
& \citep{ID8paper} & Verilog & GitHub, Verilog Textbooks & Collected \\
& \citep{IDsnow4paper} & Verilog & GitHub and HuggingFace datasets & Collected\\
& \citep{IDsnow27paper} & Verilog & Verilog Dataset on Hugging Face & Modified Existing \\
& \citep{IDsnow15paper} & Verilog & Stack, VeriGen Dataset~\citep{thakur2023verigenlargelanguagemodel} & Modified Existing \\
& \citep{IDsnow41paper} & Verilog & VerilogEval-Human~\citep{ID92paper} & Modified Existing \\
& \citep{ID99paper} & Verilog & IRIS flower dataset, MNIST handwritten digit dataset & Existing \\
& \citep{ID4paper} & Verilog & RTLLLM~\citep{ID160paper}, VerilogEval dataset~\citep{ID92paper} & Existing \\

& \citep{IDsnow38paper} & VHDL & Verilog-Eval~\citep{ID92paper}, VHDL Tutorials & Collected, Modified Existing \\
& \citep{ID49paper} & System Verilog, Verilog, VHDL & GitHub, TrustHub, CWE, CAD For Assurance & Collected \\
& \citep{ID75paper} & GitHub-YAML & Argus Dataset~\citep{Argusdataset} & Modified Existing\\
& \citep{ID6paper} & Ansible & GitHub, GitLab, Ansible Galaxy & Collected \\
& \citep{ID5paper} & Ansible & Ansible Lightspeed User Interactions and Feedback & Collected \\
& \citep{IDsnow42paper} & Ansible & Google Big Query, Ansible Galaxy, Ansible Documentation & Collected \\
& \citep{ID31paper} & Coq & Foundational Libraries, Formalized Mathematical Theorems, Computer Science Concepts, Algorithm Implementations & Collected \\
& \citep{ID65paper} & IEC-61131-3-ST & OSCAT IEC 61131-3 Library Test Files & Modified Existing \\
& \citep{IDsnow22paper} & Lean & High School and Undergraduate competition level Textbooks & Collected \\
& \citep{ID43paper} & Regex, FOL, LTL & Regex-synthetic, Regex-turk datasets\citep{locascio-etal-2016-neural}, MNLI dataset, Codesc dataset\citep{hasan-etal-2021-codesc}, Spot library, Hardware synthesis specifications\citep{schmitt2021neural} & Modified Existing \\
& \citep{IDsnow12paper} & FOL & FOLIO~\citep{Han2022FOLIONL}, ProofWriter~\citep{Tafjord2020ProofWriterGI} & Existing \\
& \citep{IDsnow25paper} & CQL & TCFL Textbook, EnWiki~\citep{enwiki} & Modified Existing \\
& \citep{ID81paper} & CAD Sketches & SketchGraphs dataset~\citep{Seff2020SketchGraphsAL} & Modified Existing \\
& \citep{ID63paper} & Verus & Diffy benchmark~\citep{chakraborty2021diffyinductivereasoningarray} & Modified Existing\\
& \citep{ID74paper} & OCL & Educational Resources, Literature, GitHub & Collected \\
& \citep{IDadd1paper} & M & Forums, Gathered from Teams & Collected \\
& \citep{ID39paper} & Lean & IMO Shortlisted Problems (2006-2021) & Collected \\
& \citep{IDsnow36paper} & Lean & ProofWiki, University courses & Collected \\
& \citep{ID12paper} & SVA & OpenTitan repository\tnote{i} & Collected \\
& \citep{IDsnow18paper} & SVA & Hack@DAC hardware security competitions, open-source silicon root of trust~\citep{comp1, comp2}, OpenTitan & Collected \\
& \citep{ID32paper} & XDL & Organic Syntheses dataset (volume 77), Chem-RnD & Existing \\
& \citep{ID46paper} & SMILES, PDDL & GeoQuery, SMCalflow and Overnight-Blk \citep{shaw-etal-2021-compositional, andreas-etal-2020-task, wang-etal-2015-building} & Existing \\
& \citep{ID78paper} & FQL & Energy Exascale Earth System Model (E3SM)~\citep{https://doi.org/10.1029/2022MS003156} & Modified Existing \\
& \citep{IDsnow32paper} & Vega-Lite & nvBench~\citep{10.1145/3448016.3457261} & Existing\\
& \citep{IDsnow26paper} & PDDL & Well-known domain planning domains such as gripper, logistics, tyreworld & Existing \\
& \citep{ID107paper} & SKILL & Proprietary SKILL Repositories, GitHub  & Collected\\
& \citep{ID42paper} & Excel Formulas & Public Excel workbooks, Excel help forums, Enron spreadsheet corpus & Collected  \\

& \citep{extend3_kon2024iac} & Infrastructure-as-Code (IaC) & Code generation benchmark for cloud IaC programs & Collected \\

\bottomrule
\end{tabular}

\end{threeparttable}
\end{table}

\textbf{Curated Datasets.}
Tables \ref{tab:lrpl_datasets} and \ref{tab:dsl_datasets} present an overview of datasets that were curated for LRPL and DSLs code generation. 
The last column of the tables shows the dataset types. 
We observe that most datasets are collected (25 in LRPL and 23 in DSL), followed by Modified Existing (6 in LRPL and 10 in DSL studies), and using Existing Datasets (3 LRPL and 7 DSL). 
GitHub is frequently used for curating LRPL datasets~\citep{ID48paper, ID157paper, ID113paper, ID129paper, ID120paper}, with other features such as commit history in COMMITPACK dataset~\citep{ID73paper}.
Other sources of collecting data include Stack Overflow \citep{ID41paper, ID82paper}, codeforces.com \citep{ID93paper, IDsnow43paper}, samples from codeforces.com \citep{ID93paper}, programming contest platforms \citep{ID105paper}, and programming contest datasets~\citep{ID105paper}. 
A dataset from 25M openly available samples from codeforces.com, covering 7,514 distinct problems, is constructed in~\cite{ID93paper}. They gathered this data to create xCodeEval, an executable multilingual multitask benchmark, to test LLMs on various LRPLs such as  Kotlin, PHP, Ruby, and Rust.
Other works \citep{ID59paper} created an IDE extension called Code4Me and collected real auto-completion usage data from more than 1,200 users over a year. This approach allowed them to assess real-world failure modes in code completion of open-source LLMs such as UniXCoder, Incoder, and CodeGPT in low-resource languages like Kotlin, Rust, Scala, and Ruby.

In a similar vein, to overcome the challenge of collecting parallel code and corresponding Intermediate Representations (IR), particularly for LRPLs like Rust and Swift that cannot be compiled at the file level, \citet{ID105paper} turned to programming contest platforms instead of relying on GitHub. These platforms provided self-contained, functional code samples across multiple languages, eliminating dependency issues and ensuring compilability. This approach allowed the authors to create SLTrans, a substantial dataset of nearly 4 million source code-IR pairs across 12 languages, including C, C++, Python, Codon, Rust, Haskell, Go, Fortran, D, Ruby, Crystal, Nim, Swift, Obj-C.
In another study, a set of mathematical kernels for high-performance computing (HPC) in Fortran is curated~\citep{ID35paper, ID20paper}. 
These included AXPY, GEMV, GEMM, SpMV, 3D Jacobi stencil computations, and CG. This carefully selected range of kernels provided a comprehensive basis for assessing the AI model's performance across various computational tasks commonly encountered in HPC applications.
Open Science Framework (OSF) is used to collect a R dataset due to its focus on research data from social sciences and psychology~\citep{ID44paper}.
Their choice was also motivated by the high quality of the OSF, as its focus is on statistical analysis of research data, often linked to research articles published in various journals.

For Bash-related datasets, TLDR, a collection of community-maintained help pages for command-line tools is used \citep{ID29paper}.
The pages are crawled to gather the TLDR dataset for natural language to Bash command translation~\cite{ID29paper}. They ensured that train and test splits had different Bash commands for fair evaluation, resulting in a high-quality dataset with natural language intents written by human users.
\citet{ID130paper} proposed a dataset for natural language to Bash continual pretraining.
They collected data from Ubuntu Man Pages for the task of man page completion, acquired users' Bash histories from real companies for command sequence completion, and used the explainshell project for command explanation tasks.
Lastly, \citet{ID56paper} adapted the NL2Bash dataset \citep{lin-etal-2018-nl2bash} to make it suitable for interactive code evaluation.
Most of these datasets contain both code and code-comments since a majority of them are extracted from GitHub. There are some datasets, such as the one curated in~\cite{IDsnow42paper} that not only contains code samples, but also accompanying documentation, usage examples, and structured schema information for each command or module in Bash.

\textbf{Domain-Specific Language Data Curation.}
The curation of high-quality datasets for domain-specific programming languages is a critical challenge in advancing code generation capabilities for specialized fields. Unlike general-purpose programming languages, DSLs often have limited resources and require targeted efforts to collect relevant data. This is due to data scarcity, which relates to various reasons such as licensing issues~\citep{ID5paper}.
For DSL, while the majority of researchers compiled their own datasets~\citep{ID2paper, ID92paper, IDsnow29paper, IDsnow31paper}, some utilized pre-existing datasets~\citep{IDsnow32paper, IDsnow26paper, ID32paper, ID46paper, IDsnow12paper}. Additionally, several studies adapted existing datasets to better align with their specific research objectives~\citep{ID63paper, ID81paper, IDsnow25paper, ID43paper, ID65paper, ID75paper}.

GitHub has served as a primary source for dataset collection in numerous studies, particularly for Verilog-related research~\citep{ID2paper, ID92paper, IDsnow29paper, IDsnow31paper, IDsnow37paper, IDsnow39paper}. A selected set from HDLBits\footnote{\url{https://hdlbits.01xz.net/wiki/Problem_sets}}, a collection of digital circuit design exercises, and an online judge for Verilog is utilized in~\cite{ID92paper}. In a similar vein, Google's BigQuery is used to gather approximately 300MB of Verilog data, supplementing this with Verilog textbooks from an online e-library~\cite{ID8paper}. Others~\cite{ID23paper} generated Verilog using a custom-built tool that produces Task/Result pairs based on predefined templates, rather than being curated from existing real-world sources or open-source Verilog code repositories. 
Expanding on these efforts, a hardware domain-specific dataset encompassing three hardware programming languages, SystemVerilog, Verilog, and VHDL, is compiled in~\cite{ID49paper}. This work also leveraged Google BigQuery and various hardware security sources, including TrustHub, CAD for Assurance, and Common Weakness Enumeration (CWE).
Several studies~\citep{IDsnow41paper, IDsnow38paper, ID4paper} have made use of datasets such as Verilog-Eval~\citep{IDsnow41paper, IDsnow38paper} and RTLLM~\citep{ID4paper}. Additionally, some researchers have sourced datasets from Hugging Face~\citep{IDsnow4paper, IDsnow27paper}, further diversifying the data sources utilized in this field of study.
Many of these works~\citep{ID2paper, ID92paper, IDsnow29paper} use this collected data to pre-train or fine-tune their own LLMs.

\citet{ID75paper} collects GitHub YAML workflow data from the ARGUS dataset. For Ansible, \citet{ID133paper} proposed Ansible LightSpeed service and collected and analyzed data from thousands of users of their service to examine various key aspects, including user behavior, retention rates, suggestion acceptance rates, user sentiment, and the acceptance rates of the user from the generated code. In another study, the authors curated NL to Ansible-YAML data from Google BigQuery and Ansible Galaxy, while also leveraging Ansible module documentation from the Galaxy API for information retrieval and schema rules~\cite{IDsnow42paper}.
To address the scarcity of COQ data, \citet{ID31paper} created a dataset from various sources, including foundational libraries, formalized mathematical theorems, computer science concepts, and algorithm implementations, as existing datasets were unusable, had licensing issues, or were outdated.

\citet{ID107paper} employed a combination of proprietary and open-source data to gather their SKILL dataset. For the PowerQuery M expression language, two datasets were created: one by scraping data from PowerQuery help forums, and another sourced internally from the PowerQuery team~\cite{IDadd1paper}.
For Lean, a dataset for automated theorem proving in the Lean formal language is curated, choosing IMO Grand Challenge shortlisted problems~\cite{ID39paper}. Other studies have used university-level textbooks and courses to gather proof data to be used in their Lean generation research~\cite{IDsnow22paper, IDsnow36paper}.
The dataset in~\cite{ID65paper} was gathered from the OSCAT IEC 61131-3 Library, which provides a collection of Structured Text (ST) files used for Programmable Logic Controller (PLC) programming.
\citet{ID43paper} used datasets for three languages: Regex (Regex-synthetic, Regex-turk), FOL (FOL-mnli, FOL-codesc), and LTL (LTL-pattern, LTL-synthesis), combining existing sources with newly created datasets translated from various inputs.
The work of \citet{IDsnow12paper} evaluates first-order logic reasoning over natural language in English, using two main datasets: FOLIO, an expert-written open-domain dataset for FOL reasoning, and ProofWriter, a synthetically generated dataset for logical reasoning.

The SketchGraphs dataset, derived from OnShape's CAD platform, was utilized in~\cite{ID81paper} to train and evaluate their generative models for CAD sketches.
Dataset used in \citep{ID63paper} consists of 20 vector-manipulating programs selected from the ``safe'' category of the Diffy benchmark, which were then translated from C to Verus (a Rust-like verification language).
\citet{ID74paper} collected 15 UML models with 168 OCL specifications compiled from various educational resources, literature, and GitHub repositories.
For SVA, \citet{ID12paper} used designs from the OpenTitan project to evaluate ChIRAAG's performance across six different hardware modules, including RV Timer, PattGen, GPIO, ROM Ctrl, sram ctrl, and adc ctrl. \citet{IDsnow18paper} used a benchmark suite of 10 hardware designs from Hack@DAC competitions and OpenTitan.

The nvBench dataset is used in~\cite{IDsnow32paper} to evaluate GPT-3.5's ability to generate Vega-Lite specifications, a declarative JSON format for creating interactive visualizations.
The Energy Exascale Earth System Model (E3SM) is used as dataset source, employing Feature Query Language (FQL) to analyze this large-scale scientific software~\cite{ID78paper}.
To test the effectiveness of their LLM-based PDDL generation, \citet{IDsnow26paper} evaluated their approach using well-known planning domains such as gripper, logistics, and tyreworld, along with custom domains like household and pizza cooking.
Finally, \citet{ID32paper} utilized chemistry protocols from the Organic Syntheses dataset and custom educational experiments to generate task plans in XDL (Chemical Description Language).

In summary, many of the DSL datasets are created for Verilog, from different resources including GitHub and textbooks. For other DSLs, forums, education resources, and specialized repositories are used to collect datasets, or existing benchmarks are modified.

\begin{table}[htbp]
\renewcommand{\arraystretch}{1.2}
\begin{threeparttable}
\centering
\footnotesize
\caption{Synthesized datasets for code generation in low-resource and domain-specific languages}
\label{tab:synthezied_datasets-lrpl-dsl}  
\begin{tabular}{p{0.05\textwidth}p{0.10\textwidth}p{0.25\textwidth}p{0.50\textwidth}}
\toprule
\multirow{19}{*}[5em]{\rotatebox[origin=c]{90}{\textbf{Synthesized}}} & \textbf{ID} & \textbf{Languages} & \textbf{Model Used} \\
\midrule
& \citep{ID58paper} & OCaml, Racket, R, Julia, Lua & StarCoder-15B \\
& \citep{ID44paper} & R & GPT-3.5 Turbo\\
& \citep{ID66paper} & PHP, Swift, Rust & GPT-3.5-turbo-1106 \\
& \citep{IDsnow34paper} & Kotlin & GPT-3.5-Turbo,  Mistral-7B-Instruct-v0.2 \\
& \citep{IDsnow33paper} & Rust & GPT-4 \\
& \citep{ID52paper} & Kotlin, PHP, Ruby, Scala, Perl, Swift & GPT-4 \\
& \citep{IDsnow8paper} & 40 PLs & GPT-4-1106-preview \\
& \citep{ID2paper} & Verilog & ChatGPT-3.5-Turbo API \\
& \citep{ID92paper} & Verilog &  ChatGPT-3.5-Turbo API\\
& \citep{ID77paper} & Verilog & GPT-3.5 \\
& \citep{IDsnow39paper} & Verilog & LLaMA2-70B-Chat, GPT-3.5-Turbo \\
& \citep{IDsnow27paper} & Verilog & GPT-3.5-Turbo\\
& \citep{IDsnow29paper} & Verilog & GPT-3.5 \\
& \citep{IDsnow31paper} & Verilog & GPT-3.5 \\
& \citep{IDsnow37paper} & Verilog & Finetuned CodeLlama, Finetuned DeepSeek, Finetuned CodeQwen \\
& \citep{IDsnow15paper} & Verilog & Claude3-Haiku \\
& \citep{ID39paper} & Lean &  GPT-4\\
& \citep{IDsnow22paper} & Lean & DeepseekMath-Base-7B (fine-tuned) \\
& \citep{IDsnow11paper} & FOL & GPT-4 \\
& \citep{IDsnow16paper} & Ansible & GPt-4 \\
\bottomrule
\end{tabular}
\end{threeparttable}
\end{table}

\textbf{Synthesized Datasets.}
Recent trends in dataset creation for code generation tasks reveal a strong preference for large, closed-source language models. As evident from Table~\ref{tab:synthezied_datasets-lrpl-dsl}, the majority of researchers have utilized closed-source models such as GPT-3.5, GPT-4, and only a few of them have used open-source models such as StarCoder-15B~\citep{ID58paper} and DeepSeekMath-Base-7B~\citep{IDsnow22paper} to synthesize datasets for various low-resource and domain-specific languages. 
These models have been employed in diverse ways to generate, augment, and categorize code-related data.

\begin{table}[htbp]
\renewcommand{\arraystretch}{1.2}
\begin{threeparttable}
\centering
\footnotesize
\caption{Manually curated datasets for code generation in LRPLs and DSLs}
\label{tab:manual_datasets-lrpl-dsl}  
\begin{tabular}{p{0.05\textwidth}p{0.10\textwidth}p{0.25\textwidth}p{0.50\textwidth}}
\toprule
\multirow{9}{*}[-5em]{\rotatebox[origin=c]{90}{\textbf{Manual}}} & \textbf{ID} & \textbf{Languages} & \textbf{Description} \\
\midrule
& \citep{ID103paper} & Bash &  A set of 50 diverse prompts covering various GNU/Linux tasks \\
& \citep{ID142paper} & IEC-61131-3-ST & Manually created 100 diverse prompts \\
& \citep{ID160paper} & Verilog & 30 common digital designs collected by the authors \\
& \citep{ID33paper} & Verilog & Manually created, 8 representative hardware design tasks \\
& \citep{ID72paper} & ODSL & Evaluation dataset of 197 test cases \\
& \citep{ID112paper} & DSL for Simulation Process & Chemists curated questions from US patent chemical reactions annotated by computer scientists  \\
& \citep{ID92paper} & Verilog & Converted 156 problem descriptions from HDLBits into a text-only format\\
& \citep{IDsnow14paper} & SVA & 20 open-source designs that cover a wide array of applications   \\
& \citep{IDsnow34paper} & Kotlin & Rewrote 164 HumanEval examples in Kotlin\\
& \citep{IDsnow8paper} & 40 Programming Languages & 10 software developers are recruited as multilingual programming annotators \\
& \citep{IDsnow3paper} & Verilog &  Dataset of 15 Verilog Problems \\
& \citep{ID48paper} & Rust &  A microbenchmark dataset called MGDMICROBENCH  \\
& \citep{IDsnow38paper} & VHDL & Generated VHDL canonical Solutions, problem descriptions, and testbenches \\
& \citep{IDsnow10paper} & LTL & Written 36 natural language and LTL pairs for evaluation \\
& \citep{IDsnow20paper} & OCL & 15 UML classes along with 168 NL specifications are prepared to evaluate LLM OCL generation capacity \\
& \citep{ID71paper} & R & Few shot demonstrations for prompting the model \\
& \citep{IDsnow2paper} & Verilog & A set of hardware design benchmarks \\
& \citep{IDsnow40paper} & Verilog & Benchmark suite consisting of complex hardware modules  \\
\bottomrule
\end{tabular}
\end{threeparttable}
\end{table}

\citet{ID2paper} utilized ChatGPT-3.5-Turbo to craft descriptions from the README files and code snippets of the repository, as well as to categorize their curated Verilog files from GitHub across four levels of complexity (Basic, Intermediate, Advanced, Expert). 
Similarly, few-shot prompting with GPT-3.5-Turbo is used to generate descriptions for Verilog modules sourced from GitHub~\cite{ID92paper}. These descriptions and code are used to fine-tune CodeGen Models.
In a different approach, GPT-3.5-Turbo is used to curate an R dataset, for two distinct tasks: to classify code-comment pairs into predefined categories during the creation of the StatCodeSearch dataset, and for binary classification evaluation to determine if a comment accurately represents the code's function~\cite{ID44paper}.

GPT-4 is utilized for auto-formalization of mathematical problems through few-shot learning with iterative refinement~\cite{ID39paper}. The authors initially prompted GPT-4 with manually rewritten examples of informal-formal statement pairs in the Lean language. To improve formalization accuracy, they implemented a feedback loop where GPT-4's outputs were verified by the Lean proof assistant. If errors were detected, GPT-4 was prompted to refine its result based on error messages from Lean or feedback from human experts. This iterative process continued until the formalization passed Lean's verification or reached a maximum of five iterations. Through this GPT-4-driven approach, they successfully formalized over 60\% of the IMO Shortlisted Problems in Algebra and Number Theory categories, resulting in the 149 problems that comprise the FIMO dataset.

LLMs are used for both cleaning the dataset and translating data from Python~\cite{IDsnow34paper}. Mistral-7B-Instruct and GPT-3.5-Turbo are used to predict the quality of a small portion of the dataset. This prediction is used to build a classifier and clean their large dataset, and obtain a high-quality dataset. They also used GPT-3.5-Turbo to translate the instruction dataset in Python code from Code Exercise dataset to Kotlin. Based on these steps, the authors created various datasets, KStack, KStack-clean, and KExercises, and fine-tuned the models on these datasets to improve the performance of LLMs on Kotlin-specific tasks.
In another study~\cite{IDsnow8paper}, researchers used GPT-4 to initially rewrite an existing code to a cleaner code snippet, then using these code snippets, the LLM is prompted to create Instruction corpora, thus creating a large-scale Instruction corpora containing 110K samples in 40 languages, which they used to fine-tune their MCODER model for multilingual code understanding and generation tasks. Some of the low-resource programming languages included in this dataset are AWK, Fortran, Pascal, Shell, VimScript, Common Lisp, Elixir, Erlang, Haskell, Racket, Scheme, and TCL.

For low-resource programming languages, StarCoder-15B is employed in the dataset generation framework, MultiPL-T~\cite{ID58paper}. StarCoder-15B was used for two critical roles: test case generation for Python functions and translating Python functions to target low-resource languages. They generated five independent test suites per function with a high temperature of 0.8 to ensure diversity, and produced 50 translations into target low-resource languages for each Python function. This large number of candidates was generated to increase the likelihood of producing at least one correct translation, which was later filtered using the previously generated tests. The low-resource programming languages targeted in this study include Julia, Lua, OCaml, R, and Racket, with the authors demonstrating significant improvements in model performance for these languages after fine-tuning on the generated data. 

In another approach, \citet{ID66paper} used GPT-3.5-Turbo to generate self-contained coding problems and their solutions from code snippets extracted from the StarCoderData dataset, a filtered version of The Stack \citep{ID90paper}. They used greedy decoding to maximize consistency between problems and solutions, repeating this process to generate approximately 75,000 problem-solution pairs for their OSS-INSTRUCT dataset. 
\citet{IDsnow27paper} used GPT-3.5-Turbo to label the unlabeled Verilog modules, creating a large dataset for fine-tuning Mistral.
\citet{IDsnow11paper} used GPT-4 to generate NL-FOL pairs. To ensure a diverse dataset, they specifically instructed the model not to produce NL-FOL pairs containing n-grams that are already frequent in the dataset created so far.
\citet{IDsnow16paper} initially employed prompt engineering to tune the prompt using 10 NL-Ansible Playbooks, using the refined prompt, they generated a dataset of 130 Ansible playbooks along with the NL prompts that can be used for Auto-remediation. 

Based on the observation that current LLMs are better at summarizing code than generating it, \citet{IDsnow31paper} used GPT-3.5 to produce descriptions to get high-quality descriptions. They used multi-level summarization in a few-shot setting, where the model first generates a fine-grained summary and then a high-level description. This dataset of descriptions and Verilog Code is used to fine-tune three base models: CodeLlama-7b-Instruct, Deepseek-Coder-6.7b-Instruct, and CodeQwen1.5-7B-Chat.
Similarly, the fine-tuned DeepseekMath-Base-7B model in~\cite{IDsnow22paper} is used to classify the quality of the formal statements. An ablation study demonstrated the effectiveness of this approach, showing that the model trained on high-score proof data outperformed the model trained on low-score proof data by 4.5\% on the miniF2F benchmark, thus highlighting the importance of quality classification in enhancing the model's theorem-proving capabilities.

\citet{IDsnow29paper} utilized GPT-3.5 for two key tasks: scoring Verilog code quality based on educational value, readability, design, scalability, and robustness; and generating problem-solution pairs for instruction data. To mitigate costs, they scored a subset of data and trained a Sentence-Ttransformer on these scores for efficient large-scale evaluation. The generated Verilog code underwent syntactic and functional correctness tests before inclusion in the dataset. This approach allowed them to create a high-quality, diverse dataset for fine-tuning Codellama-7B, DeepSeek-Coder-6.7B, and CodeQwen1.5-7B in Verilog code generation.
\citet{IDsnow37paper} finetuned models using data they gathered from GitHub. Then they used these models to get diverse Verilog modules from the Verilog definitions.
Additionally, Claude3-Haiku is used to generate Verilog Code~\cite{IDsnow15paper}. The code is then evaluated, and errors are fed back into the model for refinement. This process is repeated until the correct code is generated or it reaches a predefined number of iterations. This dataset of erroneous code and correct code is used for fine-tuning a student model. Other studies followed a similar approach for LRPLs to generate Self-reflection sequences~\cite{IDsnow33paper}. 
Lastly, \citet{ID77paper} extensively used GPT-3.5 in their dataset creation phase for digital IC design. They prompted GPT-3.5 to generate a comprehensive pool of keywords, expand these into instructions, and perform mutation operations to enhance dataset diversity. In the final stage, GPT-3.5 was employed to generate multiple Verilog code solutions for each instruction, resulting in over 27,000 instruction-code pairs covering a wide spectrum of RTL design tasks. Following the dataset creation, they leveraged this extensive collection to fine-tune pre-trained language models, resulting in their RTLCoder.

\textbf{Manual Datasets.}
The manually created datasets are small datasets mostly used for evaluation purposes.
Table \ref{tab:manual_datasets-lrpl-dsl} presents an overview of manually curated datasets for code generation in low-resource and domain-specific languages.
This table summarizes various research efforts, highlighting the diverse languages and descriptions of how they were collected. The manual curation approach has been particularly valuable in domains where large-scale, high-quality datasets are not readily available or where specific, targeted evaluations are required. Several researchers have created manually curated datasets to evaluate code generation in various domains. 

\citet{ID103paper} proposed an execution-based evaluation framework for the NL2Bash task. They executed the Bash output from the model in a podman container and compared the actual results against the expected ones. However, testing LLMs in such a setting requires carefully written prompts. As gathering high-quality prompts on a large scale from the internet is impractical, they manually wrote 50 prompts covering various tasks in Linux Operating System, including system diagnostics and administration. 164 HumanEval samples are manually translated to Kotlin to solve the issue of Genric Variable type in function signatures in existing HumanEval Kotlin benchmarks~\cite{IDsnow34paper}.

In the field of industrial automation, \citet{ID142paper} manually curated 100 prompts to assess ChatGPT's capabilities across a wide range of tasks. Using this dataset, they evaluated the syntactical correctness, functional plausibility, and overall quality of ChatGPT's control logic generation. 
For digital design, \citet{ID160paper} collected 30 common designs to systematically evaluate auto-generated RTL from LLMs. Their RTLLM benchmark comprises diverse digital designs, each accompanied by a natural language description, testbench, and reference RTL implementation. This comprehensive dataset facilitates automated evaluation of RTL generated by large language models, assessing syntax correctness, functional accuracy, and design quality metrics.

MCEval, massively multilingual benchmark, containing 16K samples and covering 40 languages, is curated in~\cite{IDsnow8paper}. Ten software developers were tasked to create clear and self-contained problem statements and corresponding solutions. 
\citet{ID33paper} prepared a test suite consisting of 8 different hardware design challenges, ranging from simple components like shift registers to more complex systems like traffic light FSMs. These tasks were chosen to represent various common hardware design scenarios, allowing for a comprehensive assessment of LLMs' performance in this domain. 
In the context of natural language commanding systems, \citet{ID72paper} developed an evaluation set comprising 197 test cases. This dataset addresses two key challenges: evaluating their custom-designed Office Domain Specific Language (ODSL) and covering the vast range of possible user utterances.

\citet{IDsnow14paper} created a comprehensive benchmark suite consisting of 20 open-source designs from various hardware domains to evaluate LLMs in generating SystemVerilog Assertions (SVAs) from natural language specifications.
For chemical process simulation, \citet{ID112paper} created their own DSL to balance expressiveness and simplicity. 
In this work, given a question that is related to a chemical equation, the model generates code in the DSL they designed to answer it. 
This approach allows for the representation of complex chemical processes while reducing the search space for code generation. As their DSL is new, manual curation and annotation were necessary. 
Lastly, the challenge of translating visual elements from the HDLBits website into text-only formats suitable for LLMs was studied in~\cite{ID92paper}. The authors manually converted circuit schematic diagrams, Boolean logic tables, Karnaugh maps, and state transition graphs into clear, unambiguous text descriptions and edge list-based formats.


\subsection{Dataset Processing}

The collected datasets include noise or irrelevant information, needs reformatting~\citep{ID129paper, ID39paper, ID75paper}, or have licensing issues~\citep{ID73paper, ID67paper}, among others. Additionally, inconsistent coding styles~\citep{ID6paper} or conventions, potential security vulnerabilities~\citep{ID84paper} or sensitive information~\citep{ID84paper}, duplicate or near-duplicate entries~\citep{ID92paper, ID49paper}, outdated or deprecated code, and lack of proper documentation or comments can further complicate the use of raw data. Processing is conducted to create high-quality data.
Inspired by the data preprocessing procedures proposed by \citet{hou2024large}, we divided the data preprocessing steps followed for LRPL and DSL datasets into 8 common stages.  
While the exact process may vary depending on the specific domain, source of the data, and language, the following list outlines the common major stages observed across various research efforts.
The major stages in dataset processing typically include (1) Data Collection, (2) Initial Filtering, (3) Deduplication, (4) Fine-grained Filtering, (5) Code Extraction and Cleaning,  (6) Quality Checks, (7) Dataset-specific Processing, and (8) Decontamination.
Table~\ref{tab:data-processing-steps} provides an overview of these stages in processing the datasets. 
Dataset collection is explained in detail in the previous sections. Therefore, in the following, we will focus on explaining the details of steps (2)-(8).

\begin{table}[htbp]
\centering
\fontsize{8pt}{11pt}\selectfont
\begin{tabular}{p{0.15\textwidth}|p{0.25\textwidth}|p{0.25\textwidth}|p{0.2\textwidth}}
\hline
\textbf{Processing Step} & \textbf{Description} & \textbf{Example} & \textbf{References} \\
\hline
Initial Filtering & Removing obviously irrelevant or low-quality data & Filtering based on file size, file extensions, and license compatibility; using PR labels for hardware designs & \citep{ID117paper,ID157paper,codegeex,ID18paper,ID19paper,ID32paper,ID59paper,ID5paper,ID67paper,ID6paper,ID73paper,ID80paper,ID84paper,ID8paper,ID93paper,IDsnow18paper,IDsnow19paper,IDsnow23paper,IDsnow31paper,IDsnow34paper,IDsnow37paper} \\
\hline
Deduplication & Preventing overrepresentation of certain data points & Using MinHash algorithm with Jaccard similarity; ROUGE-L based deduplication for text; workflow deduplication & \citep{ID113paper,ID117paper,ID157paper,ID19paper,ID57paper,ID58paper,ID5paper,ID67paper,ID6paper,ID75paper,ID8paper,ID90paper,ID92paper,IDsnow31paper,IDsnow34paper,IDsnow39paper, ID42paper} \\
\hline
Fine-grained Filtering & More precise control over dataset composition & Verilog: identifying self-contained modules; R: limiting line lengths, removing auto-generated files; Bash: filtering out non-UNIX and GUI-dependent utilities & \citep{ID18paper,ID56paper,ID6paper,ID71paper,ID75paper,ID77paper,ID92paper,IDsnow12paper,IDsnow15paper,IDsnow27paper,IDsnow29paper,IDsnow31paper,IDsnow32paper}\\
\hline
Code Extraction and Cleaning & Isolating relevant code snippets and removing extraneous information & R: converting RMD to R scripts, extracting code-comment pairs; Verilog: using OCR on textbooks; Math: converting PDF to LaTeX; Text normalization and cleaning & \citep{ID129paper,ID27paper,ID31paper,ID34paper,ID39paper,ID43paper,ID44paper,ID49paper,ID60paper,ID70paper,ID78paper,ID8paper,IDsnow20paper,IDsnow23paper,IDsnow37paper}\\
\hline
Quality Checks & Ensuring processed data meets predefined standards & SKILL: using static analysis for code quality and compilability; Verilog: verifying module structures; Bash: enhancing under-specified commands with specific file names and paths & \citep{ID107paper,ID52paper,ID56paper,ID65paper,ID92paper,IDsnow22paper} \\
\hline
Dataset-specific Processing & Customization based on unique requirements of the domain & Math: converting to Lean formal language; R: incorporating metadata like difficulty level; SKILL: mining input-output pairs & \citep{ID107paper,ID112paper,ID12paper,ID133paper,codegeex,ID34paper,ID39paper,ID46paper,ID74paper,ID81paper,IDsnow11paper,IDsnow20paper,IDsnow33paper,ID42paper} \\
\hline
Decontamination & Ensuring originality and integrity of the dataset & Removing coding problems matching popular benchmarks; checking for data leakage between train and test sets & \citep{ID66paper,ID77paper,IDsnow31paper} \\
\hline
\end{tabular}
\caption{Stages in dataset processing for code datasets in LRPLs and DSLs}
\label{tab:data-processing-steps}
\end{table}

\paragraph{I. Data Collection}
This step is discussed in detail in the previous subsections.

\paragraph{II. Initial Filtering}
Initial filtering is crucial to remove obviously irrelevant or low-quality data, thereby reducing noise and computational overhead in subsequent processing steps. This stage involves broad filtering criteria such as filtering based on file extensions to identify relevant languages~\citep{ID117paper}. For instance, \citet{ID8paper} gathered Verilog data from the Google BigQuery dataset using a query search for relevant words and the \texttt{.v} file extension. Similarly, Kotlin files were filtered using extensions such as \texttt{.kt}, \texttt{.kts}, and \texttt{.gradle.kts}~\cite{IDsnow34paper}, and \texttt{.yml} and \texttt{.yaml} extensions were used to filter YAML files~\cite{ID6paper}.
Another important criterion is removing files that are too small or too large~\citep{ID32paper, ID80paper, IDsnow23paper}. For example, \citet{ID8paper} filtered out large Verilog files having more than 20,000 characters, while \citet{IDsnow31paper} deleted files exceeding 2096 characters to avoid truncation during training. \citet{ID157paper} removed files larger than 1MB or with fewer than 100 tokens, and \citet{codegeex} removed files bigger than 100KB and smaller than 1KB. 
This ensured the exclusion of potentially non-representative data and led to more standard and readable datasets~\citep{ID18paper}. \citet{IDsnow18paper} trimmed down source code files to contain only 100 lines of code to limit the number of tokens in the prompt.
At this stage of dataset processing, auto-generated files are also removed~\citep{IDsnow37paper, ID157paper, ID19paper, codex, IDsnow19paper}. A dataset of meaningful interactions is maintained by filtering out cases where users did not write anything after calling the model (empty ground truth) or where models failed to generate predictions (empty predictions)~\cite{ID59paper}.
Additionally, filtering for commercially friendlier licenses is employed by some researchers \citep{ID73paper, ID67paper, IDsnow37paper, IDsnow34paper}. For instance, \citet{ID5paper} removed all data samples coming to their Ansible service from files or repositories with restrictive licenses to ensure ethical use of training data. The training data was also filtered to remove hate speech, 
abuse, and profanity, helping to maintain the ethical standards of the generated content.
This step was required as users' queries were used to create the dataset for pre-training.
Finally, removing Personal Identifiable Information (PII) such as names and emails from the prompts is of prime importance to ensure ethical dataset preparation practices~\citep{ID5paper, ID84paper, ID93paper, IDsnow34paper}.

\paragraph{III. Deduplication}
Duplication is common in code-related sources such as GitHub~\citep{ID117paper}, as code is often reused. However, using this duplicated code to train LLMs can adversely affect performance~\citep{ID157paper}. Deduplication is essential to prevent overrepresentation of certain data points and to avoid biasing models towards repeated information, making it crucial for maintaining dataset quality. Studies~\cite{ID90paper} demonstrated that deduplicating the training set can significantly improve LLM performance. Consequently, several researchers have employed various techniques for deduplication in both LRPL and DSL datasets~\citep{ID157paper, ID113paper, IDsnow34paper, ID19paper, IDsnow39paper, ID75paper}.
Exact deduplication is employed by several studies~\citep{ID157paper, ID6paper}, while near deduplication is used by others~\citep{ID67paper, IDsnow34paper}. Some studies~\cite{ID57paper} note that exact deduplication is relatively efficient with amortized $\mathcal{O}(1)$ complexity compared to near deduplication, which takes $\mathcal{O}(n^2)$. Additionally, deduplication based on the MinHash Algorithm using Jaccard Similarity was utilized in several studies~\citep{ID92paper,IDsnow31paper,ID8paper,IDsnow31paper}.
Other approaches include ROUGE-L based deduplication~\cite{ID58paper}, Levenshtein distance based deduplication~\cite{ID117paper}, and file-level deduplication~\cite{ID5paper} for Ansible YAML. These diverse deduplication techniques demonstrate the importance of this step in dataset preparation for training LLMs on code-related data. Others~\cite{ID42paper} deduplicated formulas by creating ``formula sketches,'' where specific values like numbers, strings, and cell references were replaced with their general token types; providing an example where the Excel formula \texttt{=SUM(A1:A10)} is converted to the sketch \texttt{=SUM(cell:cell)}.

\paragraph{IV. Fine-Grained Filtering}
Fine-grained filtering allows for more precise control over dataset composition, ensuring that the data meets specific quality or relevance criteria. This stage involves more specific filtering criteria, such as the identification of self-contained Verilog modules. Self-contained Verilog modules are identified by leveraging Abstract Syntax Trees (AST) and ensuring modules contain both \texttt{module} and \texttt{endmodule} without any module instantiation~\cite{ID92paper}. 
Other works~\cite{IDsnow15paper} also checked for the inclusion of at least one occurrence of keywords related to procedural blocks (e.g., \texttt{always}, \texttt{always\_comb}, \texttt{always\_ff}, \texttt{always\_latch}). Regular expressions is used to extract individual Verilog modules~\cite{IDsnow27paper}, while others~\cite{IDsnow31paper} excluded files with external references (containing \texttt{include} or \texttt{import} keywords) to make the modules self-contained.
Syntactic correctness was a key focus for many researchers. \citet{ID6paper} checked for valid YAML syntax in Ansible playbooks using PyYAML. Similarly, \citet{ID77paper} used PyVerilog to verify the syntactic correctness of Verilog code, and \citet{IDsnow29paper} used the Icarus Verilog tool for the same purpose. \citet{IDsnow31paper} ran syntax checks on all Verilog code and removed any that failed.
Comment handling was another important aspect of fine-grained filtering. \citet{IDsnow31paper} used regular expressions on Verilog code to delete implementation-unrelated comments, ensuring that comments were relevant to the implementation of the Verilog module. 

Similarly, another study~\cite{IDsnow15paper} removed comments within the code samples to prevent extraneous information from influencing the generation of accurate and relevant specifications.
Domain-specific filtering was also applied in various studies. \citet{ID56paper} filtered out non-UNIX, non-Linux, and unsupported utility commands (e.g., \texttt{ssh}, \texttt{sudo}, and GUI-dependent utilities) when adapting the NL2Bash dataset. \citet{ID75paper} removed workflows without natural language descriptions. \citet{ID71paper} filtered the MATH dataset to include only problems with single numerical answers, excluding interval or formula-based solutions, to ensure compatibility with the PoT approach across different programming languages.
Additionally, data quality improvements were made in several studies. \citet{IDsnow12paper} removed 22 samples from the FOLIO validation set due to various errors, including unbalanced parentheses in ground truth FOL expressions, mismatches between computed and provided labels, and inconsistencies between the number of premises and FOL expressions. 
Instances requiring multiple data table joins were excluded from the evaluation, as Vega-Lite doesn't support multiple data sources in a single visualization specification~\cite{IDsnow32paper}.
\citet{ID18paper} employed a hybrid approach combining rule-based and learning-based filtering to curate high-quality training data. The rule-based system used metrics like code structure using AST and adherence to standards, while the learning-based method utilized a BERT model trained on GPT-4-generated quality-scored examples to evaluate code correctness, readability, and security.

\paragraph{V. Code and Text Extraction and Cleaning.}
Code extraction and cleaning are vital processes to isolate relevant code snippets and remove extraneous information, thereby improving the signal-to-noise ratio in datasets. These processes, along with text processing, are essential for standardizing and cleaning data, making it more amenable to analysis or model training. Researchers have employed various techniques to achieve this.
\citet{ID129paper} erased non-code information and converted RMD files to R scripts, while \citet{ID44paper} implemented rule-based extraction to create code-comment pairs from R projects, detecting comment lines (starting with \texttt{\#}) and concatenating multiple subsequent comment lines. For Verilog data, \citet{ID8paper} applied regular expressions to identify snippets from textbook data and used optical character recognition, specifically pymuPDF, to extract text from Verilog textbooks.
In other domains, the Mathpix snipping tool is used to convert IMO problems from PDF to \LaTeX{} format~\cite{ID39paper}, and text normalization and cleaning using tools like datasketch are performed~\cite{ID49paper}. 

\citet{ID27paper} extracted text contents from Unix manual pages and segmented them into conceptually distinct paragraphs delineated by line breaks, creating a fine-grained documentation pool that facilitates more targeted retrieval and analysis of command functionalities and flag usages.
For specialized languages, the dataset is normalized on Intermediate Representation by discarding comments, debug metadata, and attributes, and standardizing whitespace while retaining newlines~\cite{ID60paper}. \citet{ID31paper} developed a customized OCaml parser to process COQ source files, handling its unique syntax by separating individual COQ sentences, removing comments, and eliminating certain directives like \texttt{\#global}.
In the realm of natural language processing, \citet{ID43paper} used the candc-boxer toolchain to translate natural language into formal logical representations, creating large-scale parallel datasets of sentences and their corresponding FOL translations. 

Similarly, custom grammars were used to generate large datasets of paired LTL formulas and their natural language descriptions, ensuring consistency and diversity in the examples.
Others~\cite{ID70paper} preprocessed Python code generation benchmarks, HumanEval, and MBPP, by modifying them for consistency and translating them to 18 additional programming languages. This compiler-driven translation process resulted in MultiPL-E, a set of parallel multi-language benchmarks that maintained the core structure of the original problems while accounting for each language's unique features.
For specific applications, XScan is employed to extract key information from E3SM source code, including lines of code, languages used, and dependencies~\cite{ID78paper}. \citet{IDsnow20paper} focused on extracting UML elements from English specifications through tokenization, part-of-speech tagging, and lemmatization, particularly identifying nouns and adjectives that typically correspond to UML elements.
Another work~\cite{IDsnow23paper} removed unnecessary information from PowerShell files such as logging and echo commands. 
\citet{ID34paper} entirely removed Package declaration lines from their dataset and implemented a probabilistic removal process for import statements, where each import line had a 50\% chance of being deleted. This strategic approach was implemented to prevent model overfitting on these often numerous declarations.
To standardize and generalize the command data, \citet{ID130paper} replaced specific elements like IP addresses, long digit sequences, and random strings with generic placeholders such as \texttt{<ip>}, \texttt{<digits>}, and \texttt{<str>}, respectively.
\citet{ID141paper} removed non-ASCII characters, comments, and extraneous spaces, replacing multiple spaces with single spaces, combining files within each corpus to ensure a consistent and representative codebase.
Lastly, \citet{IDsnow37paper} leveraged LLMs' pre-existing knowledge of C programming by creating Verilog-C pairs using a V2C tool, demonstrating an innovative approach to data preparation that capitalizes on model knowledge transfer.

\paragraph{VI. Quality Checks}
Ensuring data quality through rigorous quality checks is crucial to maintaining predefined standards of correctness, completeness, and relevance in processed data. Researchers have employed various techniques to achieve this goal. \citet{ID107paper} utilized a SKILL static analysis tool to measure code quality and compilability, removing files with low scores. Similarly, another work~\cite{ID92paper} verified that filtered Verilog code contained proper module structures, ensuring structural integrity of the data.
For command-line utilities, \citet{ID56paper} enhanced under-specified commands by adding specific file names, directory names, and paths, and updated deprecated utilities and flags, thereby improving the practical applicability of the dataset. In the realm of multilingual data, \citet{ID52paper} employed BERTScore to evaluate the similarity between back-translated English prompts and original English texts, retaining translations with a similarity score above 0.95 to ensure high-quality multilingual data.
Others~\cite{ID65paper} implemented an automated culling process using the MATIEC compiler to validate and filter the OSCAT dataset. This process involved discarding any Structured Text (ST) files that failed to compile successfully, ensuring only valid code remained in the dataset.
For synthetic proof data, the work in~\cite{IDsnow22paper} introduced a novel approach called Hypothesis rejection to clean inconsistent samples. This was accomplished by attempting to prove each formal statement with \texttt{False} as the conclusion. If such a proof succeeded, it indicated that the original statement's hypotheses were inconsistent, allowing the researchers to exclude these logically unsound statements from their dataset. This method ensured the logical validity of the synthetic proofs, enhancing the overall quality of the dataset.

\paragraph{VII. Dataset-Specific Processing}
Dataset-specific processing allows for customization based on the unique requirements of the domain or research question, enhancing the dataset's utility for specific applications. Various researchers have employed unique processing steps tailored to their datasets.
Mathematical problems are converted into propositions open to formalized proof in the Lean formal language, reframing problems asking to find variables as provable statements~\cite{ID39paper}. For SKILL data, \citet{ID107paper} mined input-output pairs for training and evaluation. \citet{ID34paper} incorporated metadata such as difficulty level, type of code to be generated (e.g., numerical, statistical, visualization), and source information in their R programming benchmark.
In the SimQA dataset preprocessing, each question was augmented with three randomly generated incorrect answers close in value to the correct answer, creating a multiple-choice format. This approach enables evaluation of both multiple-choice models and open-ended answer generation approaches on the same dataset.
\citet{ID112paper} and \citet{ID12paper} formatted design specifications from documents into a structured JSON format, categorizing information under labels such as system overview, definitions, and functional requirements. This facilitated more effective processing by the LLM for generating System Verilog Assertions (SVA).

To handle multiple tasks with a single model, \citet{ID133paper} added distinct prefixes to the input data: \texttt{ShellCodeGen} for shellcode generation and \texttt{ShellCodeSum} for summarization. Similarly, \citet{codegeex} added language-specific tags in the form of comments before each code segment to help the model distinguish between multiple programming languages.
\citet{IDsnow20paper} applied pre-processing to OCL constraint generation, specifically processing English specifications and UML models by tokenizing, lemmatizing, extracting elements, generating graph paths, and ranking them to craft focused prompts. This approach selectively augmented prompts with relevant UML classes.
In the context of grammar-based approaches, \citet{ID46paper} employed a specialized preprocessing technique to derive minimal grammars for each example. This involved using the full grammar $G$ to parse each output $y$, then extracting only the rules used in that derivation, resulting in a minimal specialized grammar $G[y]$ specific to each example.

\citet{ID74paper} used human-written OCL constraints when available in the dataset, and for specifications lacking ground truth, they utilized correctly generated constraints from their previous experiments to ensure comprehensive example coverage. For CAD sketches, a custom encoding scheme is developed to represent them as sequential data, converting each sketch into a series of tokens representing basic shapes (primitives) and the relationships between these shapes (constraints)~\cite{ID81paper}.
\citet{IDsnow11paper} performed atomic perturbations such as label change (modifying predicates, terms, or operators), insert (adding terms, negations, or formulas), and delete (removing terms, negations, or formulas) for training the Chain-of-Thought correction model. Also, the input data was structured by wrapping different components (such as user messages, assistant messages, code blocks, and execution results) with special tokens like \texttt{<|user|>}, \texttt{<|assistant|>}, \texttt{<|code|>}, and \texttt{<|execution|>}. This formatting helps the model distinguish between different parts of the input~\cite{IDsnow33paper}.

\citet{ID42paper} performed case normalization. Since Excel is case-insensitive (except for string constants), they converted all input tokens to lowercase to map differently capitalized tokens to a single token.
\citet{ID130paper} improved task-specific context by incorporating special tokens: \texttt{</s>} to denote command endings, \texttt{<sh>} to signify command sequences, and \texttt{<nl2sh>} to identify natural language queries for the NL2SH (Natural Language to Shell) task.
\citet{ID69paper} included adding \texttt{<s>} and \texttt{</s>} tokens to demarcate the beginning and end of code snippets, as well as incorporating language-specific identifiers such as \texttt{<ruby>} and \texttt{<en>} to distinguish between different programming languages and natural language docstrings.
\citet{ID23paper} paired and formatted the generated English tasks and corresponding Verilog code into a specific structure: \texttt{TASK: <English Text> RESULT: <Verilog Code>}. This formatting prepares the data for input into the GPT-2 model.

\paragraph{VIII. Decontamination}
Decontamination is crucial to ensure the originality and integrity of datasets, preventing inadvertent inclusion of test data in training sets. This process is essential for maintaining the validity of model evaluations and ensuring fair comparisons across different benchmarks.
\citet{ID66paper} took a direct approach by removing coding problems that matched popular benchmarks such as HumanEval, MBPP, DS-1000, and GSM8K to avoid direct copying. This method helps preserve the uniqueness of their dataset and prevents potential overlap with widely used evaluation metrics.
A more sophisticated approach was employed by \citet{ID77paper} in their preprocessing of the RTLCoder model's training data. They implemented a step involving similarity checking between their training dataset and benchmark test cases. The researchers utilized the ROUGE-L metric, a widely used measure in natural language processing, to quantify the similarity between training samples and test cases. Any training samples that exhibited a ROUGE-L score greater than 0.5 when compared to any test case were systematically removed from the training set. This process minimizes potential data leakage and reduces the risk of overfitting to the evaluation benchmarks.
Similarly, \citet{IDsnow31paper} employed the ROUGE-L metric to ensure the integrity of their evaluation process and mitigate potential data contamination. Their approach mirrored that of \citet{ID77paper}, systematically excluding any training samples with a ROUGE-L score exceeding 0.5 when compared to any benchmark item.

\subsection{Distinctive Features for LRPLs and DSLs}

While GitHub is the main resource for data collection for HRPLs, the sources for LRPLs and DSLs are diverse and many times from specialized sources. 
For Ansible, Bash, and Verilog, the data collection also includes repositories like GitHub. However, most other benchmarks require manual efforts \cite{IDsnow42paper}.
Compared to HRPLs, the dataset for these languages, especially DSLs, is often collected from specialized sources: programming contests, textbooks, documentation platforms, help forums, and internal corpora, and the researchers also have to consider licensing constraints. Furthermore, in the curation process, often manual intervention and expert opinion are needed. 
To support DSL modeling, some works include additional structured information (e.g., schemas, intermediate representations, formal specs) as part of the curation process, thus integrating other modalities.
For LRPLs and DSLs, synthetic data generation is more about targeted, validated, and diversified quality. Data curation is more quality-dependent, where domain knowledge is essential, and this manual curation can limit the scale. 
In LRPLs/DSLs, the quality and structure of data have a greater impact than volume due to the limited availability of real-world examples. 
Additionally, though feedback loops and diversity in the code are considered in HRPLs, these techniques are carefully considered for DSLs. 

Furthermore, for manual curation of LRPLs and DSLs, LLMs play a central role. Often, a pipeline is required for data synthesis, including stages like prompting, prompt refinement, scoring, iterative correction, and formal verification. Many approaches also focus on code diversity and semantic correctness (especially for theorem provers or hardware description languages), and need an evaluator, e.g., a verifier, in the feedback loop.
The necessity for specialized tools and approaches is also seen for data processing, in addition to data collection. While the general steps to process data are similar to the steps used in preparing data for HRPLs, the techniques are adapted to the requirements of LRPLs and DSLs, as discussed above.

\begin{tcolorbox}[colback=gray!10,colframe=gray!40,boxrule=0.5pt,title=Summary RQ3,coltitle=black]
\begin{enumerate}
    \item We identified three main approaches for dataset creation: curated datasets from sources like GitHub (most common), manually created datasets for specific evaluations, and synthesized datasets generated using large language models (mostly using closed-source models). We further categorized curated datasets into three subcategories: Existing datasets, modified existing datasets, and collected datasets.
    
    \item We summarized the sources of datasets used for LRPL and DSL datasets, along with the programming languages represented and the types of datasets employed. We observed that researchers employ a wide variety of sources to build comprehensive datasets. These range from version control platforms and educational resources to domain-specific databases, professional forums, competition datasets, industrial sources, synthetic data generation, and cross-lingual resources, reflecting the ingenuity required to overcome data limitations in these specialized fields.
    \item We summarized different data pre-processing steps that are applied to make the data suitable for LLMs and found several common pre-processing steps, i.e., Initial Filtering, Deduplication, Fine-grained Filtering, Code and Text Extraction with Cleaning, Quality Checks, Dataset-specific Processing, and Decontamination. 

    \item We discussed the distinctions of data collection and preparation for LRPL and DSLs compared to high-resource programming languages. 
\end{enumerate}
\end{tcolorbox}

\section{Challenges and Opportunities}
\label{sec:challanges}

With the goal of closing the performance gap between LRPLs/DSLs and HRPLs as shown in Figure~\ref{fig:heatmap}, we examine the challenges and opportunities from RQ1 to RQ3 and summarize the challenges and opportunities that are unique for LRPLs and DSLs.

\subsection{Evaluation}
\paragraph{\textbf{Evaluation Challenges With Expert Judgement, Specialized Infrastructure and Semantics.}} 
As we have extensively described in Section~\ref{sec:rq1-distinction}, one of the challenges of evaluating DSLs is evaluator availability, which is limited as few people can judge each domain knowledge, e.g., COQ proofs or chemistry plans. As such, extensive manual evaluation or crowdsourcing could be challenging. 
Many DSLs are niche or proprietary, lacking test suites or labeled corpora or even enough unlabeled data to train a model. Additionally, due to specific factors that need to be evaluated for these languages, such as performance or feasibility of the generated code, developing automatic metrics is not always feasible, or the automated metrics should be combined with manual evaluations to cover all needed aspects of the assessment. The evaluation of these languages is also not as trivial, as a slight change in their syntax might alter the generated code, or there can be several correct forms with different syntax that are considered correct for a given problem; thus semantic equivalence is an important factor for some DSLs like Regex or Excel formulas. 
Another challenge is the execution environment limitations for DSLs 
that require complex simulators or constrained runtimes. Additionally, the tools used for each DSL are custom builds, sometimes to each hardware, making the evaluations custom and not generalized to other languages; in contrast to other metrics like Pass@k that are used for HRPLs. 

\paragraph{\textbf{Opportunity \#1: Develop Domain-Specific and Semi-Automated Evaluation Tools}}
Though there are specific requirements within each domain, rubrics could be developed and shared in each domain. Therefore, the development of standardized manual evaluation rubrics per domain can improve consistency and is an opportunity for researchers to act on.
Another research opportunity is developing semi-automated hybrid methods, combining domain parsers with manual scoring to reduce human burden. Though these methods are domain-specific, having semi-automatic approaches in each domain, and adopting them across similar ones can reduce the effort of developing new metrics or manual evaluation.
In some DSLs, the semantic or logic equivalence plays an important role. Developing tools that adapt matching logic based on DSL semantics (e.g., symbol abstraction) could help develop automatic evaluation metrics as well as developing semi-automated hybrid evaluation methods. 
Considering executable DSL sandboxes is another research direction to provide portable, lightweight environments for automated DSL code execution or simulation, which can be used for developing benchmarks and evaluation metrics that can be used across different settings of the same domain (e.g., Verilog).


\subsection{Data and Benchmark}
\paragraph{\textbf{Data Accessibility and Availability, Quality-Quantity Trade-off Challenges.}} 
Dataset creation for LRPLs and DSLs presents fundamentally different challenges compared to HRPLs like Python or Java. The most significant distinction lies in data accessibility and availability. While mainstream languages benefit from millions of publicly available repositories on platforms like GitHub, LRPLs and DSLs face severe scarcity constraints \cite{ID67paper, ID105paper}. For instance, despite Rust having 3.5 million developers, its public code repositories are substantially fewer than Python's ecosystem \cite{ID105paper}. More critically, many DSLs are used in proprietary environments where code cannot be publicly shared, e.g., Verilog designs contain intellectual property in semiconductor companies \cite{ID92paper, IDsnow37paper}, Ansible playbooks may expose sensitive infrastructure details \cite{IDsnow42paper}, and R scripts in pharmaceutical research remain confidential \cite{ID23paper}. This creates a fundamental data access problem that requires novel collection strategies beyond simple repository scraping \cite{ID27paper, IDsnow23paper}.

The quality-quantity trade-off in LRPL/DSL datasets differs markedly from HRPLs. Public repositories for these languages disproportionately contain educational or tutorial examples rather than production-quality code. Studies reveal that over 60\% of publicly available Verilog code consists of learning exercises rather than real hardware designs \cite{ID92paper}, contrasting sharply with Python, where frameworks like Django provide extensive production examples. Additionally, LRPL/DSL dataset curation requires domain expertise that general programmers lack, e.g., validating Lean theorem proving datasets requires mathematical knowledge, while assessing Verilog quality needs hardware design expertise. This expert dependency, especially for DSLs, creates curation bottlenecks and significantly increases costs compared to crowdsourced validation approaches used for mainstream languages.

\paragraph{\textbf{Dataset Curation and Preparation Challenges for LRPLs and DSLs}}
As also stated in Section~\ref{sec:rq1-distinction}, creating evaluation benchmarks for LRPLs and DSLs requires more effort compared to HRPLs. These benchmarks, even for LRPLs, are often generated through code translation, either manually from HumanEval or automated approaches that develop multi-step, careful techniques to ensure the correctness of the translated code.
Many DSLs lack standardized benchmarks, compelling researchers to curate their own evaluation datasets tailored to the specific DSLs under investigation. A recent advancement in this area is the introduction of Verilog-Eval~\citep{ID92paper}, which has been widely adopted by the Verilog code generation research community. The adoption of such benchmarks underscores the importance of standardized evaluation tools in fostering reproducible and comparable research outcomes.
However, developing such benchmarks requires techniques specific to each domain or special setups for executing code.
Similar to evaluation pipelines for LRPLs and DSLs that must be custom-built with domain expertise, to curate meaningful benchmarking, domain expertise is essential.

In RQ3 (Section~\ref{sec:rq3}), we have examined various data sources utilized by researchers in this field. A common theme observed is the reliance on diverse, language-specific sources to curate datasets. For instance, researchers \citep{IDsnow42paper} sourced their data from Linux Man Pages, \citet{ID42paper} extracted information from Excel help forums, and \citet{ID34paper} utilized R programming textbooks. 
The curation of high-quality datasets for DSLs is a critical challenge in advancing code generation capabilities for specialized fields due to various reasons, such as licensing issues~\citep{ID5paper}. Though ethical and licensing considerations are applied in HRPL data curation, the data scarcity for LRPLs and DSLs aggravates the problem.
Other challenges echo the burdens for evaluations and benchmarking. Many LRPLs and DSLs are deeply tied to their application domain (e.g., hardware design, theorem proving), and data curation must preserve semantics and domain constraints, unlike general-purpose HRPLs. Often, alternative modalities are used. For example, the datasets require pairing code with logic, schemas, intermediate representation, or visual representations (e.g., CAD, theorem proofs). Due to a lack of scale, LRPL/DSL dataset construction often involves expert-guided curation, synthesis, or template creation.
Even when data is synthesized with prompting LLMs, human feedback, iterative processes, or verification is required. 

Furthermore, data preparation, though following similar steps as in HRPLs, requires developing and adapting new techniques. These techniques are related to requiring expert knowledge and the domain-specific needs of LRPL/DSL.
For example, as Excel is case-sensitive, case normalization should be applied to prepare data~\cite{ID42paper}. Similarly, domain-specific tools should be used to filter and check the quality of the data. In some languages, such as R, concerns regarding different R paradigms affect how the two paradigms, base-R and tidy-verse, are handled in the dataset creation~\citep{ShawnR}. 
Similar to HRPLs, concerns regarding data leakage in benchmark datasets must be meticulously addressed. Researchers implement rigorous decontamination processes to ensure that benchmark data remains free from inadvertent overlaps with training datasets, thereby guaranteeing the reliability and validity of evaluation results. 
However, these techniques require novel approaches for LRPLs and DSLs. For example, formula sketches are developed to deduplicate formulas~\cite{ID42paper}.

\paragraph{\textbf{Opportunity \#2: Develop Standardized, Multi-Domain Aware Benchmarks, Platforms and Communities}}
While MultiPL-E~\citep{ID70paper} benchmark is frequently employed to assess LLM performance across various programming languages~\citep{ ID67paper, ID58paper, ID66paper, ID7paper}, there is a notable deficiency in comprehensive, language-specific benchmarks for LRPLs. In contrast, HRPLs like Python benefit from a diverse array of specialized benchmarks, such as APPS~\citep{apps} and CodeContests~\citep{alphacode} for general coding proficiency, as well as domain-specific evaluations like DSP~\citep{chandel2022trainingevaluatingjupyternotebook}, NumpyEval~\citep{cert}, and PandasEval~\citep{cert} for data science applications. Furthermore, evaluations on proprietary libraries, exemplified by TorchDataEval~\citep{zan-etal-2022-language}, provide insights into LLM capabilities with specialized frameworks.
The introduction of Software Engineering Benchmark (SWE-bench)~\citep{jimenez2024swebench} further exemplifies the sophisticated evaluation frameworks available for Python. 
This disparity underscores the need to develop analogous, use-case-driven benchmarks for LRPLs. Such tailored evaluation frameworks would enable more nuanced and accurate assessments of LLM performance in generating code that aligns with the specific paradigms, idioms, and common applications of each LRPL. 
An opportunity for researchers arises to develop automatic data augmentation techniques for creating benchmarks, either for a specific domain or techniques that can be used in multiple domains. 

Additionally, to advance the efficacy of LLMs in DSL contexts, it is essential to develop and adopt standardized benchmark suites for a broader range of DSLs. These benchmarks should encompass the unique syntactic and semantic features of each DSL, enabling precise evaluation of LLM capabilities in specialized domains. 
Developing a leaderboard platform that covers benchmarks and models' performance for DSLs and LRPLs can be a step towards this goal. 
An opportunity for researchers and practitioners is to create leaderboards for more popular code generation LRPLs or DSLs, such as Bash, Verilog, PLC, and Ansible, similar to code generation leaderboards like HuggingFace. Though some of the industrial code generation studies are linked on GitHub repositories like Awesome Code Benchmark\footnote{\url{https://github.com/tongye98/Awesome-Code-Benchmark}}, having a collective platform for the developed benchmarks and evaluation metrics is still lacking.
Further research opportunities are developing translation and execution techniques for related domains in DSLs, such as the ones used for Verilog and RTL~\cite{ID8paper}, which facilitates developing benchmark datasets.
To maximize impact, such benchmarks should adopt best practices established in HRPLs, including well-defined tasks, standardized test splits, and clear evaluation metrics. Additionally, fostering community involvement through open contributions and releasing versioned datasets can further promote adoption and iterative improvements.

\paragraph{\textbf{Opportunity \#3: Develop Synthetic, Interactive and Meta-Resourced Frameworks for Preparing LRPLs and DSLs Data}}
The challenges for data preparation lead to several research opportunities. The current approaches for data preparation are conducted in various ways. Creating a standardized data collection pipeline for popular DSLs and LRPLs can enable replicability of the approaches and pave the path for data collection in similar domains. Similarly, having open-sourced and standard pipelines for feedback-based synthesis and error-driven correction could be a valuable approach transferable to other domains.
Moreover, designing interactive semi-automated or human-in-the-loop curation frameworks that involve minimal but targeted human supervision (e.g., via active learning, few-shot sampling) can assist experts in selecting, templating, or annotating DSL code.
Another practical opportunity for data preparation in LRPLs and DSLs is creating meta-resources that document best practices, licensing constraints, and structural requirements. Having a central platform for such metadata for languages that are studied more, e.g., Verilog, could be a first step towards this goal. Having standard curation protocols and taxonomies as part of this meta-resource can increase reproducibility and comparability across DSLs.

Finally, recent advancements in LLMs themselves offer promising avenues for dataset expansion through synthetic data generation. We covered research works using synthetic datasets in Section~\ref{sec:rq3}. 
However, techniques such as self-instruct \citep{wang2023selfinstructaligninglanguagemodels} and Evol-Instruct~\citep{luo2023wizardcoderempoweringcodelarge}, and newer techniques like agentic workflows can be adapted to generate high-quality, diverse examples of LRPL and DSL code. 
These synthetic datasets, when carefully curated and validated, can complement naturally sourced data, potentially addressing gaps in coverage and enhancing the overall quality of training data. 
Evaluating the efficacy of combining curated real-world data with synthetically generated examples, assessing the impact on model performance, and generalization capabilities are still needed.
However, these techniques need to be adapted to the domain specifications of LRPLs and especially DSLs. Another research path can be developing techniques to adapt approaches like agentic workflows for data synthesis, data curation, and even reducing the manual evaluation in LRPLs and DSLs.

\subsection{Performance-Enhancing Techniques}
\paragraph{\textbf{Methodological Challenges for Code Generation With Limited Data and Domain Sensitivity}}
DSLs face unique challenges for code generation with LLMs. Their limited data and domain-specific grammar, syntax, and schema is unseen by LLMs. Additionally, tools like advanced parsers are not common for DSLs and can provide incomplete constraints; instead, specialized tools or human guidance are used in providing feedback to the model. Limited samples and sensitivity to the prompts are among other challenges~\cite{IDsnow42paper}. Assessing the functional equivalence of the code and making them scalable is another challenge mentioned in the studied works~\cite{IDsnow38paper}. Therefore, techniques like fine-tuning, in-context learning, or prompting might not work for DSLs and should be adapted to the specialized domain or the language under study. 
Depending on domain expertise and specific toolchains for each language, adds a burden to develop approaches for code generation and its evaluation for LRPL and specifically for DSLs. For example, LLM-as-a-judge is used in many of the recent code generation and agentic workflows~\cite{jiang2024survey}. However, as the underlying LLMs are not capable of generating DSLs, adding one step further to use them as judges might not be directly applicable.

\paragraph{\textbf{Opportunity \#4: Develop Specialized and Semantically Equivalent Code Generation with Bridged Intermediate Representations}}
The above challenges lead to opportunities for researchers to develop new techniques, apart from the ones used in general code generation. The work of \citet{mora2024syntheticNIPS} introduces synthetic programming elicitation and compilation, where an intermediate language is designed that is natural to LLMs, and is automatically compiled to a target very low-resource programming language. Similarly, researchers can focus on combining user specification and formal method techniques to improve the code generation capability for LRPLs and DSLs. One recent example is a proposed pipeline to improve the semantic correctness of a formal modeling and verification language, UCLID5, with specification at variable- and line-coverage~\cite{chaudhary2025BerkeleyThesis}. These two recent works also incorporate automated repair of the generated code, which can assist in generating code in iterations, similar to general code generation. However, the repair process might require certain techniques, such as an intermediate language as used in~\cite{mora2024syntheticNIPS}. Yet, combining these techniques in an agentic framework, along with connecting to knowledge bases and specialized tools and verifiers used for LRPLs and DSLs, could be an opportunity for future research.

The semantic alignment and equivalence in DSLs is an important perspective, as we observed in the evaluation metrics. Developing formal equivalence checkers, using formal methods, and symbolic execution or theorem proving can be used for DSLs with mathematical semantics, new directions for code generation or evaluation metrics. \citet{chaudhary2025BerkeleyThesis} also discusses using structural specifications that can be used to capture user intent for code generation.
Another interesting research direction can be developing new techniques that help in cross-lingual transfer learning, from HRPLs to LRPLs, beyond the studied works in this survey.

Developing an intermediate language representation and code translation can also be a promising technique that, in combination with the reasoning abilities of LLMs and proof techniques, can be used for dataset creation or generating code in DSLs. A recent study for code transpilation uses this approach for Java into MapReduce implementations written using the Apache Spark for distributed computing, C into operators of programmable switch devices for network packet processing, C++ programs into TACO’s API (a tensor processing compiler), and C++ programs to a tensor processing IR (TensIR)~\cite{bhatia2024verified}. 
Though intermediate languages and representations are applied in some studies~\cite{IDsnow7paper}, exploring them in other LRPLs and DSLs could be a future research direction.

\section{Threats to Validity}\label{sec:threats}

A primary threat to the external validity of our study stems from our database selection process. We carefully selected ArXiv, IEEE Xplore, Web of Science, and ACM Digital Library for their comprehensive coverage and relevance to the field of Large Language Models and code generation.
To mitigate the limitation due to database choice, we employed additional strategies such as backward and forward snowballing, including those using Google Scholar, from our core set of papers to capture any significant works that may have been missed in our primary database search.
We also excluded SQL but acknowledge the potential for future research to explore synergies between NL2SQL techniques and approaches for LRPL/DSL.
To further enhance the coverage of our study, we extended our search in summer 2025 and included 5 additional papers published in a number of AI, NLP, and SE venues. For this extension, though, due to restricted time and resources, we did not cover initial resources thoroughly; thus, we might have missed some published papers (e.g., Arxiv).

Based on our exploration of the papers to answer each of the RQs, we noted that there is even a difference among LRPLs and DSLs. The domain-specific languages still face more challenges compared to LRPLs, due to their specialty of the domain or their narrow applications. Though we have tried to separate the techniques used for LRPLs and DSLs when possible, this distinction is an important note to consider by the readers when interpreting the results in this paper. 

\section{Conclusion} \label{sec:conclusion}

We conducted a systematic literature review, filtering over $27,000$ papers to investigate the usage of LLMs for code generation in low-resource and domain-specific programming languages.
This research fills the gap in the literature by focusing on a variety of LRPLs and DSLs and their large community of developers to understand i) metrics and benchmarks used, ii) methodologies proposed by researchers to improve the code generation capability of LLMs for these languages, and iii) strategies used to curate and pre-process the datasets in LRPL and DSL domains. 
Our findings show that the studies addressing code generation in these domains have increased recently. 
While there are multiple benchmarks in high-resource languages, most LRPLs and DSLs lack a benchmark for evaluation. 
Several metrics and strategies are proposed in the literature to evaluate and improve the results. 
However, there are several constraints researchers face for code generation in LRPLs and DSLs.
Based on our findings, we further discussed the challenges for the studied criteria (i.e., evaluation metrics and benchmarks, improvement methodologies, and data collection techniques). These challenges are more pronounced for DSLs, based on their special domain needs and required expertise. Additionally, we discussed various related research opportunities.
This research can pave the path for a comprehensive overview of the existing approaches while opening up new directions for future research, improving the code generation in LRPLs and DSLs.

\label{section:conclusion}



\begin{acks}

This research is supported by a grant from the Natural Sciences and Engineering Research Council of Canada RGPIN-2019-05175. 

\end{acks}

\bibliographystyle{ACM-Reference-Format}
\bibliography{References}


\end{document}